\documentclass[british]{elsarticle}

\usepackage{lmodern}
\usepackage[T1]{fontenc}
\usepackage{babel}
\usepackage{array}
\usepackage{verbatim}
\usepackage{longtable}
\usepackage{url}
\usepackage{multirow}
\usepackage{rotfloat}
\usepackage{tabularx}
\usepackage{ragged2e}
\usepackage{standalone}
\usepackage{subcaption}
\usepackage{xltabular}

\makeatletter

\providecommand{\tabularnewline}{\\}

\usepackage{etoolbox}
\usepackage{morefloats}
\usepackage{hyperref}
\usepackage{tikz}
\usetikzlibrary{arrows,shapes,positioning,shadows,trees,calc,chains}
\usepackage{fullpage}
\usepackage{pdflscape}
\usepackage{listings}
\usepackage[scaled=0.85]{beramono}
\lstdefinelanguage{AnB}{
    keywords={Protocol,Knowledge,Types,Actions,Goals,secret,between,share,weakly,authenticates,on,Agent,Number,Certified,Function,SymmetricKey,PublicKey},
    sensitive=true,
	tabsize=4
}

\usepackage{nomencl}
\makenomenclature

\extrafloats{100}
\usepackage{versions}

\includeversion{LONG}
\excludeversion{SHORT}

\makeatletter
\def\ps@pprintTitle{%
	\let\@oddhead\@empty
	\let\@evenhead\@empty
	\let\@oddfoot\@empty
	\let\@evenfoot\@oddfoot
}
\makeatother
\begin{document}
	
	\title{Bridging the Gap: A Survey and Classification of\\ Research-informed Ethical Hacking Tools\footnote{This is the extended version of the published paper \cite{Modesti2024EH} is available at \url{https://doi.org/10.3390/jcp4030021}.}}

	\author[]{Paolo Modesti}
	\ead{p.modesti@tees.ac.uk}
	\author[]{Lewis Golightly}
	\ead{l.golightly@tees.ac.uk}
	\author[]{Louis Holmes}
	\ead{b1445121@tees.ac.uk}
	\author[]{Chidimma Opara}
	\ead{c.opara@tees.ac.uk}
	\author[]{Marco Moscini}
	\ead{m.moscini@tees.ac.uk}

	\address{Teesside University, Middlesbrough, United Kingdom}
	
	\begin{abstract}
	The majority of Ethical Hacking (EH) tools utilised in penetration testing are developed by practitioners within the industry or underground communities. Similarly, academic researchers have also contributed to developing security tools. However, there appears to be limited awareness among practitioners of academic contributions in this domain, creating a significant gap between industry and academia's contributions to EH tools. This research paper aims to survey the current state of EH academic research, primarily focusing on research-informed security tools. We categorise these tools into process-based frameworks (such as PTES and Mitre ATT\&CK) and knowledge-based frameworks (such as CyBOK and ACM CCS). This classification provides a comprehensive overview of novel, research-informed tools, considering their functionality and application areas.  The analysis covers licensing, release dates, source code availability, development activity, and peer review status, providing valuable insights into the current state of research in this field. 
		
	\end{abstract}
	
		\begin{keyword}
			Ethical Hacking; Tools and Techniques; Research-informed, Classification; PTES; Mitre ATT\&CK; CyBOK; ACM CCS
		\end{keyword}

	\maketitle

\section{Introduction\label{sec:Introduction}}

In the domain of Ethical Hacking (EH), developing innovative tools is essential to tackle emerging threats and vulnerabilities. Ethical Hacking tools are designed mainly by industry practitioners, occasionally by underground communities~\cite{duque2020investigating}, and sometimes even by state actors~\cite{doi:10.1080/13523260.2023.2216112}. However, even experienced security developers may overlook critical requirements for such applications. An intriguing example is provided by Valenza et al.~\cite{valenza2020never}, challenging the conventional belief that remote scanning carries negligible risk. Their methodology, which transformed the scanning system into a target for counterattacks, revealed vulnerabilities in widely deployed tools, including Metasploit Pro. Overall, the researchers identified weaknesses in 36 out of 78 scanning applications.

The existing divide between industry and academia in developing EH tools reflects differing goals and approaches, highlighting a significant awareness gap. Industry practitioners are often insufficiently informed about the outcomes and insights generated by academic research in this field. Driven by immediate operational requirements, the industry tends to favour established tools and practices that promptly address real-time threats. However, this emphasis on practical application can result in a lack of awareness regarding significant academic contributions, such as novel methodologies and solutions for emerging threats or advancements in theoretical frameworks. Consequently, research findings may remain underutilised by industry practitioners.

One way to bridge the gap between industry and academia in developing EH tools is by producing in-depth survey papers that detail the tools created by both communities. However, existing surveys primarily assess and compare tools used by industry practitioners, with only occasional consideration of research-informed tools \cite{denis2016penetration,Yaacoub2023,aarya2018web}. This focus overlooks the innovative contributions from the research community.

Additionally, the quantity and breadth of tools reviewed by state-of-the-art surveys in EH tools are limited. For instance, the work by Altulaihan et al. \cite{Altulaihan} covered 15 papers for web application penetration testing, while Yaacoub et al. \cite{Yaacoub2023} reviewed 13 tools specifically applied to IoT. This limited scope restricts the comprehensive evaluation of EH tools.
Moreover, existing surveys that classify EH methodologies or frameworks compare existing frameworks such as PTES or other industry methodologies like the Information Systems Security Assessment Framework (ISAF) \cite{shanley2015selection}. However, they do not discuss the specific tools that fall under each category. This narrow focus fails to provide a holistic view of the EH tools landscape.

\subparagraph*{Research Contributions}
In light of these limitations, this paper makes two significant contributions:
		\begin{enumerate}
			\item \textit{Survey of Research-Informed EH Tools}: this study surveys 100 research-informed EH tools developed in the last decade. It highlights key areas such as licensing, release dates, source code availability, development activity level, and peer review status. This analysis aims to provide insights into the state-of-the-art EH tools developed by the research community.
			\item \textit{Alignment with Recognised Frameworks}: this study categorises the tools into \textit{process-based} frameworks, such as the \textit{Penetration Testing Execution Standard} (PTES) \cite{ptes_2014}, and the \textit{Mitre ATT\&CK framework} \cite{strom2018mitre} and \textit{knowledge-based} frameworks like the \textit{National Cyber Security Centre's Cyber Security Body Of Knowledge} (CyBOK) \cite{CyBOK} and the \textit{Association for Computing Machinery's Computing Classification System} (ACM CCS) \cite{Rous2012}. Combining these four classifications offers an informative view of the landscape of novel and research-informed ethical tools, their functionality, and application domain for the benefit of scholars, researchers, and practitioners.			
		\end{enumerate}
This comprehensive approach not only bridges the gap between industry and academia but also ensures that Ethical Hacking tools evolve in tandem with the ever-changing cyber threat landscape.

\subparagraph*{Outline of the Paper}

Section \ref{sec:Background} introduces the background of EH and the methodologies used by practitioners, Section \ref{sec:Methodology} presents our research methodology, and Section \ref{sec:Classification-criteria} discusses the classification criteria applied to the EH tools.
 Section \ref{sec:Classification} discusses the tool categorisation into process-based and knowledge-based frameworks. Section \ref{sec:Evaluation} presents the systematic evaluation of research-informed EH tools, while Section \ref{sec:Conclusion-and-Future} concludes the paper.

\section{Background\label{sec:Background}} 
In this section, we discuss the background and fundamentals of Ethical Hacking, including the motivations behind hacking systems and the different motivations of hackers categorised and represented using \textit{hats}. Additionally, we introduce methodologies used in EH.

\subsection{(Unethical) Hacking Landscape and Motivations}

Cyberattacks, intrusion techniques, social engineering, and information manipulation are increasingly becoming more sophisticated, targeting individuals and organisations. The objective of each attack, regardless of its nature, is to circumvent the three primary principles of security:  
\textit{Confidentiality}, \textit{Integrity}, and \textit{Availability} \cite{bishop2018computer}.
There is a wide range of motivations for cyberattacks, and many factors interplay. According to \cite{Yaacoub2021,walker2021ceh,hald2012updated,DBLP:journals/jcis/OliverR22}, the motivations for these attacks can be grouped into:

\begin{itemize}

\item \textit{Economic Gain}: cybercriminals often target individuals, businesses, or organisations to extort money through ransomware \cite{nhsWannaCry} or financial fraud. Financial institutions such as banks and related services can be a target, as in the case of the attack on the Swift international transaction system \cite{swiftAttack}.

\item \textit{Competitive Advantage and Sabotage}: competing companies, state-sponsored actors, and individuals can steal and reveal industrial secrets and intellectual properties to gain a competitive edge and compromise the data integrity and accessibility in businesses. While the WannaCry ransomware was used primarily to extort money from the victims, the attack on the UK  National Health Service (NHS) could have also been done to demonstrate the business complacency and lack of digital transformation \cite{nhsWannaCry}.
\item \textit{Personal Revenge}: cyberattacks driven by personal revenge are often perpetrated by disgruntled insiders or individuals with a vendetta against specific targets. These attacks leverage insider knowledge or access to inflict damage, disrupt operations, or steal sensitive data.

\item \textit{Political}: the attack is carried out as groups of hackers engaged in politics, sponsored-stated hacking teams aiming at damaging specific targets. This includes governmental institutions, political parties, social society organisations and other public subjects. Examples are the alleged interference in the US presidential elections by Russian state-sponsored cyber actors in 2016 \cite{usElection}, and the \textit{Operation Socialist} in 2010--2013 against Belgacom attributed to the UK's GCHQ \cite{Steffens2020}, a case of an attack perpetrated by a NATO member state against another one.
\end{itemize}

The activities described above broadly fall into the category of cybercrime and involve hacking, data theft, identity theft, financial fraud, and malware distribution. However, when cyberattacks are carried out by state-sponsored actors against other nations or entities, they are often called cyber warfare. The distinction can be blurred in some cases, as the direct involvement of government organisations can have surprising ramifications and side effects. 

The \textit{EternalBlue} exploit \cite{liu2022} was developed by the United States National Security Agency (NSA) targeting a vulnerability in Microsoft's Windows operating system, specifically in the Server Message Block (SMB) protocol. The NSA utilised the exploit for years without reporting the vulnerability to Microsoft. However, it became widely known when a hacking group called the Shadow Brokers leaked the exploit in April 2017. The most notorious incident involving EternalBlue was the aforementioned  WannaCry attack in May 2017. The exploit allowed the rapid spread of malware across networks, affecting hundreds of thousands of computers in over 150 countries and causing hundreds of millions of USD of damage worldwide.

This demonstrates the potential for unintended consequences and collateral damage as malicious actors can weaponise an offensive security tool developed by a government agency for large-scale cybercrime. This case also highlights the importance of responsible handling and disclosure of vulnerabilities by any entity, including government intelligence agencies.

\subsection{Ethical Hacking}

Ethical Hacking, also known as penetration testing, aims to identify vulnerabilities in computer systems, networks, and software applications before real-world attackers can exploit them. By uncovering weaknesses and providing recommendations for mitigation, EH helps organisations enhance their defences, protect sensitive data, and prevent unauthorised access. Ethical hackers utilise their skills to simulate potential cyberattacks and assess the security of a system. In fact, such specialists essentially utilise the same techniques as cyber attackers, with the important difference being that the system's owner authorises them and agrees on the scope of the penetration testing exercise. As individuals capable of compromising systems, any misuse of their skills is criminally punishable according to the laws of various countries. The Budapest Convention on Cybercrime of 2001 (Article 6) \cite{BudapestConvention}, the EU Directive 2013/40 (Article 7) \cite{EuDirective201340}, and the UK Computer Misuse Act of 1990 (Section 3A) \cite{UkMisuseAct} are some of the legislations that regulate cybersecurity activities in terms of the improper use of personal capabilities, software, and hardware dedicated to unauthorised access to third-party information.

The key issue in determining the legality of hacking activities is avoiding actions that contravene the law. Hacking professionals must prioritise legal compliance to avoid prosecution. While the term \textit{Legal Hacking} may more accurately describe this focus on legality, it is essential to recognise that legality does not always equate to ethical behaviour. Nevertheless, the term \textit{Ethical Hacking} remains widely used, emphasising the importance of both legal compliance and ethical conduct in the profession.

\subsection{Ethical Hackers}

Traditional media often portrays hackers as mysterious figures, typically depicted wearing hoodies in dimly lit rooms, perpetuating a stereotype prevalent in pop culture. Hackers are commonly seen as computer pirates who infiltrate systems for personal or financial gain. However, historical context reveals a more nuanced understanding. According to the classic definition reported by Gehring \cite{internetPublicLife}, hackers enjoy the intellectual challenge of overcoming programming limitations and seeking to extend their capabilities. This definition, prevalent until the 1980s and intertwined with notions of democracy and freedom, has evolved, as discussed by Jaquet-Chiffelle and Loi \cite{jaquet2020ethical}.
\textit{Hats} of different colours are broadly used as a symbolic representation of individuals based on their intentions and actions related to hacking \cite{walker2021ceh}:

\begin{itemize}
	\item \textit{White Hat} (ethical): Embodies the principles of hacker culture by employing technical skills to proactively enhance system security measures. These individuals focus on identifying vulnerabilities and developing defensive strategies to mitigate potential risks.
	\item \textit{Black Hat} (malicious): Represents individuals who maliciously exploit vulnerabilities within systems for personal gain or disruptive purposes. Their actions typically involve unauthorised access, data theft, and system manipulation, often resulting in financial losses or reputational damage for targeted entities.
	\item \textit{Grey Hat} (undecided): Occupies an intermediary role, engaging in activities that blur the line between ethical and malicious hacking. These individuals engage in operations as both Black Hat and White Hat, depending on the circumstances \cite{jaquet2020ethical}.
\end{itemize}

\processifversion{LONG}{
According to \cite{vishnuram2022ethical,hald2012updated}, further categories have been identified based on distinctive motivations and competencies from the original definition of hacker:

\begin{itemize}
	\item \textit{Red Hat}: Individuals dedicated to countering black hat hackers using offensive techniques rather than defensive ones \cite{filiol2021method}.
	\item \textit{Blue Hat}: People employed in companies that test software before release to find bugs \cite{chandrika2014ethical}.
	\item \textit{Green Hat}: A category composed of people new to the hacking world \cite{vishnuram2022ethical}.
	\item \textit{Script Kiddie}: People with few hacking skills who use software already created by real hackers.
	\item \textit{Hacktivist}: People who use hacking to promote political, social, or ideological manifestos \cite{egho2024big}.
	\item \textit{Professional Criminals}: People with professional skills in security who use hacking to steal financial information.
	\item \textit{Internals or Insiders}: People who usually attack their workplace for revenge or financial gain.
\end{itemize}
}

In recent years, cybersecurity and privacy protection have emerged as central themes for all organisations, and professional roles have arisen to address these needs. Penetration testing and malware analysis are among the sought-after roles in the cybersecurity job market, falling under the main category of EH.

\subsection{Ethical Hacking Methodologies}
Penetration Testing takes different forms and can cover various areas. Yaacoub et al. \cite{Yaacoub2021} describe the process of conducting an attack in five main phases:

\begin{itemize}
	\item \textit{Reconnaissance}: The hacker gathers information on systems and users through passive or active techniques. This includes physical methods like social engineering and analysing network packets to identify details such as network configuration, hardware, and security measures.

\item \textit{Scanning}: The hacker searches for vulnerabilities in systems through simulated tests, including identifying open ports, active hosts, and weak firewall configurations. Enumeration is then carried out to gather further information while maintaining an active connection.

\item \textit{Gaining Access}: The hacker attempts to access the system using penetration testing tools and techniques, aiming to bypass security measures.

\item \textit{Maintaining Access}: The hacker establishes backdoors or rootkits to maintain remote access with elevated privileges.

\item \textit{Covering Tracks}: The hacker eliminates evidence that could reveal their identity or traces of the attack.

\end{itemize}

Each phase is complex and crucial for the success of a cyber attack. Due to the unique nature of each system, there are no strict rules for systematically executing an attack or penetration test. However, various frameworks and methodologies have been developed to guide the penetration testing process in planning and executing cyber attack simulations.

These frameworks can be categorized into three main areas: open source, maintained by non-profit organisations or security institutes; industrial/governmental, maintained by government entities such as the National Institute of Standards and Technology (NIST); and proprietary, maintained by private companies and accessible through payment of a usage license.

\subsubsection{PTES}

\textit{The Penetration Testing Execution Standard} \cite{ptes_2014} was created in 2009 by a group of practitioners who developed this framework to provide both businesses and security service providers with a common language for conducting penetration tests. It comprises 7 phases: \textit{Pre-engagement Interactions}, \textit{Intelligence Gathering}, \textit{Threat Modelling}, \textit{Vulnerability Analysis}, \textit{Exploitation}, \textit{Post Exploitation}, and \textit{Reporting}. The methodology is presented in detail in Section \ref{subsec:ptes}.

\subsubsection{Mitre ATT\&CK}
\textit{Mitre Adversarial Tactics, Techniques, and Common Knowledge} \cite{strom2018mitre} is a matrix that describes the behaviour of attackers throughout the life cycle of an operation. It covers \textit{tactics, techniques, and procedures (TTP)} used by threat actors to achieve their objectives. Mitre is a non-profit American company involved in numerous cybersecurity standards and frameworks, such as \textit{CVE (Common Vulnerabilities and Exposures)} \cite{cve} to identify and classify disclosed security vulnerabilities. \textit{CWE (Common Weakness Enumeration)} \cite{CWE} is used as a common language for weakness identification, prevention and mitigation. See Section \ref{subsec:mitre} for the details of this methodology.

\subsubsection{PCI DSS Penetration Testing guidance}

\textit{Payment Card Industry Data Security Standard} (PCI DSS) \cite{pcidss} is a set of security requirements designed to protect payment card information during transactions. Developed by the PCI Security Standards Council, it applies to all organisations that accept, process, store, or transmit payment card data. PCI DSS establishes requirements for data security, network management, application protection, and other measures to prevent credit card fraud. The Penetration Testing Guide \cite{pcidsspt} is divided into four parts: \textit{Penetration Tester Components}, \textit{Qualification of a Penetration Tester}, \textit{Methodology} and \textit{Reporting and Documentation}. Organisations must comply with these requirements to ensure the security of credit card transactions and protect cardholders' sensitive data.

\subsubsection{ISSAF}
The \textit{Information Systems Security Assessment Framework} \cite{issaf} is a standard supported by the Open Information System Security Group (OISSG). It incorporates all possible attack domains, and the main feature is that the penetration testing activity is divided into three phases: \textit{Planning and Preparation}, \textit{Assessment} and \textit{Reporting, Cleanup, and Artefact Destruction}.

\subsubsection{OSSTMM} 

The \textit{Open Source Security Testing Methodology Manual} \cite{osstmm3} is a set of guidelines and procedures for conducting security tests and assessing the security of information systems. Developed by the Institute for Security and Open Methodologies (ISECOM), OSSTMM aims to provide an open-source standardised methodology for cybersecurity professionals. It focuses on testing operational security through five channels: \textit{Human Security}, \textit{Physical Security}, \textit{Wireless Communications}, \textit{Telecommunications}, and \textit{Data Networks}. 

\subsubsection{NIST800-115} 

\textit{NIST800-115} \cite{NIST800} is a methodology published by the National Institute of Standards and Technology (NIST) in the United States. This standard provides detailed guidelines for conducting tests and assessments of information security in computer environments. It covers a wide range of security testing and assessment activities. The standard includes planning, information gathering, vulnerability analysis, test execution, risk assessment, and documentation. It offers practical and detailed recommendations for performing various tests, including tool selection and management of information collected during the testing process. Although published by NIST, the standard is designed to be adopted in both the public and private sectors, providing a flexible framework that can be applied to different environments.%

\subsubsection{OWASP} 

The \textit{Open Worldwide Application Security Project} \cite{owasp} was launched in 2001 as an open-source project that provides guidelines, tools, and methodologies to improve the security of applications, collected into a guide named OTG 4.0 (\textit{Owasp Testing Guide}) \cite{owaspotg}. This document is divided into five parts: \textit{Before development begins}, \textit{During definition and design}, \textit{During development}, \textit{During deployment} and \textit{Maintenance and operations}. By completing these procedures, developers can significantly reduce the risk of data breaches caused by attacks facilitated by poor code quality.

\section{Survey Methodology\label{sec:Methodology}}

\begin{figure}[ht]
	\centering
	\tikzset{
		basic/.style  = {draw, text width=110mm, drop shadow, font=\footnotesize\sffamily, rectangle, node distance = 1.0cm},
		root/.style   = {basic, rounded corners=2pt, thin, align=center, fill=blue!30},
		level 2/.style = {basic, rounded corners=6pt, thin, align=center, fill=red!30,
			text width=7.0em,font=\scriptsize\sffamily,minimum height=10mm},
		level 3/.style = {basic, thin, align=center, fill=yellow!30, text width=7.0em,
			font=\scriptsize\sffamily,minimum height=10mm, node distance = 0.2cm}
	}
	
	\centering
	\begin{tikzpicture}[
		level 1/.style={sibling distance=20mm},
		>=latex
		]

		\begin{scope}[every node/.style={level 2}]
			\node (criteria) [] {Criteria for Inclusion};
			\node (identification) [right=of criteria] {Identification of Papers and Tools};
			\node (categorisation) [right=of identification] {Categorisation of Tools};
		\end{scope}
		
		\begin{scope}[every node/.style={level 3}]
			\node [below=of criteria] (academic) {Academic and research context};
			\node [below=of academic] (peer) {Peer-reviewed research papers};
			\node [below=of peer] (offensive) {Potential for offensive use};
			\node [below=of offensive] (authorship) {Authorship by tool developers};
			\node [below=of authorship] (opensource) {Open source availability};
			
			\node [below=of categorisation] (PTES) {PTES};
			\node [below=of PTES] (MITRE) {Mitre ATT\&CK};
			\node [below=of MITRE] (CyBOK) {CyBOK};
			\node [below=of CyBOK] (ACM) {ACM CCS};
		\end{scope}
		
		\draw[->] (criteria) -- (identification);
		\draw[->] (identification) -- (categorisation);
		
	\end{tikzpicture}
	
	\caption{Survey Methodology}
	\label{fig:methodology}
\end{figure}
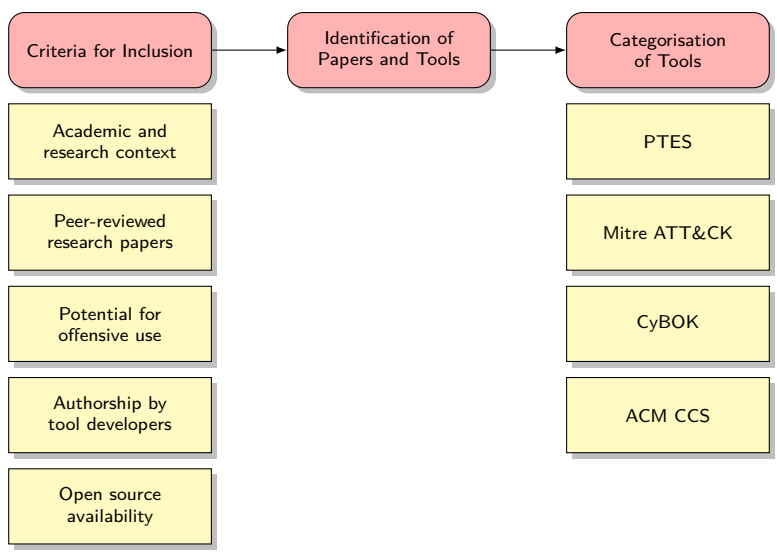

A three-step approach was devised to investigate Ethical Hacking tools developed by the research community over the past decade (Figure \ref{fig:methodology}). First, clear guidelines were established to determine the inclusion of tools in the survey. Second, relevant papers and tools satisfying the above criteria were identified. Finally, these tools were categorised based on established cybersecurity frameworks.

\subsection{Criteria for Inclusion of Ethical Hacking Tools in the Paper}
\label{sub:inclusion-criteria}
This survey established specific conditions to determine the inclusion of state-of-the-art EH tools in the paper. The following criteria were adhered to:

\begin{itemize}
\item \textit{Academic and research context}: The tool has been developed within an academic/research project: this excludes any tools developed primarily as practitioner tools (e.g. they are included in a popular EH distribution, like Kali Linux).
\item \textit{Peer-reviewed research papers}: Each EH tool included in the survey must be published in a peer-reviewed venue. Peer review validates the tool's architecture, functionalities, and relevance.
\item \textit{Potential for offensive use}: The tool has at least the potential to be used in an offensive context even if authors do not state that explicitly, as the tool could have been developed for another purpose (e.g. software testing, supporting software or system development).
\item \textit{Authorship by tool developers}: The survey also requires that the authors of the papers have designed/developed the tool. This criterion ensures credibility and depth of insight, as the creators are directly involved in its conception and development.
\item \textit{Open source availability}: The tool should be open source, and the source code (or distribution package) should be freely available. However, this requirement was relaxed throughout the research as we acknowledge that some tools may not be open-source for various reasons, such as proprietary nature, pending patents, or limited accessibility.
\end{itemize}

\subsection{Collating Research-Informed Ethical Hacking Tools} %
The inclusion of 100 EH tools was driven by the aim of achieving a balance between depth and breadth in our analysis. The selection process was systematic and rigorous, inspired by the PRISMA methodology \cite{moher2015preferred}, with identification, screening and inclusion phases.

Initially, we initiated a collaborative research project involving cohorts of students from the MSc Cybersecurity program at Teesside University (United Kingdom). Despite the absence of a dedicated module on EH in their curriculum, the students showed considerable interest in working within this domain. This project allowed them to integrate EH professionalism with their research interests.

The students' initial submissions yielded over 200 academic references. However, after a careful review process conducted by the authors, approximately 30 tools aligned with the research scope and were included in this paper. Many tools found by the students were excluded for these reasons:

\begin{itemize}
\item Difficulty of the students in distinguishing between research-informed and practitioner tools.
\item Confusion between papers describing the design and implementation of a tool and those describing its application.
\item The approach of identifying tools first and then searching for papers to support the findings leads to the above misconceptions.
\end{itemize}

Following the criteria outlined in Section \ref{sub:inclusion-criteria}, the authors expanded the total count of EH tools to 185. These tools were then resampled, and 100 tools were finally selected. The final selection was based not only on adherence to the criteria in Section \ref{sub:inclusion-criteria} but also on the fact that these tools were not merely applications of existing methodologies, frameworks, or aggregations of practitioner tools.

In fact, among the 85 candidates excluded in the final round, 28 focused on mitigation tools and techniques, 18 on methodologies and frameworks, 14 on the application of practitioner tools, 13 were surveys, 5 addressed socio-technical aspects, 4 were simulation tools, and 3 focused on education.

 For details of the 100 tools surveyed in this paper, see Table \ref{tab:tools-availability}.

Moreover, as discussed in Section \ref{sec:Evaluation}, to include a significant number of tools that could reflect the current state of the art, we had to relax on the criteria of availability of the source code. Therefore, in this survey, we have 41 tools that satisfy all other criteria, but no source code has been published.

\processifversion{LONG}{
For a detailed description of each tool included in this survey, please refer to \ref{sec:Review-of-Research}.}

\subsection{Classification of Identified Ethical Hacking Tools}
In the second phase of the research, the identified EH tools were classified according to established cybersecurity frameworks. This task was undertaken by the authors, who have extensive expertise in EH from years of teaching, research, and professional experience in the field.

\begin{figure}[ht!]
	\centering
	\tikzset{
		basic/.style  = {draw, text width=3cm, drop shadow, font=\footnotesize\sffamily, rectangle, node distance = 0.5cm},
		root/.style   = {basic, rounded corners=2pt, thin, align=center,
			fill=blue!20},
		level 2/.style = {basic, rounded corners=6pt, thin,align=center, fill=red!30},
		level 3/.style = {basic, thin, align=left, fill=yellow!30, text width=7.9em}
	}
	
	\begin{tikzpicture}[
		level 1/.style={sibling distance=40mm},
		edge from parent/.style={->,draw},
		>=latex]
		
		\node[root] {Ethical Hacking Tools}
		child {node[level 2] (c1) {Process Based Frameworks}}
		child {node[level 2] (c2) {Knowledge Based Frameworks}};
		
		\begin{scope}[every node/.style={level 3}]
			\node [below=of c1, xshift=15pt] (c11) {Penetration Testing Execution Standard (PTES)};
			\node [below=of c11] (c12) {Mitre Adversarial Tactics, Techniques, and Common Knowledge (ATT\&CK)};
			
			\node [below=of c2, xshift=15pt] (c21) {NCSC Cyber Security Body of Knowledge (CyBOK)};
			\node [below=of c21] (c22) {ACM Computing Classification System (CCS)};
		\end{scope}
		
		\foreach \value in {1,...,2}
		\draw[->] (c1.195) |- (c1\value.west);
		
		\foreach \value in {1,...,2}
		\draw[->] (c2.195) |- (c2\value.west);
		
	\end{tikzpicture}
	\caption{Classification Criteria applied in this survey}
	\label{fig:EH_Classification}
\end{figure}
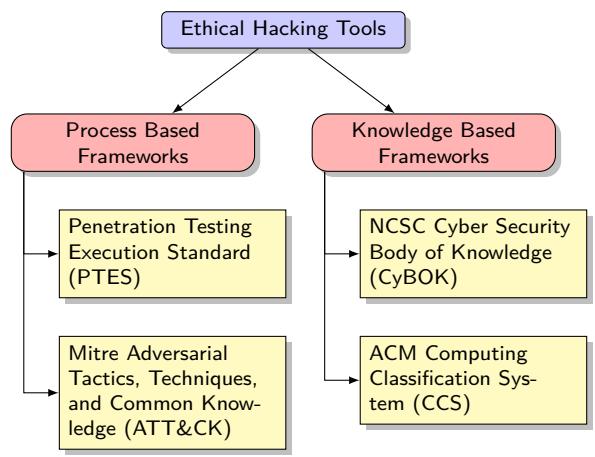

\noindent All 100 identified tools were categorised according to the following classifications (Figure \ref{fig:EH_Classification}):

\begin{enumerate}
	\item Penetration Testing Execution Standard (PTES) \cite{ptes_2014}
	\item Mitre ATT\&CK Framework \cite{strom2018mitre}
	\item NCSC CyBOK \cite{CyBOK} 
	\item ACM Computing Classification System (CCS) \cite{Rous2012}

\end{enumerate}

Incorporating \textbf{process-based} classifications such as the PTES and the Mitre ATT\&CK Frameworks ensures that the survey covers the practical aspects of EH tools. On the other hand, \textbf{knowledge-based} classifications such as NCSC CyBOK and ACM CCS focus on the theoretical and conceptual aspects of computing and cybersecurity domains, thereby exploring the underlying theoretical bases of EH tools.

The next section provides an in-depth discussion of these frameworks.

\section{Cybersecurity Frameworks used for Tools Classification\label{sec:Classification-criteria}}

This section delves into a detailed examination of the cybersecurity frameworks used to categorise the Ethical Hacking tools surveyed in this paper. These frameworks include PTES, MITRE ATT\&CK framework, CyBOK, and ACM CCS).

\subsection{Penetration Testing Execution Standard\label{subsec:ptes}} %

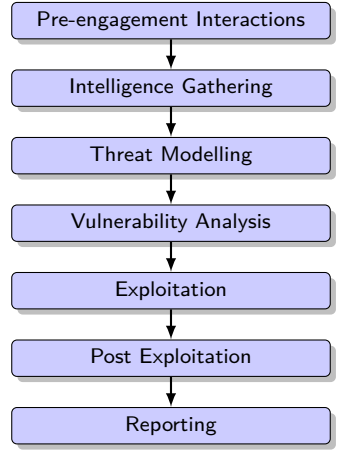
\begin{figure}
	\centering
	\tikzset{
		basic/.style  = {draw, text width=4cm, drop shadow, font=\footnotesize\sffamily, rectangle, node distance = 0.40cm},
		phase/.style   = {basic, rounded corners=2pt, thin, align=center,
		fill=blue!20},
		arrow/.style = {->, >=latex, thick}
	}
	
	\begin{tikzpicture}
		
		\node[phase] (1) {Pre-engagement Interactions};
		\node[phase, below=of 1] (2) {Intelligence Gathering};
		\node[phase, below=of 2] (3) {Threat Modelling};
		\node[phase, below=of 3] (4) {Vulnerability Analysis};
		\node[phase, below=of 4] (5) {Exploitation};
		\node[phase, below=of 5] (6) {Post Exploitation};
		\node[phase, below=of 6] (7) {Reporting};
		
		\draw[arrow] (1) -- (2);
		\draw[arrow] (2) -- (3);
		\draw[arrow] (3) -- (4);
		\draw[arrow] (4) -- (5);
		\draw[arrow] (5) -- (6);
		\draw[arrow] (6) -- (7);
		
	\end{tikzpicture}
	\caption{Penetration Testing phases according to PTES methodology}
	\label{fig:PTES_Phases}
\end{figure}

\textit{PTES \cite{ptes_2014}} is a standardised methodology for planning, executing, and reporting security tests and it is widely used within the cybersecurity industry as one of the most significant standards for conducting penetration tests.
PTES was proposed by a group of penetration testers and security professionals to provide guidance and best practices for conducting effective penetration tests within legal and ethical boundaries. It consists of seven phases (Figure \ref{fig:PTES_Phases}):

\begin{enumerate}
	\item \textit{Pre-engagement Interactions}: In this phase, the scope and rules of engagement are defined through an agreement between the pen-testing team and the system's owner. The system's owner must provide permissions and authorizations, and communication lines must be established between the testers and the target organization.

\item \textit{Intelligence Gathering}: Information about the target organization or system is collected using techniques such as open-source intelligence (OSINT) gathering, reconnaissance, and network scanning. Active and passive information gathering methods are distinguished based on direct interaction with the target system.

\item \textit{Threat Modelling}: This phase identifies potential vulnerabilities and threats specific to the target organization or system. It involves analysing collected information, understanding infrastructure and architecture, prioritizing attack vectors, and assigning risks to threats to inform vulnerability mitigation.

\item \textit{Vulnerability Analysis}: Vulnerabilities and weaknesses in the target's systems and applications are identified and assessed, typically using classification systems like the \textit{Common Vulnerability Scoring System} (CVSS). Manual and automated testing, configuration analysis, and examination of insecure application design may be involved.

\item \textit{Exploitation}: Vulnerabilities previously identified are exploited to compromise the target system, gain unauthorized access, or execute malicious activities. The goal is to demonstrate the impact of vulnerabilities and their potential exploitation, bypassing security mechanisms.

\item \textit{Post-Exploitation}: After successful exploitation, the focus shifts to determining the value of the compromised system, maintaining access, escalating privileges, and pivoting to other systems within the network. This simulates an attacker's post-compromise activities, considering the data's importance and the advantage provided for further attacks.

\item \textit{Reporting}: the final phase involves documenting the findings, including identified vulnerabilities, their impact, and recommendations for remediation. The report should be clear, concise, and actionable for the target organization, tailored to various audiences ranging from senior managers to technical staff.
\end{enumerate}

The PTES methodology can be applied to various systems to assess their security, including networks and critical infrastructures \cite{astrida2022analysis}.

\subsection{Mitre Att\&ck Framework\label{subsec:mitre}} %

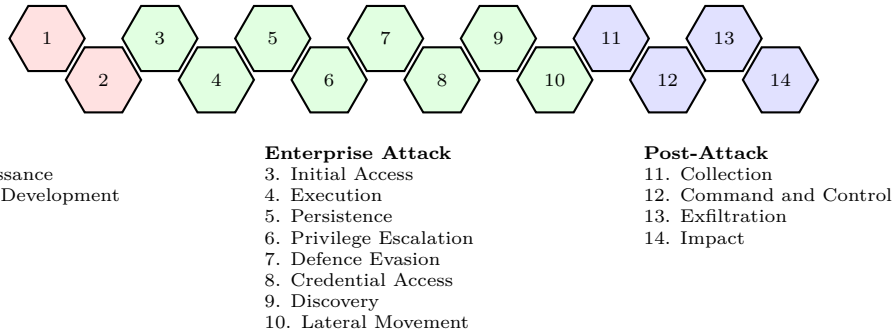
\begin{figure}
\centering
\begin{tikzpicture}[hexa/.style= {shape=regular polygon, regular polygon sides=6, draw, minimum size=1cm, inner sep=0, outer sep=0, rotate=0, thick, align=center}]

\definecolor{preattack}{RGB}{255, 225, 225}
\definecolor{enterpriseattack}{RGB}{225, 255, 225}
\definecolor{postattack}{RGB}{225, 225, 255}

\foreach \n/\l/\c in {1/Reconnaissance/preattack, 3/Initial Access/enterpriseattack, 5/Persistence/enterpriseattack, 7/Defense Evasion/enterpriseattack, 9/Discovery/enterpriseattack, 11/Collection/postattack, 13/Exfiltration/postattack, 2/Resource Development/preattack, 4/Execution/enterpriseattack, 6/Privilege Escalation/enterpriseattack, 8/Credential Access/enterpriseattack, 10/Lateral Movement/enterpriseattack, 12/Command and Control (C\&C)/postattack, 14/Impact/postattack}{\scriptsize
    \ifodd\n
        \node[hexa, fill=\c] (hex\n) at ({\n*1.5*cos(60)},0) {\n};
\else
\node[hexa, fill=\c] (hex\n) at ({(\n-1)*1.5*cos(60)+0.75},-0.55) {\n}; %
\fi
}

\node[align=left, anchor=north, yshift=-0.3cm] at (hex8.south) {
\scriptsize
\begin{tabular}{p{0.28\linewidth}p{0.28\linewidth}p{0.28\linewidth}}
\textbf{Pre-attack} & \textbf{Enterprise Attack} & \textbf{Post-Attack}\\
1. Reconnaissance & 3. Initial Access & 11. Collection \\
2. Resource Development & 4. Execution & 12. Command and Control\\
& 5. Persistence & 13. Exfiltration \\
& 6. Privilege Escalation & 14. Impact \\
& 7. Defence Evasion & \\
& 8. Credential Access & \\
& 9. Discovery & \\
& 10. Lateral Movement & \\
\end{tabular}
    };

\end{tikzpicture}
\caption{Mitre ATT\&CK Framework: Phases and Tactics (inspired by \cite{duque2020investigating})}
\label{fig:Mitre}
\end{figure}

\textit{Mitre ATT\&CK (Adversarial Tactics, Techniques, and Common Knowledge)} \cite{strom2018mitre} is a framework that categorises the tactics, techniques, and procedures (TTPs) used by real-world threat actors during cyberattacks (Figure \ref{fig:Mitre}). It provides a standardised and comprehensive mapping of the various stages of an attack and consists of a matrix that categorises adversary behaviours across different stages of the attack lifecycle. %
Within each tactic, specific techniques and procedures are listed, which outline the specific actions and methods used by adversaries to accomplish their objectives.

Mitre ATT\&CK consists of fourteen phases \cite{rajesh2022analysis}:

\begin{enumerate}
	\item \textit{Reconnaissance}: Collecting information on the target to plan and execute attacks. Methods include: \textit{Active Scanning}, \textit{Passive Scanning}, \textit{Social Engineering} and \textit{OSINT}.	
	
	\item \textit{Resource Development}: Acquiring resources required for further exploitation and maintaining access. Methods include: \textit{Developing Tools} and \textit{Developing and Executing Malware}.
	
	\item \textit{Initial Access}: Techniques performed to gain access to the target environment. Methods to achieve this include: \textit{Spear-Phishing}, \textit{Exploiting Vulnerabilities} and \textit{Stolen Credentials}.
	
	\item \textit{Execution}: Techniques performed executing Malicious Software (Malware) on a target system. Methods include: \textit{Executing Binaries}, \textit{Scripts} and \textit{System Tools}.
		
	\item \textit{Persistence}: Techniques performed around maintaining system access over a significant period of time. Methods include: \textit{Backdoor Creation} and \textit{Scheduled Tasks}.
	
	\item \textit{Privilege Escalation}: Increasing the Access Control levels in the compromised environment. Methods include: \textit{Vulnerability Exploitation}, \textit{Configuration Manipulation} and \textit{Credential Theft}.
	
	\item \textit{Defence Evasion}: Techniques to avoid detection or target defensive mechanisms. Methods include: \textit{Anti-Virus Evasion},\textit{Obfuscation} and \textit{Living-off-the-land Techniques}.
	
	\item \textit{Credential Access}: Techniques for stealing credentials for unauthorised access. Methods include: \textit{Credential Dumping}, \textit{Keylogging} and \textit{Brute-Force Attacks}.
	
	\item \textit{Discovery}: Techniques for identifying information about the target system. Methods include: \textit{Network Scanning}, \textit{System Enumeration} and \textit{Querying Systems}.
	
	\item \textit{Lateral Movement}: Methods for moving through the network for accessing additional systems by using: \textit{RDP}, \textit{Trust Relationships} and \textit{Lateral Tool Transfer}.
	
	\item \textit{Collection}: Acquiring and consolidating target system information. Methods include: \textit{Data Mining}, \textit{Scraping} and \textit{Information Capture}.
	
	\item \textit{Command and Control}: Creating and Maintaining communication channels between the attacker and compromised systems. Methods include: \textit{Command and Control (C2)},  \textit{Covert Channels} and \textit{Network Protocols}.
	
	\item \textit{Exfiltration}: Techniques around the unauthorised data transfer external to the target environment. Methods include: \textit{Network Data Exfiltration}, \textit{Encryption Channels} and \textit{Scheduled Transfer}.
	
	\item \textit{Impact}: Achieving the desired outcome or effect could involve damaging a target. Methods include: \textit{Destroying Data}, \textit{System Operation Disruption} and \textit{Deploying Malware}.
\end{enumerate}

The Mitre ATT\&CK framework can be applied to broad kinds of targets, including, financial systems \cite{georgiadou2021assessing}, healthcare \cite{messinis2024enhancing} and Industrial Control Systems (ICS) \cite{alexander2020mitre}.

\subsection{NCSC CyBOK} %

\begin{figure}[ht]
	\centering
	\tikzset{
		basic/.style  = {draw, text width=110mm, drop shadow, font=\footnotesize\sffamily, rectangle, node distance = 1.0cm},
		root/.style   = {basic, rounded corners=2pt, thin, align=center, fill=blue!30},
		level 2/.style = {basic, rounded corners=6pt, thin, align=center, fill=red!30,
			text width=4.4em,font=\scriptsize\sffamily,minimum height=15mm},
		level 3/.style = {basic, thin, align=center, fill=yellow!30, text width=4.4em,
			font=\scriptsize\sffamily,minimum height=10mm, node distance = 0.2cm}
	}
	
	\centering
	\begin{tikzpicture}
		[
		level 1/.style={sibling distance=19mm},
		edge from parent/.style={},
		>=latex
		]
		
		\node[root] (cybok) {CyBOK \\ Knowledge Areas}
		
		child {node[level 2] (intro) {Introductory Concepts}}
		child {node[level 2] (hoa) {Human,\\ Organisational \& Regulatory Aspects}}
		child {node[level 2] (ad) {Attacks \& Defences}}
		child {node[level 2] (syssec) {Systems Security}}
		child {node[level 2] (softplat) {Software and Platform Security}}
		child {node[level 2] (infrasec) {Infrastructure Security}};
		
		\begin{scope}[every node/.style={level 3}]
			
			\node [below=of hoa] (risk) {Risk Mgmt \& Governance};
			\node [below=of risk] (law) {Law and Regulation};
			\node [below=of law] (human) {Human Factors};
			\node [below=of human] (privacy) {Privacy and Online Rights};
			\node [below=of ad] (malware) {Malware and Attack Technologies};
			\node [below=of malware] (adv) {Adversarial Behaviours};
			\node [below=of adv] (secops) {Security Operations and Incident Management};
			\node [below=of secops] (forensics) {Forensics};
			\node [below=of syssec] (crypto) {Cryptography};
			\node [below=of crypto] (os) {Operating Systems and Virtualisation Security};
			\node [below=of os] (distsys) {Distributed Systems Security};
			\node [below=of distsys] (formal) {Formal Methods for Security};
			\node [below=of formal] (auth) {Authentication, Authorisation, and Accountability};
			\node [below=of softplat] (softsec) {Software Security};
			\node [below=of softsec] (webmob) {Web and Mobile Security};
			\node [below=of webmob] (sslc) {Secure Software Lifecycle};
			\node [below=of infrasec] (appcrypto) {Applied\\Cryptography};
			\node [below=of appcrypto] (network) {Network Security};
			\node [below=of network] (hardware) {Hardware Security};
			\node [below=of hardware] (cps) {Cyber Physical Systems};
			\node [below=of cps] (physlayer) {Physical Layer and Telecom Security};
			
		\end{scope}

	\end{tikzpicture}
	
	\caption{Cyber Security Body Of Knowledge (CyBOK) Knowledge Areas}
	\label{fig:Cybok}
\end{figure}
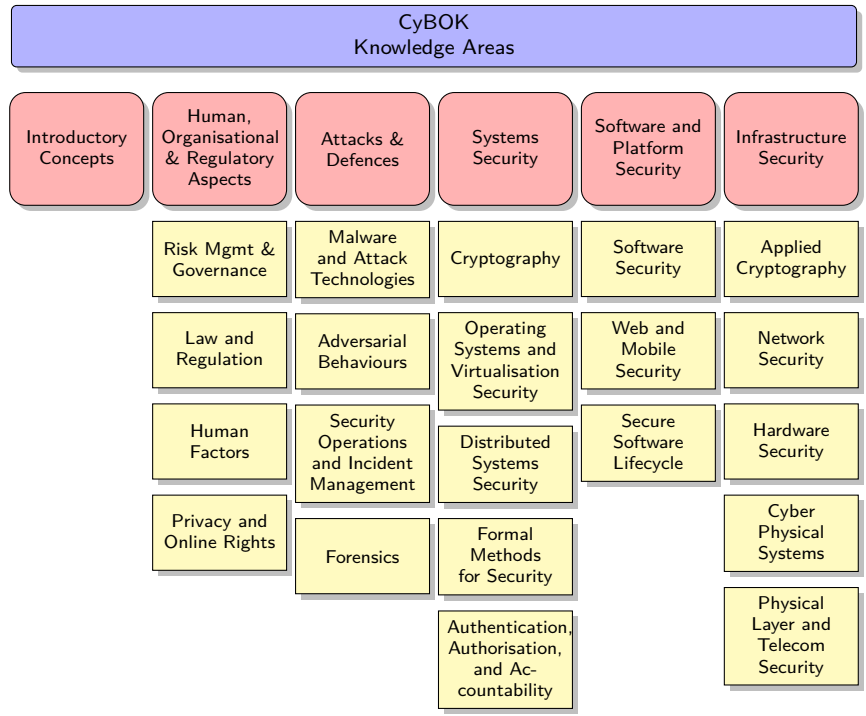

The \textit{Cyber Security Body of Knowledge} (CyBOK) \cite{CyBOK} is a project developed by the United Kingdom's National Cyber Security Centre (NCSC), a child agency of the Government Communications Headquarters (GCHQ), in collaboration with academia, industry, and government partners. CyBOK aims to provide a comprehensive and authoritative reference for the knowledge, skills, and competencies on which various educational programmes and job descriptions may be aligned. The CyBOK is divided into 21 top-level \textit{Knowledge Areas} (KAs) and five broad categories, along with the introductory concepts, as shown in Figure \ref{fig:Cybok}. These categories, while orthogonal, are not entirely separate from each other, reflecting the interdisciplinary nature of cyber security. 

In summary, Knowledge Areas in CyBOK version 1.1 \cite{cybok-ssl} are organised as follows:

	\begin{enumerate}
		\item Introductory Concepts: \textit{Introduction to CyBOK}.
		\item Human, Organisational \& Regulatory Aspects: (a) \textit{Risk Management and Governance}, (b) \textit{Law and Regulation}, (c) \textit{Human Factors} and (d) \textit{Privacy and Online Rights}.
		\item Attacks \& Defences: (a) \textit{Malware and Attack Technologies}, (b) \textit{Adversarial Behaviours}, (c) \textit{Security Operations and Incident Management} and  (d) \textit{Forensics}. 
		\item Systems Security:  (a) \textit{Cryptography},  (b) \textit{Operating Systems and Virtualisation Security},  (c)  \textit{Distributed Systems Security},  (d) \textit{Formal Methods for Security} and  (e)  \textit{Authentication, Authorisation, and Accountability}.
		\item Software and Platform Security:  (a)  \textit{Software Security},  (b) \textit{Web and Mobile Security} and  (c) \textit{Secure Software Lifecycle}.
		\item Infrastructure Security:  (a) \textit{Applied Cryptography},  (b) \textit{Network Security},  (c)  \textit{Hardware Security},  (d) \textit{Cyber Physical Systems} and  (e) \textit{Physical Layer and Telecommunications Security}.
	\end{enumerate}

The CyBOK can be applied in various ways to enhance the security posture of businesses. It can be used to assess skills, develop workforces, curriculum design in higher education, and certification programs \cite{nautiyal2020uk, attwood2023exploring}.

\subsection{ACM Computing Classification System (CCS)}%

The \textit{Computing Classification System} (CCS) \cite{Rous2012} is a taxonomy developed by the Association for Computing Machinery (ACM). It is designed to categorise and organise the various areas of research and practice within the field of computing. The CCS provides a hierarchical structure that classifies research papers, articles, conference proceedings, and other scholarly works in computing. Authors use appropriate CSS categories when submitting publication manuscripts to journals and conferences for classification and organisation. This system helps to locate relevant literature, understand the structure of the field, and facilitate communication within the computing community. The root concepts of ACM CSS include \cite{Rous2012}:

\begin{enumerate}
    \item \textit{General and Reference:} Fundamental concepts and cross-disciplinary topics in computing.
    \item \textit{Hardware:} Physical components and architecture of computing systems.
    \item \textit{Computer Systems Organisation:} Organisation and structure of computer systems.
    \item \textit{Networks:} Communication and connectivity in computing environments.
    \item \textit{Software and Its Engineering:} Development, design, and maintenance of software systems.
    \item \textit{Theory of Computation:} Mathematical and theoretical aspects of computation.
    \item \textit{Mathematics of Computing:} Mathematical foundations of algorithms and computation.
    \item \textit{Information Systems:} Management, retrieval, and processing of information in computing.
    \item \textit{Security and Privacy:} Protection of computing systems and data privacy concerns.
    \item \textit{Human-Centred Computing:} Interaction between humans and computing technologies.
    \item \textit{Computing Methodologies:} Methodological approaches in computing research and practice.
    \item \textit{Applied Computing:} Application of computing techniques in various domains.
    \item \textit{Social and Professional Topics:} Ethical, legal, and social aspects of computing.
\end{enumerate}

\section{Classification\label{sec:Classification}} %

The complete list of identified tools is available in Table \ref{tab:tools-availability}. We include the availability of source code, the license type, the source code repository and the year of publication.
For readability reasons, we put the other classification tables in \ref{app:classification}.
\processifversion{LONG}{The description of individual tools is presented in \ref{sec:Review-of-Research}.}

\footnotesize
\begin{longtable}{|>{\RaggedRight}p{0.25\textwidth}|>{\centering}p{0.04\textwidth}|>{\RaggedRight}p{0.15\textwidth}|>{\RaggedRight}p{0.45\textwidth}|}	

\hline
\normalsize
\textit{Tool Name} & \normalsize \textit{Year} & \normalsize \textit{License Type} & \normalsize \textit{Source Code Repository} \tabularnewline

\hline
\endhead

ADaMs \cite{pasquini2021reducing} & 2021  & MIT License & \url{https://github.com/TheAdamProject/adams} \tabularnewline
AIBugHunter \cite{Fu2023} & 2023  & MIT License & \url{https://github.com/awsm-research/aibughunter} \tabularnewline
ARMONY \cite{chen2013armory} & 2013  & Not Available & Not Available \tabularnewline
Autosploit \cite{Moscovich2020} & 2020  & Not Available & Not Available \tabularnewline
AVAIN \cite{Egert2019} & 2019  & MIT License & \url{https://github.com/ra1nb0rN/Avain} \tabularnewline
Bbuzz \cite{DBLP:conf/milcom/BlumbergsV17} & 2017  & MIT License & \url{https://github.com/lockout/Bbuzz} \tabularnewline
Black Ostrich \cite{Eriksson2023} & 2023  & Not Available & Not Available \tabularnewline
Black Widow \cite{DBLP:conf/sp/ErikssonPS21} & 2021  & Not Specified & \url{https://github.com/SecuringWeb/BlackWidow} \tabularnewline
Bleem \cite{Luo2023} & 2023  & Not Available & Not Available \tabularnewline
Cairis \cite{Faily2020} & 2020  & Apache 2.0 & \url{https://github.com/cairis-platform/cairis} \tabularnewline
Censys \cite{censys15} & 2015  & Apache 2.0 + ISC & \url{https://github.com/zmap/zgrab2} \tabularnewline
Chainsaw \cite{DBLP:conf/ccs/AlhuzaliEGV16} & 2016  & Not Available & Not Available \tabularnewline
Chucky \cite{DBLP:conf/ccs/YamaguchiWGR13} & 2013  & GPLv3 & \url{https://github.com/a0x77n/chucky-ng/} \tabularnewline
Commix \cite{DBLP:journals/ijisec/StasinopoulosNX19} & 2019  & GPLv3 & \url{https://github.com/commixproject/commix} \tabularnewline
CryptoGuard \cite{DBLP:conf/ccs/RahamanXASTFKY19} & 2019  & GPLv3 & \url{https://github.com/CryptoGuardOSS/cryptoguard} \tabularnewline
CuPerFuzzer \cite{DBLP:conf/sp/LiDLDG21} & 2021  & Not Specified & \url{https://github.com/little-leiry/CuPerFuzzer} \tabularnewline
Deemon \cite{DBLP:conf/ccs/PellegrinoJ0BR17} & 2017  & GPLv3 & \url{https://github.com/tgianko/deemon} \tabularnewline
Delta \cite{lee2017delta} & 2017  & Not Specified & \url{https://github.com/seungsoo-lee/DELTA} \tabularnewline
DFBC \cite{ng2021dfbc} & 2021  & Not Available & Not Available \tabularnewline
Diane \cite{redini2021diane} & 2021  & Not Specified & \url{https://github.com/ucsb-seclab/diane} \tabularnewline
EBF \cite{Aljaafari2021} & 2021  & MIT License & \url{https://github.com/fatimahkj/EBF} \tabularnewline
ELAID \cite{DBLP:journals/cybersec/XuXLH20} & 2020  & Not Available & Not Available \tabularnewline
ESASCF \cite{Ghanem2023} & 2023  & Available upon request & Available upon request \tabularnewline
ESRFuzzer \cite{DBLP:journals/cybersec/ZhangHJSLZZL21} & 2021  & Not Available & Not Available \tabularnewline
ESSecA \cite{rak2022esseca} & 2022  & Not Specified & \url{https://github.com/DanieleGranata94/SlaGenerator} \tabularnewline
Firmaster \cite{8457340} & 2018  & Not Available & Not Available \tabularnewline
FUGIO \cite{DBLP:conf/uss/ParkKJS22} & 2022  & Not Specified & \url{https://github.com/WSP-LAB/FUGIO} \tabularnewline
FUSE \cite{DBLP:conf/ndss/LeeWLS20} & 2020  & Not Specified & \url{https://github.com/WSP-LAB/FUSE} \tabularnewline
Gail-PT \cite{chen2023gail} & 2023  & Not Specified & \url{https://github.com/Shulong98/GAIL-PT/} \tabularnewline
GNPassGAN \cite{Yu2022} & 2022  & MIT License & \url{https://github.com/fangyiyu/GNPassGAN/} \tabularnewline
HARMer \cite{DBLP:journals/access/EnochHMLAK20} & 2020  & MIT License & \url{https://github.com/whistlebee/harmat} \tabularnewline
HILTI \cite{Sommer2014} & 2014  & Not Specified & \url{https://github.com/rsmmr/hilti} \tabularnewline
IoTFuzzer \cite{DBLP:conf/ndss/ChenDZZL0LSYZ18} & 2018  & Not Specified & \url{https://github.com/zyw-200/IOTFuzzer_Full} \tabularnewline
JCOMIX \cite{stallenberg_jcomix_2019} & 2019  & Not Specified & \url{https://github.com/SERG-Delft/JCOMIX} \tabularnewline
LAID \cite{DBLP:conf/cisc/XuLXLHMLH18} & 2018  & Not Available & Not Available \tabularnewline
Link \cite{lee2022link} & 2022  & Not Specified & \url{https://github.com/WSP-LAB/Link} \tabularnewline
Lore \cite{DBLP:journals/tdsc/Holm23} & 2023  & Not Available & Not Available \tabularnewline
LTESniffer \cite{10.1145/3558482.3590196} & 2023  & Not Specified & \url{https://github.com/SysSec-KAIST/LTESniffer} \tabularnewline
Mace \cite{monshizadeh2014mace} & 2014  & Not Available & Not Available \tabularnewline
MAIT \cite{yucel2021mait} & 2021  & Not Available & Not Available \tabularnewline
MAL \cite{johnson2018meta} & 2018  & Apache 2.0 & \url{https://github.com/mal-lang/malcompiler/} \tabularnewline
MaliceScript \cite{liu2018malicescript} & 2018  & Not Available & Not Available \tabularnewline
Masat \cite{mjihil2015masat} & 2015  & Not Available & Not Available \tabularnewline
Mirage \cite{cayre2019mirage} & 2019  & MIT License & \url{https://github.com/RCayre/mirage} \tabularnewline
Mitch \cite{DBLP:conf/eurosp/CalzavaraCFRT19} & 2019  & Not Specified & \url{https://github.com/alviser/mitch} \tabularnewline
MoScan \cite{Wei2021} & 2021  & UPL 1.0 & \url{https://github.com/baigd/moscan} \tabularnewline
NAUTILUS \cite{DBLP:conf/uss/DengZLL00YW23} & 2023  & Apache 2.0 & \url{https://github.com/chenleji/nautilus} \tabularnewline
NAVEX \cite{DBLP:conf/uss/AlhuzaliGEV18} & 2018  & GPLv3 & \url{https://github.com/aalhuz/navex} \tabularnewline
NetCAT \cite{Kurth2020} & 2020  & Not Available & Not Available \tabularnewline
NeuralNetworkCracking \cite{melicher2016fast} & 2016  & Apache 2.0 & \url{https://github.com/cupslab/neural_network_cracking} \tabularnewline
No Name (CSRF) \cite{9357029} & 2020  & Not Available & Not Available \tabularnewline
No Name (TTCN-3) \cite{DBLP:conf/isncc/LealT18} & 2018  & Not Available & Not Available \tabularnewline
NoCrack \cite{Chatterjee2015} & 2015  & MIT License & \url{https://github.com/rchatterjee/nocrack} \tabularnewline
NodeXP \cite{ntantogian2021nodexp} & 2021  & Not Specified & \url{https://github.com/esmog/nodexp} \tabularnewline
ObjectMap \cite{DBLP:conf/pci/Koutroumpouchos19} & 2019  & MIT License & \url{https://github.com/georlav/objectmap} \tabularnewline
OMEN \cite{durmuth2015omen} & 2015  & MIT License & \url{https://github.com/RUB-SysSec/OMEN} \tabularnewline
OSV \cite{kasemsuwan2017osv} & 2017  & GPLv3 & \url{https://github.com/Emoform/OSV} \tabularnewline
Owfuzz \cite{10.1145/3558482.3590174} & 2023  & GPLv3 & \url{https://github.com/alipay/Owfuzz} \tabularnewline
PassGAN \cite{DBLP:conf/acns/HitajGAP19} & 2019  & MIT License & \url{https://github.com/brannondorsey/PassGAN} \tabularnewline
PassGPT \cite{rando2023passgpt} & 2023  & CC BY-NC 4.0 & \url{https://github.com/javirandor/passgpt} \tabularnewline
PasswordCrackingTraining \cite{DBLP:conf/esorics/CampiFL22} & 2022  & MIT License & \url{https://github.com/focardi/PasswordCrackingTraining} \tabularnewline
PenQuest \cite{DBLP:journals/virology/LuhTTSJ20} & 2020  & Proprietary & \url{https://www.pen.quest/} \tabularnewline
PentestGPT \cite{DBLP:journals/corr/abs-2308-06782} & 2023  & MIT License & \url{https://github.com/GreyDGL/PentestGPT} \tabularnewline
PhpSAFE \cite{DBLP:conf/dsn/NunesFV15} & 2015  & GPLv2 & \url{https://github.com/JoseCarlosFonseca/phpSAFE} \tabularnewline
PJCT \cite{jain2015pjct} & 2015  & Not Available & Not Available \tabularnewline
Project Achilles \cite{DBLP:conf/kbse/SaccenteDDCX19} & 2019  & LGPLv3 & \url{https://github.com/secure-software-engineering/achilles-benchmark-depscanners} \tabularnewline
PURITY \cite{bozic2015purity} & 2015  & Proprietary & Not Available \tabularnewline
Pyciuti \cite{muralidharan2023pyciuti} & 2023  & Not Available & Not Available \tabularnewline
RAT \cite{Amouei2022} & 2022  & Available upon request & Available upon request \tabularnewline
Revealer \cite{DBLP:conf/sp/LiuZM21} & 2021  & GPLv2 & \url{https://github.com/cuhk-seclab/Revealer} \tabularnewline
RiscyROP \cite{10.1145/3545948.3545997} & 2022  & Not Available & Not Available \tabularnewline
Robin \cite{DBLP:journals/corr/abs-2007-06629} & 2020  & Not Specified & \url{https://github.com/olmps/Robin} \tabularnewline
ROSploit \cite{DBLP:conf/irc/RiveraLS19} & 2019  & MIT License & \url{https://github.com/seanrivera/rosploit} \tabularnewline
RT-RCT \cite{fagroud2021rt} & 2021  & Not Available & Not Available \tabularnewline
Scanner++ \cite{10.1145/3517036} & 2023  & Not Available & Not Available \tabularnewline
SemanticGuesser \cite{veras2014semantic} & 2014  & Not Specified & \url{https://github.com/vialab/semantic-guesser} \tabularnewline
SerialDetector \cite{DBLP:conf/ndss/ShcherbakovB21} & 2021  & Not Specified & \url{https://github.com/yuske/SerialDetector} \tabularnewline
ShoVAT \cite{genge2016shovat} & 2016  & Not Available & Not Available \tabularnewline
Snout \cite{Mikulskis2019} & 2019  & Not Specified & \url{https://github.com/nislab/snout/} \tabularnewline
SOA-Scanner \cite{antunes2013soa} & 2013  & Not Available & Not Available \tabularnewline
Spicy \cite{DBLP:conf/acsac/SommerAH16} & 2016  & MIT License & \url{https://github.com/zeek/spicy/} \tabularnewline
SuperEye \cite{DBLP:conf/icccsec/LiYWLYH19} & 2019  & Not Available & Not Available \tabularnewline
SVED \cite{DBLP:conf/milcom/HolmS16} & 2016  & Not Available & Not Available \tabularnewline
TAMELESS \cite{DBLP:journals/tdsc/ValenzaKSL23} & 2023  & Not Specified & \url{https://github.com/FulvioValenza/TAMELESS} \tabularnewline
TChecker \cite{DBLP:conf/ccs/Luo0022} & 2022  & Not Available & Not Available \tabularnewline
TORPEDO \cite{olivo2015detecting} & 2015  & Not Available & Not Available \tabularnewline
UE Security Reloaded \cite{10.1145/3558482.3590194} & 2023  & Not Available & Not Available \tabularnewline
Untangle \cite{DBLP:conf/dimva/BertaniBBCZP23} & 2023  & Not Specified & \url{https://github.com/untangle-tool/untangle} \tabularnewline
VAPE-BRIDGE \cite{Vimala2022} & 2022  & Not Available & Not Available \tabularnewline
VERA \cite{blome2013vera} & 2013  & Not Available & Not Available \tabularnewline
VUDDY \cite{DBLP:conf/sp/KimWLO17} & 2017  & MIT License & \url{https://github.com/squizz617/vuddy} \tabularnewline
Vulcan \cite{kamongi2013vulcan} & 2013  & Not Available & Not Available \tabularnewline
VulCNN \cite{10.1145/3510003.3510229} & 2022  & Not Specified & \url{https://github.com/CGCL-codes/VulCNN} \tabularnewline
VulDeePecker \cite{DBLP:conf/ndss/LiZXO0WDZ18} & 2018  & Apache 2.0 & \url{https://github.com/CGCL-codes/VulDeePecker} \tabularnewline
Vulnet \cite{8922605} & 2019  & Not Available & Not Available \tabularnewline
Vulnsloit \cite{10.1007/978-3-030-64881-7_6} & 2020  & Available upon request & Available upon request \tabularnewline
VulPecker \cite{DBLP:conf/acsac/LiZXJQH16} & 2016  & Not Specified & \url{https://github.com/vulpecker/Vulpecker} \tabularnewline
WAPTT \cite{dhuric2014waptt} & 2014  & Not Available & Not Available \tabularnewline
WebFuzz \cite{DBLP:conf/esorics/RooijCKPA21} & 2021  & GPLv3 & \url{https://github.com/ovanr/webFuzz} \tabularnewline
WebVIM \cite{DBLP:conf/iciis/RankothgeRS20} & 2020  & Not Available & Not Available \tabularnewline

\hline
\caption{Classified tools, licence type and source code availability\label{tab:tools-availability}}
		\end{longtable}

\normalsize
\subsection{Process Based Classification: PTES and Mitre ATT\&CK} %

\begin{table}[ht]
	\centering
	\begin{subtable}{0.48\textwidth}
		\centering
		\small
		\begin{tabular}{|l|r|}
			\hline
			\textit{PTES Phase} & \textit{No.} \\
			\hline
			Vulnerability Analysis & 80 \\
			Exploitation & 39 \\
			Post Exploitation & 21 \\
			Intelligence Gathering & 20 \\
			Threat Modelling & 6 \\
			Reporting & 4 \\
			Pre-engagement Interactions & 0 \\
			\hline
		\end{tabular}
		\caption{Number of Tools identified according to PTES phases}
		\label{subtab:tool_count_ptes}
	\end{subtable}
	\hfill
	\begin{subtable}{0.48\textwidth}
		\centering
		\small
		\begin{tabular}{|l|r|}
			\hline
			\textit{Mitre ATTA\&CK Phase} & \textit{No.} \\
			\hline
			Reconnaissance & 84 \\
			Initial Access & 48 \\
			Resource Development & 21 \\
			Discovery & 11 \\
			Execution & 9 \\
			Credential Access & 9 \\
			Collection & 2 \\
			Impact & 1 \\
			Persistence & 0 \\
			Privilege Escalation & 0 \\
			Defense Evasion & 0 \\
			Lateral Movement & 0 \\
			Command and Control & 0 \\
			Exfiltration & 0 \\
			\hline
		\end{tabular}
		\caption{Number of Tools identified according to Mitre ATT\&CK phases}
		\label{subtab:tool_counts_mitre}
	\end{subtable}
	\caption{Tool counts for process-based classification}
	\label{tab:tool_counts}
\end{table}

Table \ref{tab:ptes} shows the tools identified and classified for the different PTES phases. The tools distribution according to steps in the Ethical Hacking process is reported in Table \ref{subtab:tool_count_ptes}. The absence of tools in the \textit{Pre-engagement Interactions} phase aligns with expectations, considering its non-technical nature, which typically involves scoping, planning, and agreement on the terms of engagement between the penetration tester and the client. This may potentially explain the lack of interest from the research community. 

The significant presence of tools in the \textit{Vulnerability Analysis} phase (80 tools) reflects the importance of identifying and assessing vulnerabilities within target systems, which is essential for any security assessment activity. In particular, many scanners were developed.

Additionally, 20 tools possess \textit{Intelligence Gathering} capabilities, primarily because this phase sometimes overlaps with vulnerability analysis attackers interact with target systems.

\textit{Exploitation} (39 tools) has a substantial number of tools designed to exploit identified vulnerabilities to gain unauthorised access to systems. Post Exploitation has slightly fewer tools than other phases (21 tools). We found 6 tools for \textit{Threat Modelling}. However, other researchers have developed some methodologies which are not implemented as tools that we discuss in Section \ref{sub:threat_modelling}.

The Mitre ATT\&CK classification table (Table \ref{tab:mitre}) shows tools associated with different stages of the attack process. The \textit{Reconnaissance} (84 tools) and \textit{Initial Access} (48 tools) stages exhibit a higher concentration of tools (Table \ref{subtab:tool_counts_mitre}), indicating the significance of these phases. This aligns with PTES findings, where most research effort seems to be put into vulnerability analysis. \textit{Resource Development} (21) and \textit{Discovery} (11) are also well represented. In contrast, stages such as \textit{Persistence} and \textit{Privilege Escalation} appear to have no tools directly associated with them, implying potential areas of development of novel research-informed tools. Further details on the classification, with sub areas, are presented in Table \ref{tab:mitre-details}.

Overall, researchers seem to have focused more on the technical aspects of the penetration testing process, and most of the tools have vulnerability analysis capability.

\subsection{Knowledge Based Classification: NCSC CyBOK and ACM CCS}

\begin{table}[ht]
	\centering
	\small
	\begin{tabular}{|p{0.62\textwidth}|r|}
		\hline
		\textit{CyBOK Knowledge Area} & \textit{No.} \\
		\hline
		Software and Platform Security: Software Security & 77 \\
		Software and Platform Security: Web \& Mobile Security & 38 \\
		Infrastructure Security: Network Security & 26 \\
		Attacks \& Defences: Adversarial Behaviours & 9 \\
		Systems Security: Authentication, Authorisation \& Accountability & 9 \\
		Systems Security: Distributed Systems Security & 3 \\
		Infrastructure Security: Applied Cryptography & 2 \\
		Human, Organisational \& Regulatory Aspects: Human Factors & 2 \\
		Attacks \& Defences: Malware \& Attack Technology & 1 \\
		Infrastructure Security: Physical Layer \& Telecommunications Security & 1 \\
		Human, Organisational \& Regulatory Aspects: Privacy \& Online Rights & 1 \\
		\hline
	\end{tabular}
	\caption{Number of Tools identified according to CyBOK, for KAs with at least 1 tool}
	\label{tab:tool_counts_cybok}
\end{table}

Table \ref{tab:cybok} presents the classification of tools according to NCSC CyBOK. The distribution of tools across different \textit{Knowledge Areas} (KAs) reflects the range of cybersecurity domains and disciplines covered by penetration testing activities. From the categorisation in Table \ref{tab:tool_counts_cybok}, it is evident that certain areas, such as \textit{Software \& Platform Security} and \textit{Networks Security}, are unsurprisingly more prominent, indicating areas of emphasis within cybersecurity practice. It should also be noted that while each category addresses specific aspects of cybersecurity, many tools may span multiple categories. 

Most of the tools are classified under \textit{Software \& Platform Security: Software Security: Detection of Vulnerabilities} (57 tools), which is a subcategory of \textit{Software \& Platform Security: Software Security}. The significant number of tools in this area reflects the recognition of software as a primary attack vector and demonstrates the research community's effort. Moreover, 38 tools are classified in the \textit{Software \& Platform Security: Web \& Mobile Security}, highlighting the research work done to address the challenges posed by the development and deployment of web and mobile applications.

The 26 tools falling under \textit{Infrastructure Security: Network Security} demonstrate the academic efforts in this area, ranging from network traffic monitoring and anomaly detection to implementing robust encryption protocols.

Nine tools classified under the \textit{Attacks \& Defences: Adversarial Behaviours} category indicate research aimed at understanding and simulating the techniques used by attackers.

Among the various categories within the ACM CCS, EH tools in Table \ref{tab:acm-ccs} predominantly fall into the \textit{Security and Privacy} root category, specifically within the subcategories of \textit{Systems Security}, \textit{Software and Application Security}, and \textit{Network Security}. In general, ACM CSS categories are too coarse to capture certain peculiarities of the tools. As CyBOK is specific to cybersecurity, it is more granular than ACM CSS for our purpose. We discuss the limitations in the classification in Section \ref{sub:classification_limitations}.

\textit{Software And Application Security: Vulnerability Management: Vulnerability Scanners} has the highest number of tools:  73. This indicates the proactive measures the research community is taking to detect various issues, from misconfigurations and missing patches to software flaws and weak passwords. The categorisation of 35 tools under the \textit{Software and Application Security: Web Applications Security} highlights the focus on developing specialised tools designed to test and secure web applications. These tools analyse web applications for vulnerabilities like SQL injection, cross-site scripting (XSS), and security misconfigurations. 

22 tools were categorised under the \textit{Network Security} domain and sub-domains, focusing on protecting the data during its transmission across networks. These tools are essential for detecting intrusions, monitoring network traffic for suspicious activities, and implementing preventive measures such as firewalls and encryption. EH tools within this category enable security professionals to simulate attacks on the network to identify vulnerabilities and assess the network's resilience against cyber threats.

\subsection{A note on Threat Modelling tools and methodologies}\label{sub:threat_modelling}

The PTES classification shows that the number of tools identified for \textit{Threat Modelling} is relatively small. However, the research field is somewhat active, but some contributions only propose new methodologies without implementing specific tools, so we did not include them in the classification. Some threat modelling methodologies discussed here cover different frameworks, each designed to improve security in cyber-physical systems (CPS), information technology, and critical infrastructure areas.

Ding et al.'s \cite{ding2017cps} framework integrates vulnerability assessment with reliability and threat analysis (both external and internal) within a unified model focused on critical infrastructures integrated with CPS. Similarly, Agadakos et al. \cite{agadakos2017jumping} present a novel method for modelling cyber and physical interactions within IoT networks. The study emphasises the identification of unexpected event chains that could lead to security vulnerabilities. Additionally, Castiglione et al. \cite{castiglione2020hazard} proposed a hazard-driven threat modelling methodology tailored for CPS, focusing on the interplay between security, reliability, and safety. 

To highlight the critical role of human factors in information security, Evans et al. \cite{evans2019heart} introduce a methodology that systematically evaluates information security incidents caused by human error, adopting the HEART methodology from high-reliability sectors like aviation and energy. Also, David et al. \cite{david2015modelling} propose using timed automata to model socio-technical attacks, offering a method that incorporates time and cost into analysing socio-technical systems and attacks. 

Furthermore, using formal methods for security analysis, Malik et al. \cite{malik2022towards} introduce an algorithm that transforms Attack Trees into Markov Decision Process models, aiming to address the limitations of scalability, state explosion, and manual interaction inherent in Attack Trees. 

Collectively, these methodologies demonstrate a shift towards integrating diverse analytical tools and perspectives, from human factors to formal methods and system theory, to address the increasingly complex and interconnected nature of modern systems. 

\subsection{Limitations surrounding the classification of tools}\label{sub:classification_limitations} %

The four classification systems identified in this study were chosen due to their overall topic coverage and relevance to computing topics and concepts. When combined, the classifications can give a precise idea of what a tool can do and the cybersecurity field it falls under.

Although these four classifications are fit for purpose when considered within the scope and the goal of this survey and its goals, there are some challenges and potential limitations in classifying tools in this manner.
The first issue is an inconsistency within the specificity of the tools. 

For example, within \textit{MITRE} \cite{strom2018mitre}, despite having hundreds of individual attack techniques and vectors, such as \textit{Enterprise: Privilege Escalation: Access Token Manipulation: SID-History injection} and \textit{Enterprise: Defence Evasion: Hijack Execution Flow: Path Interception by PATH Environment Variable}, which are very specific vectors down to the operating system architecture within individual target systems, MITRE lumps the entire concept of compromising a web application under \textit{Enterprise: Initial Access: Exploit Public-Facing Application}. There are other means of gaining specificity within this classification system, such as the \textit{Reconnaissance: Vulnerability Scanning} field. All together, these give a more specific view of the tool. However, both fields do not offer the specificity of the \textit{SID-History Injection} or \textit{Path Interception by PATH Environment Variable} fields. %

Another issue relates to tools that can be used for activities potentially unintended by the application designer. Deciding whether to include the ACM CSS \textit{Security and privacy: Network security: Denial-of-service attacks} field when considering a web application fuzzer that could potentially crash the target web application (and making similar decisions throughout the classification of the many tools within this survey), posed a significant challenge, as opinions on whether to include the field may vary between researchers.

Another issue was found when trying to discern the exact scope of any given tools. There were many instances when, in the paper, a tool would present itself as one thing, for example, being capable of completing a specific task within the abstract and majority of the discussion in the associated paper, only to reveal that the tool itself is a \textit{proof-of-concept}, with limited capabilities than was assumed initially.

To fully understand each tool's potential, an in-depth evaluation involving running the tools and testing their capabilities would be necessary. However, this is beyond the scope of this survey. Future work could focus on specific subject clusters to provide an in-depth comparison of the tools.

\section{Evaluation\label{sec:Evaluation}}

This section evaluates 100 research-informed Ethical Hacking tools developed within the past decade and included in this study. The discussion will focus on several key aspects: their licensing, release dates, availability of source code, the activity level of their development, whether the papers publishing them underwent peer review and their alignment with recognised cybersecurity frameworks.

\subsection{Peer Review Analysis and Date of Publication}
Of the 100 tools discussed in this study, 96\% were disseminated through peer-reviewed journals and conferences (Table \ref{subtab:classification-peer-reviewed}). 

\begin{table}[ht]
	\centering
	\begin{subtable}{0.48\textwidth}
		\centering
		\small
		\begin{tabular}{|l|r|r|}
			\hline
			\textit{Peer Reviewed} & \textit{No.} & \textit{\%} \tabularnewline 
			\hline
			Y & 96 & 96.00\% \tabularnewline 
			\hline
			N & 4 & 4.00\% \tabularnewline 
			\hline
			\textit{Total} & \textit{100} & \textit{100.00\%} \tabularnewline 
			\hline
		\end{tabular}
		\caption{Peer Reviewed}
		\label{subtab:classification-peer-reviewed}
	\end{subtable}
	\hfill
	\begin{subtable}{0.48\textwidth}
		\centering
		\small
		\begin{tabular}{|l|r|r|}
			\hline
			\textit{Source Code Avail.} & \textit{No.} & \textit{\%} \tabularnewline 
			\hline
			Y & 59 & 59.00\% \tabularnewline 
			\hline
			N & 41 & 41.00\% \tabularnewline 
			\hline
			\textit{Total} & \textit{100} & \textit{100.00\%} \tabularnewline 
			\hline
		\end{tabular}
		\caption{Source Code Available}
		\label{subtab:classification-source-code-available}
	\end{subtable}
	\caption{Classification}
	\label{tab:classification}
\end{table}

This indicates that the proposed tools have undergone rigorous validation, guaranteeing their effectiveness and reliability. This emphasis on peer-reviewed tools in our study reflects our commitment to ensuring readers have confidence in the credibility and utility of the tools presented. The remaining tools, which were not yet peer-reviewed at the time of our survey but are available as pre-print (e.g., \cite{DBLP:journals/corr/abs-2308-06782,Moscovich2020}), potentially under review or to be submitted in the near future.

\begin{figure}[ht]
\includegraphics[width = 0.65\textwidth]{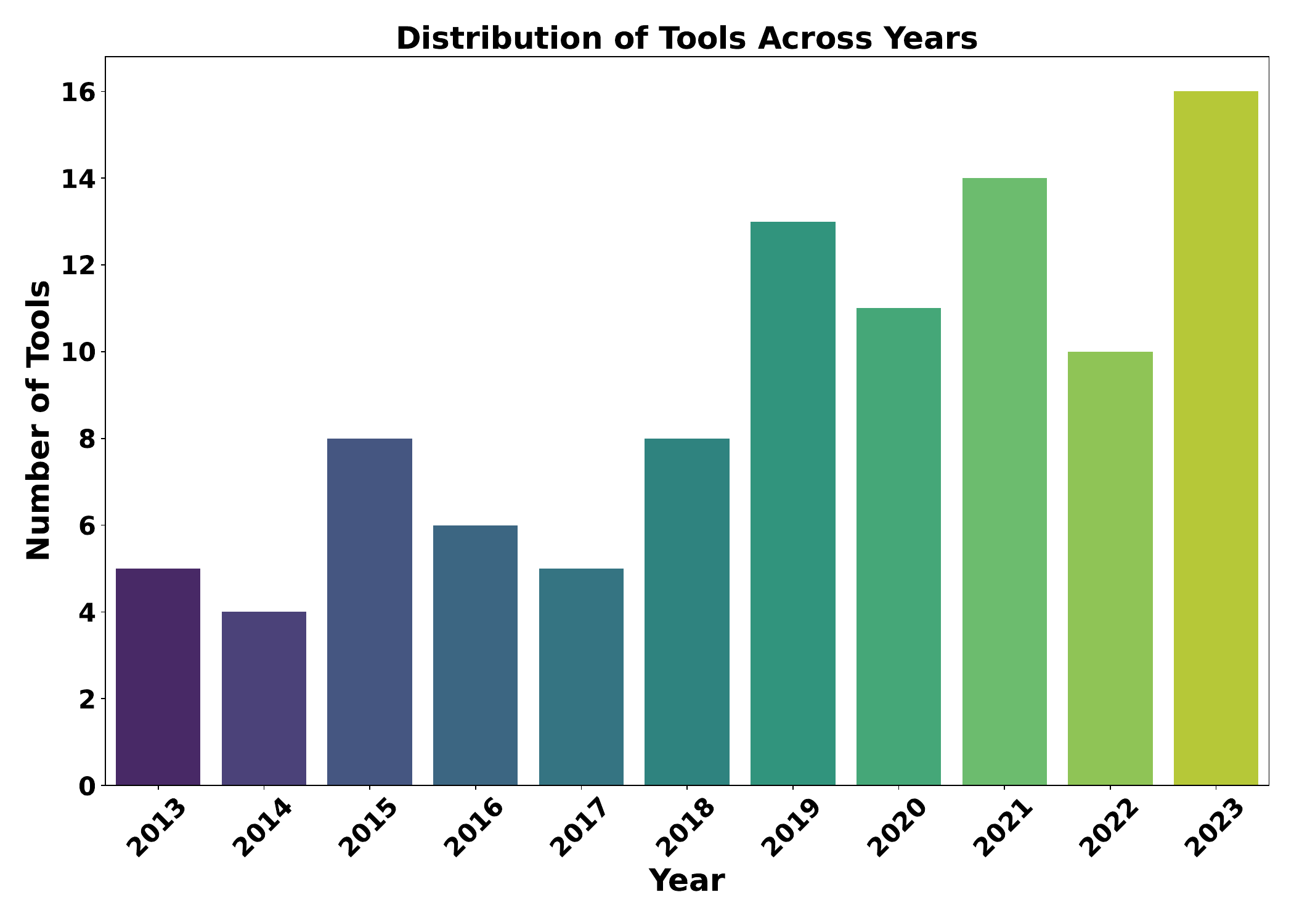}
\centering
    \caption{Distribution of Tool Releases Over Last Decade}
    \label{fig:tools_distribution_across_years}
\end{figure}

Furthermore, the distribution of tool releases over the years, as illustrated in Figure \ref{fig:tools_distribution_across_years}, shows an increase in development activities in recent years, with 16 tools released in 2023, 14 in 2021, and 13 in 2019. This trend mirrors the evolution of cybersecurity threats and the response from the research community to address these problems.

\subsection{Types of Licensing and Source Code Availability}
This section delves into the licensing types, development activity, and source code availability for the tools discussed in this paper. The evaluation found that out of the 100 tools included in this study, 59 have their source code readily available on GitHub, while 41 are unavailable (Table \ref{subtab:classification-source-code-available}).

\begin{figure}[ht]
\includegraphics[width = 0.65\textwidth]{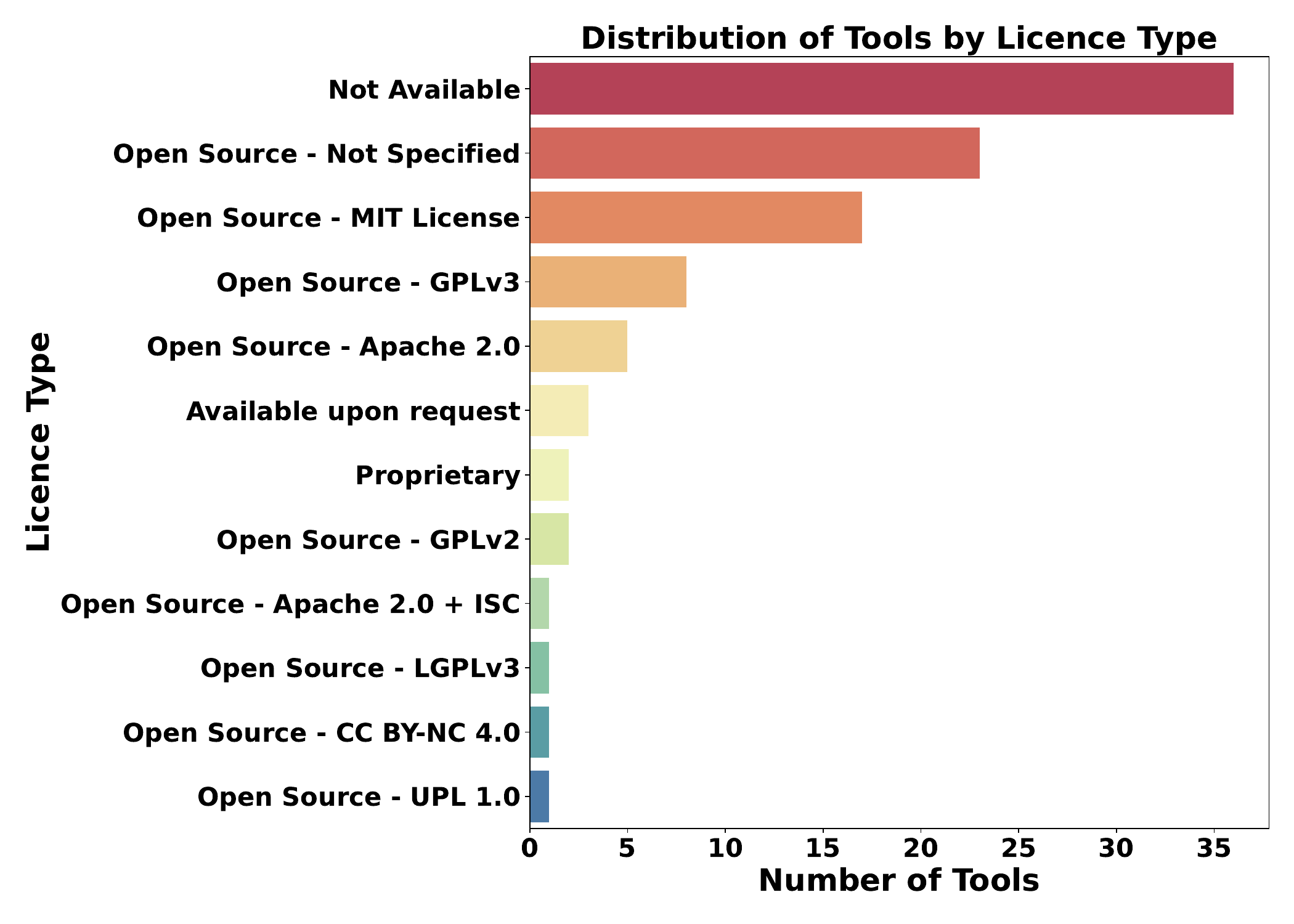}
\centering
    \caption{Distribution of Tools by License Type}
    \label{fig:tools_distribution_by_licence_type}

\end{figure}

The types of licenses for the tools available on GitHub vary widely (Figure \ref{fig:tools_distribution_by_licence_type}), mostly open source licences, ranging from the \textit{MIT License} (17 tools), \textit{GPLv3} (8 tools). However, for 23 tools, the licence is \textit{Not Specified}. The lack of clear licensing information could be an oversight by developers regarding the importance of transparent communication of usage rights. This ambiguity may potentially hinder adoption and adaptability. In the absence of a license, default copyright laws apply, meaning that authors retain all rights to their source code and reproduction, distribution, or creation of derivative works is prohibited.

Overall, the fact that source code is available for more than half of the tools (59 tools) demonstrates the cybersecurity researchers' dedication to openness and active community participation. However, 41 inaccessible tools in this study highlight an ongoing debate: the need to balance transparency with security, privacy, and commercial interests.

\subsection{Tools Development and Maintenance}
We examined the GitHub repositories of the 59 publicly available tools to understand specific features related to tool development and maintenance. Specifically, we collected data on the number of \textit{Commits} and the dates of the first and the last commits. We believe analysing commit activity in GitHub repositories provides insights into the development intensity and duration. However, we must consider that projects can move from one repository to another and that a private repository is used alongside a private one, and the public one is used only for dissemination purposes. Therefore, we can only attempt to capture some trends with this analysis, but we must be cautious about making statements regarding specific projects.

\begin{table}[ht]
	\centering
	\begin{subtable}{0.48\textwidth}
		\centering
		\small
		\begin{tabular}{|l|r|r|}
			\hline
			\textit{Period (Months)} & \textit{No.} & \textit{\%} \tabularnewline 
			\hline
			0-3 & 19 & 32.20\% \tabularnewline 
			\hline 
			4-6 & 3 & 5.08\% \tabularnewline 
			\hline 
			6-12 & 6 & 10.17\% \tabularnewline 
			\hline 
			13-24 & 6 & 10.17\% \tabularnewline 
			\hline 
			25-36 & 10 & 16.95\% \tabularnewline 
			\hline 
			37-60 & 8 & 13.56\% \tabularnewline 
			\hline 
			61-inf & 7 & 11.86\% \tabularnewline 
			\hline 
			\textit{Total} & \textit{59} & \textit{100.00\%} \tabularnewline 
			\hline
		\end{tabular}
		\caption{Distribution of Period of Project Activity (last -- first commit)}
		\label{subtab:distribution-of-period-of-project-activity}
	\end{subtable}
	\hfill
	\begin{subtable}{0.48\textwidth}
		\centering
		\small
		\begin{tabular}{|l|r|r|}
			\hline
			\textit{Commits Range} & \textit{No.} & \textit{\%} \tabularnewline 
			\hline
			1-10 & 17 & 28.81\% \tabularnewline 
			\hline 
			11-50 & 19 & 32.20\% \tabularnewline 
			\hline 
			51-100 & 4 & 6.78\% \tabularnewline 
			\hline 
			101-250 & 6 & 10.17\% \tabularnewline 
			\hline 
			251-500 & 5 & 8.47\% \tabularnewline 
			\hline 
			501-inf & 8 & 13.56\% \tabularnewline 
			\hline 
			\textit{Total} & \textit{59} & \textit{100.00\%} \tabularnewline 
			\hline
		\end{tabular}
		\caption{Distribution of Number of Project's Commits}
		\label{subtab:distribution-of-number-of-project-commits}
	\end{subtable}
	\caption{Distribution of Project Activity and Commits}
	\label{tab:distribution-of-period-and-commits}
\end{table}

\begin{table}[ht]
\centering
\small
\begin{tabular}{|l|r|r|}
\hline
\textit{\% of Project Activity} & \textit{No} & \textit{\%} \tabularnewline 
\hline
0-0 & 5 & 8.47\% \tabularnewline 
\hline 
1-25 & 12 & 20.34\% \tabularnewline 
\hline 
26-50 & 15 & 25.42\% \tabularnewline 
\hline 
51-75 & 5 & 8.47\% \tabularnewline 
\hline 
76-99 & 2 & 3.39\% \tabularnewline 
\hline 
100-100 & 20 & 33.90\% \tabularnewline 
\hline 
\textit{Total} & \textit{59} & \textit{100.00\%} \tabularnewline 
\hline
\end{tabular}
\caption{Distribution of Project Activity within 3 Years after Release}
\label{tab:distribution-of-project-activity-within-3-years-after-release}
\end{table}

Table \ref{subtab:distribution-of-period-of-project-activity} provides an overview of the distribution of project activity periods. We measured the difference in months between the first and last commit for the considered projects. The data shows that approximately one-third of the projects have a relatively short activity span of less than 3 months. This may indicate that the publication of the source code has likely been instrumental to the publication of the paper.

Comparing the year of publication of the paper and the date of the first commit, we can see that around 90\% of the projects have been active sometime in the year before or after the publication. This is not surprising. However, fewer than 10\% of the projects have been active 2 or more years before the publication, according to data publicly available.

Another parameter we considered is the number of commits (Table \ref{subtab:distribution-of-number-of-project-commits}). The data highlights the diversity in project engagement and development intensity within the examined dataset. Around 30\% of the projects have just up to 10 commits. As the commit ranges increase, the percentage of projects gradually decreases. Projects with a higher number of commits (500+) account for 13.33\% of the total, indicating a smaller but notable proportion of projects with an extensive development history.

Table \ref{tab:distribution-of-project-activity-within-3-years-after-release} represents the activity level as a percentage of the time between a project's release and the present. \textit{100--100} indicates continuous activity throughout the period, while \textit{0--0} signifies no activity at all. For example, \textit{26--50} means that the project was active for at least 26\% up to 50\% of the considered time interval (9--18 months).

The data shows that one-third of the projects remained active within three years after their release. However, over half of these projects ceased activity after just 1.5 years. This trend sheds light on the development lifecycle of these projects, indicating a high initial engagement that tends to taper off relatively quickly for a significant number of projects.

\subsection{Recommendations} %
To improve the dissemination and enhance the impact of research on the wider community of practitioners, we suggest that researchers should:

\begin{itemize}
\item Distribute the software open source without exception and keep the software repository alive. Otherwise, it would be impossible for any dissemination within the practitioner community \cite{kalliamvakou2016depth}.
\item Clearly specify the licence type and adopt standard FOSS licences \cite{metzger2015free}, like GNU GPLv3, so that users may know precisely what they can do with the tools.
\item Produce comprehensive documentation and tutorials on how to use the tools. Currently, this is partially done, but the existing documentation is primarily intended to support the peer-review process, as noted by Mirhosseini (2020) \cite{mirhosseini2020docable}.
\item Try to maintain the software by implementing bug fixes and improvements after publishing the paper. This is particularly challenging for academic projects as they operate with limited availability of human resources and funding. Once the project ends or the paper is published, the interest of the researcher tends to move to new projects \cite{kalliamvakou2016depth}.
\item Some tools may become obsolete due to several reasons: incompatibility with more recent versions of other software (OSs, libraries, applications, etc.) or the vulnerability covered by the tool being patched. In those cases, authors should update the documentation and clearly specify the requirements, scope, context and limitations of the tool.
\item Try to implement their solutions in modular tools utilised by practitioners like Metasploit and Nmap. While this can be possible for certain solutions, in general, some tools are so different and innovative that they cannot fit into the API of existing tools.
\item Consider that public dissemination mitigates the risk of weaponizing tools by promoting a level playing field approach.
\end{itemize}

Another question is: what can practitioners and industry do? We cannot expect many individual practitioners to engage directly with the research outputs except when driven by intellectual curiosity. However, the IT and cybersecurity industry should try to incentivise collaboration with academia. Industry and venture capitalists likely monitor academic research to understand the state-of-the-art and gain inspiration for new ideas. However, more effective engagement from the industry may help academic research to enhance its impact.

For example, the industry currently invests in bug bounty programs, providing monetary incentives to security researchers to identify and report vulnerabilities. This informs bug fixes and improves the quality of the product overall. However, this process is typically ex-post and not something an academic researcher would be directly involved in, except if finding a bug is a by-product of the research work. However, in many cases, academic researchers would likely engage in a responsible disclosure process.

Concretely, the industry could redirect some funding from bug bounty programs \cite{walshe2020empirical} to grant schemes supporting open source projects, for example, the Google Summer of Code, which could enable researchers to develop and enhance their tools. This is likely something that medium to large companies could be interested in. Still, it requires a shift in perspective beyond the immediate rewards and limited risks and commitments of current bug bounty programs.

Finally, an important aspect is that most research-informed tools are developed by small teams, sometimes even by a single individual, for non-profit reasons. Given the working conditions in many higher education institutions, especially in countries where the sector is very competitive and commercialised \cite{lynch2015academic}, it is often the case that, unless a research grant supports the project, the developers end up working significant hours during their own free time \cite{ucu2022}.

\subsection{Related Work} 
Existing reviews of Ethical Hacking tools typically focus on industry practitioner tools, with occasional consideration of research-informed tools. Many popular practitioner tools included in these reviews (e.g.,~\cite{Yaacoub2021,duque2020investigating,alhamed2023systematic,sarker2023penetration}) are recurrent: Nmap, Metasploit Framework, OpenVAS/GVM, Nessus, Burp Suite, OWASP ZAP, SQLMap, BeEF, Nikto, W3AF, and others.

Yaacoub et al.~\cite{Yaacoub2021} survey and classify around 40 practitioner tools and OSs (e.g., Kali Linux and ParrotOS), focusing on challenges and issues associated with EH activities. The paper maps the tools and techniques for vulnerability assessment, network scanning tools, crimeware toolkits, etc., considering different attack types and application domains. Duque Anton et al.~\cite{duque2020investigating} include in their review around 25 popular practitioner tools, and their capabilities are evaluated using criteria such as active maintenance, licensing, commercial aspects (paid vs. free), and technical elements like programming language and interaction with other technology.

Moreover, Alhamed et al.~\cite{alhamed2023systematic} analyse around 20 mostly practitioner tools, with good coverage of network vulnerability and exploitation in particular. However, they consider existing research proposals for mitigating techniques. Additionally, Sarker et al.~\cite{sarker2023penetration} reviewed penetration testing frameworks, processes, tools, and scoring methods, encompassing around 15 practitioner EH tools.

In some cases, authors restrict their focus to a specific domain. For example, Yaacoub et al.~\cite{Yaacoub2023} provide good coverage of practitioner commercial and open-source solutions for EH in IoT, while Altulaihan et al.~\cite{Altulaihan} review and compare industry practitioner tools for web application penetration testing. Similarly, Shahid et al.~\cite{shahid2022comparative} provide a comparative analysis of commercial and open-source tools for Web Application Security with a focus on accuracy and precision. Alzahrani et al.~\cite{alzahrani2017web} and Ravindran et al.~\cite{ravindran2022review} compare many EH tools, including both industry practitioner tools and a few research-informed tools for web vulnerability assessment and exploitation, e.g., XSS and SQL injection. Kowta et al.~\cite{kowta2021analysis} analysed a variety of reconnaissance and information-gathering tools and techniques including Google Dorking, Shodan, Web Crawler, Recon-ng, Photon, Final Recon, and Sherlock. The tools are compared with criteria such as update frequency, languages used, and supported OSs, with some research-informed tools also included in the review.

In a few cases, authors systematically classify the tools according to some methodology or taxonomy. Duque Anton et al.~\cite{duque2020investigating} compared and classified practitioner tools, mapping them to the Mitre ATT\&CK framework. Moreover, Zilberman et al.~\cite{zilberman2020sok} provide a review of threat emulators while mapping to the Mitre ATT\&CK matrix tactics.

Shanley et al.~\cite{shanley2015selection} review and compare several methodologies and frameworks, including PTES, Building Security in Maturity Model (BSIMM), Metasploit Framework (MSF), OWASP Testing Guide (OTG), Information Systems Security Assessment Framework (ISSAF), and the Open Source Security Testing Methodology Manual (OSSTMM). However, no tools are reviewed; therefore, no classification is attempted.

Our study significantly differs from previously published papers in both the number of tools covered and its exclusive focus on research-informed EH tools. By categorising the tools into process-based and knowledge-based classifications, we organise them according to specific phases, demonstrating where and when they are utilised in EH processes. While other reviews include classifications, the main contribution of our work is a more comprehensive and unique exploration. We surveyed 100 tools and classified them according to four different frameworks: PTES, Mitre ATT\&CK, CyBOK, and ACM CCS. Additionally, we identify and analyze trends in developing, maintaining, and disseminating novel research-informed tools.

\section{Conclusion and Future work\label{sec:Conclusion-and-Future}}%

Addressing emerging cyber security threats requires developing Ethical Hacking tools to identify vulnerabilities in networks, systems, and applications. While practitioners design most EH tools for immediate use in the industry, academic researchers have also significantly contributed to developing security tools. However, there is a noticeable gap in awareness among practitioners about academic contributions in this domain. %
This paper evaluates 100 research-informed tools, examining aspects such as licensing, release dates, source code availability, development activity, and peer review status. These tools are then aligned with established frameworks like PTES, the Mitre ATT\&CK framework, CyBOK, and ACM CCS. 

Key findings indicate that 96\% of these tools originate from peer-reviewed research, with 59\% having their source code readily accessible on GitHub. Activity analysis shows that 90\% of projects were active around their publication year, yet activity dwindles significantly within 1.5 years post-release. Under the PTES framework classification, most tools are designed for vulnerability analysis, whereas threat modelling tools are relatively few. The CyBOK and ACM CCS classifications emphasise tools for detecting vulnerabilities, particularly under the \textit{Software \& Platform Security} and \textit{Security And Privacy}, respectively. For the Mitre ATT\&CK framework, most tools primarily focus on reconnaissance, highlighting the vital role of information gathering in identifying network and system details. 
Future directions involve experimental evaluations and comparisons of specific tools, integration of existing practitioner tools, and exploration of using large language models in penetration testing. This approach aims to bridge the gap between industry and academia, enhancing the development and effectiveness of Ethical Hacking tools.

\subsection*{Acknowledgement}
The authors express their gratitude to Zia Ush Shamszaman, Sachin Manakkil Jolly, and the Advanced Practice students of the MSc Cybersecurity course at Teesside University for engaging in constructive discussions and recommending certain tools for inclusion in this survey.

\subsection*{Author Contributions}
Conceptualization, P.M.; methodology, P.M .; software, P.M.; validation, P.M. and L.H.; formal analysis, P.M.; investigation, P.M., L.G., L.H, C.O and M.M.; resources, P.M.; data curation, P.M and L.H.; writing---original draft preparation, P.M., L.G., L.H, C.O and M.M.; writing---review and editing, P.M., L.G., L.H, C.O and M.M.; visualization, P.M., L.G., and C.O.; project administration, P.M., L.G., and C.O.; funding acquisition, N/A. All authors have read and agreed to the published version of the manuscript.

\appendix
\normalsize
\section{List of Abbreviations}\label{app:abbreviations}
\nomenclature[A]{\textbf{EH}}{Ethical Hacking}
\nomenclature[A]{\textbf{PTES}}{Penetration Testing Execution Standard}
\nomenclature[A]{\textbf{CVSS}}{Common Vulnerability Scoring System}
\nomenclature[A]{\textbf{CyBOK}}{Cyber Security Body of Knowledge}
\nomenclature[A]{\textbf{CSS}}{(ACM) Computing Classification System}
\nomenclature[A]{\textbf{ATT\&CK}}{(Mitre) Adversarial Tactics, Techniques, and Common Knowledge}
\nomenclature[A]{\textbf{OSINT}}{Open-Source INTelligence}
\nomenclature[A]{\textbf{WCMS}}{Web Content Management Systems}
\nomenclature[A]{\textbf{SSJI}}{Server-Side Javascript Injection}
\nomenclature[A]{\textbf{DoS}}{Denial of Service}
\nomenclature[A]{\textbf{XSS}}{Cross Site Scripting}
\nomenclature[A]{\textbf{CSRF}}{Cross Site Request Forgery}
\nomenclature[A]{\textbf{SQLIA}}{SQL Injection Attacks}
\nomenclature[A]{\textbf{DFBC}}{Digital Footprint and Breach Check}
\nomenclature[A]{\textbf{CLI}}{Command Line Interface}
\nomenclature[A]{\textbf{GUI}}{Graphical User Interface}
\nomenclature[A]{\textbf{OS}}{Operating System}
\nomenclature[A]{\textbf{SDN}}{Software Defined Networking}
\nomenclature[A]{\textbf{VM}}{Virtual Machine}
\nomenclature[A]{\textbf{NVD}}{National Vulnerability Database}
\nomenclature[A]{\textbf{E2E}}{End-to-End}
\nomenclature[A]{\textbf{DPI}}{Deep Packet Inspection}
\nomenclature[A]{\textbf{SDR}}{Software Defined Radio}
\nomenclature[A]{\textbf{IO2BO}}{Integer-Overflow-to-Buffer-Overflow}
\nomenclature[A]{\textbf{XMLi}}{XML injection}
\nomenclature[A]{\textbf{CI}}{Continuous Integration}
\nomenclature[A]{\textbf{GAIL}}{Generative Adversarial Imitation Learning}
\nomenclature[A]{\textbf{RL}}{Reinforcement Learning}
\nomenclature[A]{\textbf{DRL}}{Deep Reinforcement Learning}
\nomenclature[A]{\textbf{IoT}}{Internet of Things}
\nomenclature[A]{\textbf{OSPF}}{Open Shortest Path First}
\nomenclature[A]{\textbf{ETSI}}{European Telecommunications Standards Institute}
\nomenclature[A]{\textbf{LFA}}{Link Flooding Attacks}
\nomenclature[A]{\textbf{SOHO}}{Small Office and Home Office}
\nomenclature[A]{\textbf{C2}}{Command and Control}
\nomenclature[A]{\textbf{RDP}}{Remote Desktop Protocol}
\nomenclature[A]{\textbf{VAPT}}{Vulnerability Assessment and Penetration Testing}
\nomenclature[A]{\textbf{AP}}{Access Point}
\nomenclature[A]{\textbf{MITM}}{Man-In-The-Middle}
\nomenclature[A]{\textbf{OSSTMM}}{Open-Source Security Testing Methodology Manuel}
\nomenclature[A]{\textbf{SP}}{Special Publication}
\nomenclature[A]{\textbf{ISAAF}}{Information System Security Assessment Framework}
\nomenclature[A]{\textbf{FTP}}{File Transfer Protocol}
\nomenclature[A]{\textbf{SET}}{Social Engineering Toolkit}
\nomenclature[A]{\textbf{TTP}}{Tactics, Techniques, and Procedures}
\nomenclature[A]{\textbf{CVE}}{Common Vulnerabilities and Exposures}
\nomenclature[A]{\textbf{CWE}}{Common Weakness Enumeration}
\nomenclature[A]{\textbf{APT}}{Advanced Persistent Threats}
\nomenclature[A]{\textbf{PCI DSS}}{Payment Card Industry Data Security Standard}
\nomenclature[A]{\textbf{OWASP}}{Open Web Application Security Project}
\nomenclature[A]{\textbf{CPE}}{Common Platform Enumeration}
\nomenclature[A]{\textbf{DFD}}{Data Flow Diagrams}
\nomenclature[A]{\textbf{CTI}}{Cyber Threat Intelligence}
\nomenclature[A]{\textbf{POI}}{PHP Object Injection}
\nomenclature[A]{\textbf{UFU}}{Unrestricted File Upload}
\nomenclature[A]{\textbf{UEFU}}{Unrestricted Executable File Upload}
\nomenclature[A]{\textbf{GAIL}}{Generative Adversarial Imitation Learning}
\nomenclature[A]{\textbf{GAN}}{Generative Adversarial Network}
\nomenclature[A]{\textbf{HARM}}{Hierarchical Attack Representation Model}
\nomenclature[A]{\textbf{RL}}{Reinforcement Learning}
\nomenclature[A]{\textbf{IoMT}}{Internet of Medical Things}
\nomenclature[A]{\textbf{ICS}}{Industrial Control Systems}
\nomenclature[A]{\textbf{ACL}}{Access Control Lists}
\nomenclature[A]{\textbf{RBAC}}{Role-Based Access Control}
\nomenclature[A]{\textbf{ABAC}}{Attribute-Based Access Control}
\nomenclature[A]{\textbf{CBAC}}{Code-Based Access Control}
\nomenclature[A]{\textbf{MAC}}{Message Authentication Code}
\nomenclature[A]{\textbf{AE}}{Authenticated Encryption}
\nomenclature[A]{\textbf{NHS}}{National Health Service}
\nomenclature[A]{\textbf{TPM}}{Trusted Platform Module}
\nomenclature[A]{\textbf{NFC}}{Near-Field Communications}

\setlength{\nomitemsep}{-\parsep}
\printnomenclature[2.0cm]

\clearpage
\section{Classification}\label{app:classification}

		\begin{longtable}{|>{\RaggedRight}p{0.21\textwidth}|>{\RaggedRight}p{0.79\textwidth}|}
			
			\hline 
			\textit{PTES Phase} & \textit{Tools}\tabularnewline
\hline
\endhead
Pre-Engagement Interactions &  \tabularnewline 
\hline
Intelligence Gathering & Bbuzz \cite{DBLP:conf/milcom/BlumbergsV17}, DFBC \cite{ng2021dfbc}, ESASCF \cite{Ghanem2023}, ESRFuzzer \cite{DBLP:journals/cybersec/ZhangHJSLZZL21}, Firmaster \cite{8457340}, IoTFuzzer \cite{DBLP:conf/ndss/ChenDZZL0LSYZ18}, LTESniffer \cite{10.1145/3558482.3590196}, Lore \cite{DBLP:journals/tdsc/Holm23}, MaliceScript \cite{liu2018malicescript}, Owfuzz \cite{10.1145/3558482.3590174}, Pyciuti \cite{muralidharan2023pyciuti}, RT-RCT \cite{fagroud2021rt}, SVED \cite{DBLP:conf/milcom/HolmS16}, Scanner++ \cite{10.1145/3517036}, ShoVAT \cite{genge2016shovat}, SuperEye \cite{DBLP:conf/icccsec/LiYWLYH19}, TORPEDO \cite{olivo2015detecting}, UE Security Reloaded \cite{10.1145/3558482.3590194}, Vulcan \cite{kamongi2013vulcan}, Vulnsloit \cite{10.1007/978-3-030-64881-7_6} \tabularnewline 
\hline
Threat Modelling & Cairis \cite{Faily2020}, ESSecA \cite{rak2022esseca}, HARMer \cite{DBLP:journals/access/EnochHMLAK20}, MAL \cite{johnson2018meta}, PenQuest \cite{DBLP:journals/virology/LuhTTSJ20}, TAMELESS \cite{DBLP:journals/tdsc/ValenzaKSL23} \tabularnewline 
\hline
Vulnerability Analysis & AIBugHunter \cite{Fu2023}, ARMONY \cite{chen2013armory}, AVAIN \cite{Egert2019}, Autosploit \cite{Moscovich2020}, Bbuzz \cite{DBLP:conf/milcom/BlumbergsV17}, Black Ostrich \cite{Eriksson2023}, Black Widow \cite{DBLP:conf/sp/ErikssonPS21}, Bleem \cite{Luo2023}, Censys \cite{censys15}, Chainsaw \cite{DBLP:conf/ccs/AlhuzaliEGV16}, Chucky \cite{DBLP:conf/ccs/YamaguchiWGR13}, Commix \cite{DBLP:journals/ijisec/StasinopoulosNX19}, CryptoGuard \cite{DBLP:conf/ccs/RahamanXASTFKY19}, CuPerFuzzer \cite{DBLP:conf/sp/LiDLDG21}, Deemon \cite{DBLP:conf/ccs/PellegrinoJ0BR17}, Delta \cite{lee2017delta}, Diane \cite{redini2021diane}, EBF \cite{Aljaafari2021}, ELAID \cite{DBLP:journals/cybersec/XuXLH20}, ESASCF \cite{Ghanem2023}, ESRFuzzer \cite{DBLP:journals/cybersec/ZhangHJSLZZL21}, FUGIO \cite{DBLP:conf/uss/ParkKJS22}, FUSE \cite{DBLP:conf/ndss/LeeWLS20}, Firmaster \cite{8457340}, Gail-PT \cite{chen2023gail}, HILTI \cite{Sommer2014}, IoTFuzzer \cite{DBLP:conf/ndss/ChenDZZL0LSYZ18}, JCOMIX \cite{stallenberg_jcomix_2019}, LAID \cite{DBLP:conf/cisc/XuLXLHMLH18}, Link \cite{lee2022link}, Lore \cite{DBLP:journals/tdsc/Holm23}, Mace \cite{monshizadeh2014mace}, MaliceScript \cite{liu2018malicescript}, Masat \cite{mjihil2015masat}, Mirage \cite{cayre2019mirage}, Mitch \cite{DBLP:conf/eurosp/CalzavaraCFRT19}, MoScan \cite{Wei2021}, NAUTILUS \cite{DBLP:conf/uss/DengZLL00YW23}, NAVEX \cite{DBLP:conf/uss/AlhuzaliGEV18}, No Name (CSRF) \cite{9357029}, No Name (TTCN-3) \cite{DBLP:conf/isncc/LealT18}, NodeXP \cite{ntantogian2021nodexp}, OSV \cite{kasemsuwan2017osv}, ObjectMap \cite{DBLP:conf/pci/Koutroumpouchos19}, Owfuzz \cite{10.1145/3558482.3590174}, PJCT \cite{jain2015pjct}, PURITY \cite{bozic2015purity}, PentestGPT \cite{DBLP:journals/corr/abs-2308-06782}, PhpSAFE \cite{DBLP:conf/dsn/NunesFV15}, Project Achilles \cite{DBLP:conf/kbse/SaccenteDDCX19}, Pyciuti \cite{muralidharan2023pyciuti}, RAT \cite{Amouei2022}, ROSploit \cite{DBLP:conf/irc/RiveraLS19}, RT-RCT \cite{fagroud2021rt}, Revealer \cite{DBLP:conf/sp/LiuZM21}, RiscyROP \cite{10.1145/3545948.3545997}, Robin \cite{DBLP:journals/corr/abs-2007-06629}, SOA-Scanner \cite{antunes2013soa}, SVED \cite{DBLP:conf/milcom/HolmS16}, Scanner++ \cite{10.1145/3517036}, SerialDetector \cite{DBLP:conf/ndss/ShcherbakovB21}, ShoVAT \cite{genge2016shovat}, Snout \cite{Mikulskis2019}, Spicy \cite{DBLP:conf/acsac/SommerAH16}, SuperEye \cite{DBLP:conf/icccsec/LiYWLYH19}, TChecker \cite{DBLP:conf/ccs/Luo0022}, TORPEDO \cite{olivo2015detecting}, UE Security Reloaded \cite{10.1145/3558482.3590194}, VAPE-BRIDGE \cite{Vimala2022}, VERA \cite{blome2013vera}, VUDDY \cite{DBLP:conf/sp/KimWLO17}, VulCNN \cite{10.1145/3510003.3510229}, VulDeePecker \cite{DBLP:conf/ndss/LiZXO0WDZ18}, VulPecker \cite{DBLP:conf/acsac/LiZXJQH16}, Vulcan \cite{kamongi2013vulcan}, Vulnet \cite{8922605}, Vulnsloit \cite{10.1007/978-3-030-64881-7_6}, WAPTT \cite{dhuric2014waptt}, WebFuzz \cite{DBLP:conf/esorics/RooijCKPA21}, WebVIM \cite{DBLP:conf/iciis/RankothgeRS20} \tabularnewline 
\hline
Exploitation & Chainsaw \cite{DBLP:conf/ccs/AlhuzaliEGV16}, Commix \cite{DBLP:journals/ijisec/StasinopoulosNX19}, ELAID \cite{DBLP:journals/cybersec/XuXLH20}, ESASCF \cite{Ghanem2023}, FUGIO \cite{DBLP:conf/uss/ParkKJS22}, Firmaster \cite{8457340}, Gail-PT \cite{chen2023gail}, LAID \cite{DBLP:conf/cisc/XuLXLHMLH18}, LTESniffer \cite{10.1145/3558482.3590196}, Lore \cite{DBLP:journals/tdsc/Holm23}, MAIT \cite{yucel2021mait}, Mace \cite{monshizadeh2014mace}, MaliceScript \cite{liu2018malicescript}, Mirage \cite{cayre2019mirage}, Mitch \cite{DBLP:conf/eurosp/CalzavaraCFRT19}, NAUTILUS \cite{DBLP:conf/uss/DengZLL00YW23}, NAVEX \cite{DBLP:conf/uss/AlhuzaliGEV18}, NetCAT \cite{Kurth2020}, No Name (TTCN-3) \cite{DBLP:conf/isncc/LealT18}, NodeXP \cite{ntantogian2021nodexp}, OSV \cite{kasemsuwan2017osv}, Owfuzz \cite{10.1145/3558482.3590174}, PURITY \cite{bozic2015purity}, PentestGPT \cite{DBLP:journals/corr/abs-2308-06782}, Pyciuti \cite{muralidharan2023pyciuti}, ROSploit \cite{DBLP:conf/irc/RiveraLS19}, Revealer \cite{DBLP:conf/sp/LiuZM21}, RiscyROP \cite{10.1145/3545948.3545997}, Robin \cite{DBLP:journals/corr/abs-2007-06629}, SOA-Scanner \cite{antunes2013soa}, SVED \cite{DBLP:conf/milcom/HolmS16}, SerialDetector \cite{DBLP:conf/ndss/ShcherbakovB21}, Snout \cite{Mikulskis2019}, TORPEDO \cite{olivo2015detecting}, Untangle \cite{DBLP:conf/dimva/BertaniBBCZP23}, VAPE-BRIDGE \cite{Vimala2022}, Vulnsloit \cite{10.1007/978-3-030-64881-7_6}, WAPTT \cite{dhuric2014waptt}, WebVIM \cite{DBLP:conf/iciis/RankothgeRS20} \tabularnewline 
\hline
Post Exploitation & ADaMs \cite{pasquini2021reducing}, AVAIN \cite{Egert2019}, Delta \cite{lee2017delta}, Diane \cite{redini2021diane}, ESRFuzzer \cite{DBLP:journals/cybersec/ZhangHJSLZZL21}, GNPassGAN \cite{Yu2022}, HILTI \cite{Sommer2014}, IoTFuzzer \cite{DBLP:conf/ndss/ChenDZZL0LSYZ18}, Mirage \cite{cayre2019mirage}, NeuralNetworkCracking \cite{melicher2016fast}, NoCrack \cite{Chatterjee2015}, OMEN \cite{durmuth2015omen}, OSV \cite{kasemsuwan2017osv}, PassGAN \cite{DBLP:conf/acns/HitajGAP19}, PassGPT \cite{rando2023passgpt}, PasswordCrackingTraining \cite{DBLP:conf/esorics/CampiFL22}, Pyciuti \cite{muralidharan2023pyciuti}, SemanticGuesser \cite{veras2014semantic}, Snout \cite{Mikulskis2019}, Spicy \cite{DBLP:conf/acsac/SommerAH16}, Untangle \cite{DBLP:conf/dimva/BertaniBBCZP23} \tabularnewline 
\hline
Reporting & ESASCF \cite{Ghanem2023}, Firmaster \cite{8457340}, No Name (TTCN-3) \cite{DBLP:conf/isncc/LealT18}, Pyciuti \cite{muralidharan2023pyciuti} \tabularnewline 
 
\hline
		\caption{PTES classification\label{tab:ptes}}
		\end{longtable}

\clearpage
					\begin{longtable}{|>{\RaggedRight}p{0.21\textwidth}|>{\RaggedRight}p{0.79\textwidth}|}
				\hline 
				\textit{Mitre ATT\&CK} & \textit{Tools}\tabularnewline
\hline
\endhead
Reconnaissance & AIBugHunter \cite{Fu2023}, ARMONY \cite{chen2013armory}, AVAIN \cite{Egert2019}, AVAIN \cite{Egert2019}, Autosploit \cite{Moscovich2020}, Bbuzz \cite{DBLP:conf/milcom/BlumbergsV17}, Black Ostrich \cite{Eriksson2023}, Black Widow \cite{DBLP:conf/sp/ErikssonPS21}, Bleem \cite{Luo2023}, Cairis \cite{Faily2020}, Censys \cite{censys15}, Chainsaw \cite{DBLP:conf/ccs/AlhuzaliEGV16}, Chucky \cite{DBLP:conf/ccs/YamaguchiWGR13}, Commix \cite{DBLP:journals/ijisec/StasinopoulosNX19}, CryptoGuard \cite{DBLP:conf/ccs/RahamanXASTFKY19}, CuPerFuzzer \cite{DBLP:conf/sp/LiDLDG21}, DFBC \cite{ng2021dfbc}, Deemon \cite{DBLP:conf/ccs/PellegrinoJ0BR17}, Delta \cite{lee2017delta}, Delta \cite{lee2017delta}, Diane \cite{redini2021diane}, EBF \cite{Aljaafari2021}, ELAID \cite{DBLP:journals/cybersec/XuXLH20}, ESASCF \cite{Ghanem2023}, ESRFuzzer \cite{DBLP:journals/cybersec/ZhangHJSLZZL21}, ESSecA \cite{rak2022esseca}, FUGIO \cite{DBLP:conf/uss/ParkKJS22}, FUSE \cite{DBLP:conf/ndss/LeeWLS20}, Firmaster \cite{8457340}, Gail-PT \cite{chen2023gail}, Gail-PT \cite{chen2023gail}, HILTI \cite{Sommer2014}, HILTI \cite{Sommer2014}, IoTFuzzer \cite{DBLP:conf/ndss/ChenDZZL0LSYZ18}, JCOMIX \cite{stallenberg_jcomix_2019}, LAID \cite{DBLP:conf/cisc/XuLXLHMLH18}, LTESniffer \cite{10.1145/3558482.3590196}, Link \cite{lee2022link}, Lore \cite{DBLP:journals/tdsc/Holm23}, Mace \cite{monshizadeh2014mace}, MaliceScript \cite{liu2018malicescript}, MaliceScript \cite{liu2018malicescript}, Masat \cite{mjihil2015masat}, Mirage \cite{cayre2019mirage}, Mirage \cite{cayre2019mirage}, Mitch \cite{DBLP:conf/eurosp/CalzavaraCFRT19}, MoScan \cite{Wei2021}, NAUTILUS \cite{DBLP:conf/uss/DengZLL00YW23}, NAVEX \cite{DBLP:conf/uss/AlhuzaliGEV18}, No Name (CSRF) \cite{9357029}, No Name (TTCN-3) \cite{DBLP:conf/isncc/LealT18}, No Name (TTCN-3) \cite{DBLP:conf/isncc/LealT18}, NodeXP \cite{ntantogian2021nodexp}, OSV \cite{kasemsuwan2017osv}, ObjectMap \cite{DBLP:conf/pci/Koutroumpouchos19}, Owfuzz \cite{10.1145/3558482.3590174}, PURITY \cite{bozic2015purity}, PenQuest \cite{DBLP:journals/virology/LuhTTSJ20}, PentestGPT \cite{DBLP:journals/corr/abs-2308-06782}, PhpSAFE \cite{DBLP:conf/dsn/NunesFV15}, Pyciuti \cite{muralidharan2023pyciuti}, Pyciuti \cite{muralidharan2023pyciuti}, RAT \cite{Amouei2022}, ROSploit \cite{DBLP:conf/irc/RiveraLS19}, RT-RCT \cite{fagroud2021rt}, RT-RCT \cite{fagroud2021rt}, Revealer \cite{DBLP:conf/sp/LiuZM21}, RiscyROP \cite{10.1145/3545948.3545997}, Robin \cite{DBLP:journals/corr/abs-2007-06629}, SOA-Scanner \cite{antunes2013soa}, SVED \cite{DBLP:conf/milcom/HolmS16}, Scanner++ \cite{10.1145/3517036}, SerialDetector \cite{DBLP:conf/ndss/ShcherbakovB21}, ShoVAT \cite{genge2016shovat}, ShoVAT \cite{genge2016shovat}, Snout \cite{Mikulskis2019}, Snout \cite{Mikulskis2019}, Spicy \cite{DBLP:conf/acsac/SommerAH16}, Spicy \cite{DBLP:conf/acsac/SommerAH16}, SuperEye \cite{DBLP:conf/icccsec/LiYWLYH19}, TAMELESS \cite{DBLP:journals/tdsc/ValenzaKSL23}, TChecker \cite{DBLP:conf/ccs/Luo0022}, TORPEDO \cite{olivo2015detecting}, UE Security Reloaded \cite{10.1145/3558482.3590194}, VAPE-BRIDGE \cite{Vimala2022}, VERA \cite{blome2013vera}, VUDDY \cite{DBLP:conf/sp/KimWLO17}, VulCNN \cite{10.1145/3510003.3510229}, VulDeePecker \cite{DBLP:conf/ndss/LiZXO0WDZ18}, VulPecker \cite{DBLP:conf/acsac/LiZXJQH16}, Vulcan \cite{kamongi2013vulcan}, Vulnet \cite{8922605}, Vulnsloit \cite{10.1007/978-3-030-64881-7_6}, WAPTT \cite{dhuric2014waptt}, WebFuzz \cite{DBLP:conf/esorics/RooijCKPA21}, WebVIM \cite{DBLP:conf/iciis/RankothgeRS20} \tabularnewline 
\hline
Resource Development & AIBugHunter \cite{Fu2023}, Autosploit \cite{Moscovich2020}, Chucky \cite{DBLP:conf/ccs/YamaguchiWGR13}, CuPerFuzzer \cite{DBLP:conf/sp/LiDLDG21}, ELAID \cite{DBLP:journals/cybersec/XuXLH20}, ESASCF \cite{Ghanem2023}, HARMer \cite{DBLP:journals/access/EnochHMLAK20}, HILTI \cite{Sommer2014}, LAID \cite{DBLP:conf/cisc/XuLXLHMLH18}, MAIT \cite{yucel2021mait}, MAL \cite{johnson2018meta}, Owfuzz \cite{10.1145/3558482.3590174}, PJCT \cite{jain2015pjct}, PJCT \cite{jain2015pjct}, Project Achilles \cite{DBLP:conf/kbse/SaccenteDDCX19}, Revealer \cite{DBLP:conf/sp/LiuZM21}, Spicy \cite{DBLP:conf/acsac/SommerAH16}, UE Security Reloaded \cite{10.1145/3558482.3590194}, Untangle \cite{DBLP:conf/dimva/BertaniBBCZP23}, VUDDY \cite{DBLP:conf/sp/KimWLO17}, VulCNN \cite{10.1145/3510003.3510229}, VulPecker \cite{DBLP:conf/acsac/LiZXJQH16} \tabularnewline 
\hline
Initial Access & Black Ostrich \cite{Eriksson2023}, Black Widow \cite{DBLP:conf/sp/ErikssonPS21}, Censys \cite{censys15}, Chainsaw \cite{DBLP:conf/ccs/AlhuzaliEGV16}, Commix \cite{DBLP:journals/ijisec/StasinopoulosNX19}, Deemon \cite{DBLP:conf/ccs/PellegrinoJ0BR17}, ESASCF \cite{Ghanem2023}, ESSecA \cite{rak2022esseca}, FUGIO \cite{DBLP:conf/uss/ParkKJS22}, FUSE \cite{DBLP:conf/ndss/LeeWLS20}, Firmaster \cite{8457340}, Gail-PT \cite{chen2023gail}, JCOMIX \cite{stallenberg_jcomix_2019}, Link \cite{lee2022link}, Lore \cite{DBLP:journals/tdsc/Holm23}, MAL \cite{johnson2018meta}, Mace \cite{monshizadeh2014mace}, MaliceScript \cite{liu2018malicescript}, Masat \cite{mjihil2015masat}, Mitch \cite{DBLP:conf/eurosp/CalzavaraCFRT19}, NAUTILUS \cite{DBLP:conf/uss/DengZLL00YW23}, NAVEX \cite{DBLP:conf/uss/AlhuzaliGEV18}, NetCAT \cite{Kurth2020}, No Name (CSRF) \cite{9357029}, NodeXP \cite{ntantogian2021nodexp}, OSV \cite{kasemsuwan2017osv}, ObjectMap \cite{DBLP:conf/pci/Koutroumpouchos19}, PURITY \cite{bozic2015purity}, PentestGPT \cite{DBLP:journals/corr/abs-2308-06782}, PhpSAFE \cite{DBLP:conf/dsn/NunesFV15}, Pyciuti \cite{muralidharan2023pyciuti}, RAT \cite{Amouei2022}, Revealer \cite{DBLP:conf/sp/LiuZM21}, Robin \cite{DBLP:journals/corr/abs-2007-06629}, SOA-Scanner \cite{antunes2013soa}, SVED \cite{DBLP:conf/milcom/HolmS16}, Scanner++ \cite{10.1145/3517036}, SerialDetector \cite{DBLP:conf/ndss/ShcherbakovB21}, ShoVAT \cite{genge2016shovat}, TChecker \cite{DBLP:conf/ccs/Luo0022}, TORPEDO \cite{olivo2015detecting}, VAPE-BRIDGE \cite{Vimala2022}, VERA \cite{blome2013vera}, Vulcan \cite{kamongi2013vulcan}, Vulnet \cite{8922605}, WAPTT \cite{dhuric2014waptt}, WebFuzz \cite{DBLP:conf/esorics/RooijCKPA21}, WebVIM \cite{DBLP:conf/iciis/RankothgeRS20} \tabularnewline 
\hline
Execution & Bbuzz \cite{DBLP:conf/milcom/BlumbergsV17}, ESASCF \cite{Ghanem2023}, Lore \cite{DBLP:journals/tdsc/Holm23}, Mirage \cite{cayre2019mirage}, PentestGPT \cite{DBLP:journals/corr/abs-2308-06782}, ROSploit \cite{DBLP:conf/irc/RiveraLS19}, RiscyROP \cite{10.1145/3545948.3545997}, SVED \cite{DBLP:conf/milcom/HolmS16}, Vulnsloit \cite{10.1007/978-3-030-64881-7_6} \tabularnewline 
\hline
Persistence &  \tabularnewline 
\hline
Privilege Escalation &  \tabularnewline 
\hline
Defense Evasion &  \tabularnewline 
\hline
Credential Access & ADaMs \cite{pasquini2021reducing}, Firmaster \cite{8457340}, GNPassGAN \cite{Yu2022}, LTESniffer \cite{10.1145/3558482.3590196}, NeuralNetworkCracking \cite{melicher2016fast}, NoCrack \cite{Chatterjee2015}, OMEN \cite{durmuth2015omen}, PassGAN \cite{DBLP:conf/acns/HitajGAP19}, PassGPT \cite{rando2023passgpt}, PasswordCrackingTraining \cite{DBLP:conf/esorics/CampiFL22}, SemanticGuesser \cite{veras2014semantic} \tabularnewline 
\hline
Discovery & AVAIN \cite{Egert2019}, Cairis \cite{Faily2020}, Firmaster \cite{8457340}, HILTI \cite{Sommer2014}, Masat \cite{mjihil2015masat}, PenQuest \cite{DBLP:journals/virology/LuhTTSJ20}, RT-RCT \cite{fagroud2021rt}, Snout \cite{Mikulskis2019}, Spicy \cite{DBLP:conf/acsac/SommerAH16}, TAMELESS \cite{DBLP:journals/tdsc/ValenzaKSL23}, Vulcan \cite{kamongi2013vulcan} \tabularnewline 
\hline
Lateral Movement &  \tabularnewline 
\hline
Collection & HILTI \cite{Sommer2014}, Spicy \cite{DBLP:conf/acsac/SommerAH16} \tabularnewline 
\hline
Command And Control &  \tabularnewline 
\hline
Exfiltration &  \tabularnewline 
\hline
Impact & Revealer \cite{DBLP:conf/sp/LiuZM21}, TORPEDO \cite{olivo2015detecting} \tabularnewline  
\hline
				\caption{Mitre ATT\&CK classification\label{tab:mitre}}
				\end{longtable}

\clearpage
					\begin{longtable}[h!]{|>{\RaggedRight}p{0.25\textwidth}|>{\RaggedRight}p{0.75\textwidth}|}
				\hline 
				\textit{Mitre ATT\&CK} & \textit{Tools}\tabularnewline
\hline
\endhead
Collection: Adversary-In-The-Middle & HILTI \cite{Sommer2014}, Spicy \cite{DBLP:conf/acsac/SommerAH16} \tabularnewline 
\hline
Credential Access: Brute Force: Password Cracking & GNPassGAN \cite{Yu2022}, PassGAN \cite{DBLP:conf/acns/HitajGAP19}, PasswordCrackingTraining \cite{DBLP:conf/esorics/CampiFL22} \tabularnewline 
\hline
Discovery: Cloud Infrastructure Discovery & MASAT \cite{mjihil2015masat}, VULCAN \cite{kamongi2013vulcan} \tabularnewline 
\hline
Discovery: Network Service Discovery & AVAIN \cite{Egert2019}, Firmaster \cite{8457340}, HILTI \cite{Sommer2014}, RT-RCT \cite{fagroud2021rt}, Snout \cite{Mikulskis2019}, Spicy \cite{DBLP:conf/acsac/SommerAH16} \tabularnewline 
\hline
Enterprise: Credential Access: Brute Force & Firmaster \cite{8457340} \tabularnewline 
\hline
Enterprise: Credential Access: Network Sniffing & LTESniffer \cite{10.1145/3558482.3590196} \tabularnewline 
\hline
Enterprise: Impact: Service Stop & TORPEDO \cite{olivo2015detecting} \tabularnewline 
\hline
Enterprise: Initial Access: External Remote Services & NetCAT \cite{Kurth2020} \tabularnewline 
\hline
Execution & Bbuzz \cite{DBLP:conf/milcom/BlumbergsV17}, Lore \cite{DBLP:journals/tdsc/Holm23}, Mirage \cite{cayre2019mirage}, PentestGPT \cite{DBLP:journals/corr/abs-2308-06782}, ROSploit \cite{DBLP:conf/irc/RiveraLS19}, SVED \cite{DBLP:conf/milcom/HolmS16}, Vulnsloit \cite{10.1007/978-3-030-64881-7_6} \tabularnewline 
\hline
Execution: Inter-Process Communication & RiscyROP \cite{10.1145/3545948.3545997} \tabularnewline 
\hline
Gather Victim Network Information & Lore \cite{DBLP:journals/tdsc/Holm23}, PentestGPT \cite{DBLP:journals/corr/abs-2308-06782}, SVED \cite{DBLP:conf/milcom/HolmS16} \tabularnewline 
\hline
Impact: Endpoint Denial Of Service & Revealer \cite{DBLP:conf/sp/LiuZM21} \tabularnewline 
\hline
Initial Access & Gail-PT \cite{chen2023gail}, Lore \cite{DBLP:journals/tdsc/Holm23}, OSV \cite{kasemsuwan2017osv}, PentestGPT \cite{DBLP:journals/corr/abs-2308-06782}, SVED \cite{DBLP:conf/milcom/HolmS16} \tabularnewline 
\hline
Initial Access: Exploit Public Facing Application & Commix \cite{DBLP:journals/ijisec/StasinopoulosNX19}, JCOMIX \cite{stallenberg_jcomix_2019}, Mitch \cite{DBLP:conf/eurosp/CalzavaraCFRT19}, No Name (CSRF) \cite{9357029}, PURITY \cite{bozic2015purity}, Puciuty \cite{muralidharan2023pyciuti}, Robin \cite{DBLP:journals/corr/abs-2007-06629}, Vulnet \cite{8922605}, WebVIM \cite{DBLP:conf/iciis/RankothgeRS20}, ZGrab \cite{censys15} \tabularnewline 
\hline
Initial Access: Exploit Public Facing-Application & WAPTT \cite{dhuric2014waptt} \tabularnewline 
\hline
Initial Access: Exploit Public-Facing Application & Black Ostrich \cite{Eriksson2023}, Black Widow \cite{DBLP:conf/sp/ErikssonPS21}, Chainsaw \cite{DBLP:conf/ccs/AlhuzaliEGV16}, Deemon \cite{DBLP:conf/ccs/PellegrinoJ0BR17}, FUGIO \cite{DBLP:conf/uss/ParkKJS22}, FUSE \cite{DBLP:conf/ndss/LeeWLS20}, Firmaster \cite{8457340}, Link \cite{lee2022link}, MASAT \cite{mjihil2015masat}, Mace \cite{monshizadeh2014mace}, MaliceScript \cite{liu2018malicescript}, NAUTILUS \cite{DBLP:conf/uss/DengZLL00YW23}, NAVEX \cite{DBLP:conf/uss/AlhuzaliGEV18}, NodeXP \cite{ntantogian2021nodexp}, ObjectMap \cite{DBLP:conf/pci/Koutroumpouchos19}, PhpSAFE \cite{DBLP:conf/dsn/NunesFV15}, Revealer \cite{DBLP:conf/sp/LiuZM21}, SOA-Scanner \cite{antunes2013soa}, Scanner++ \cite{10.1145/3517036}, SerialDetector \cite{DBLP:conf/ndss/ShcherbakovB21}, ShoVAT \cite{genge2016shovat}, TChecker \cite{DBLP:conf/ccs/Luo0022}, TORPEDO \cite{olivo2015detecting}, VAPE-BRIDGE \cite{Vimala2022}, VERA \cite{blome2013vera}, VULCAN \cite{kamongi2013vulcan}, WebFuzz \cite{DBLP:conf/esorics/RooijCKPA21} \tabularnewline 
\hline
Reconnaissance: Active Scanning & LTESniffer \cite{10.1145/3558482.3590196}, TORPEDO \cite{olivo2015detecting} \tabularnewline 
\hline
Reconnaissance: Active Scanning Vulnerability Scanning & NodeXP \cite{ntantogian2021nodexp} \tabularnewline 
\hline
Reconnaissance: Active Scanning: Vulnerability Scanning & AIBugHunter \cite{Fu2023}, ARMONY \cite{chen2013armory}, AVAIN \cite{Egert2019}, Autosploit \cite{Moscovich2020}, Bbuzz \cite{DBLP:conf/milcom/BlumbergsV17}, Black Ostrich \cite{Eriksson2023}, Black Widow \cite{DBLP:conf/sp/ErikssonPS21}, Chainsaw \cite{DBLP:conf/ccs/AlhuzaliEGV16}, Chucky \cite{DBLP:conf/ccs/YamaguchiWGR13}, Commix \cite{DBLP:journals/ijisec/StasinopoulosNX19}, CryptoGuard \cite{DBLP:conf/ccs/RahamanXASTFKY19}, CuPerFuzzer \cite{DBLP:conf/sp/LiDLDG21}, DELTA \cite{lee2017delta}, DIANE \cite{redini2021diane}, Deemon \cite{DBLP:conf/ccs/PellegrinoJ0BR17}, EBF \cite{Aljaafari2021}, ELAID \cite{DBLP:journals/cybersec/XuXLH20}, ESRFuzzer \cite{DBLP:journals/cybersec/ZhangHJSLZZL21}, FUGIO \cite{DBLP:conf/uss/ParkKJS22}, FUSE \cite{DBLP:conf/ndss/LeeWLS20}, Firmaster \cite{8457340}, Gail-PT \cite{chen2023gail}, HILTI \cite{Sommer2014}, IoTFuzzer \cite{DBLP:conf/ndss/ChenDZZL0LSYZ18}, JCOMIX \cite{stallenberg_jcomix_2019}, LAID \cite{DBLP:conf/cisc/XuLXLHMLH18}, Link \cite{lee2022link}, Lore \cite{DBLP:journals/tdsc/Holm23}, MASAT \cite{mjihil2015masat}, Mace \cite{monshizadeh2014mace}, MaliceScript \cite{liu2018malicescript}, Mirage \cite{cayre2019mirage}, Mitch \cite{DBLP:conf/eurosp/CalzavaraCFRT19}, NAUTILUS \cite{DBLP:conf/uss/DengZLL00YW23}, NAVEX \cite{DBLP:conf/uss/AlhuzaliGEV18}, No Name (CSRF) \cite{9357029}, No Name (TTCN-3) \cite{DBLP:conf/isncc/LealT18}, OSV \cite{kasemsuwan2017osv}, ObjectMap \cite{DBLP:conf/pci/Koutroumpouchos19}, Owfuzz \cite{10.1145/3558482.3590174}, PURITY \cite{bozic2015purity}, PentestGPT \cite{DBLP:journals/corr/abs-2308-06782}, PhpSAFE \cite{DBLP:conf/dsn/NunesFV15}, Puciuty \cite{muralidharan2023pyciuti}, ROSploit \cite{DBLP:conf/irc/RiveraLS19}, RT-RCT \cite{fagroud2021rt}, Revealer \cite{DBLP:conf/sp/LiuZM21}, RiscyROP \cite{10.1145/3545948.3545997}, Robin \cite{DBLP:journals/corr/abs-2007-06629}, SOA-Scanner \cite{antunes2013soa}, SVED \cite{DBLP:conf/milcom/HolmS16}, Scanner++ \cite{10.1145/3517036}, SerialDetector \cite{DBLP:conf/ndss/ShcherbakovB21}, ShoVAT \cite{genge2016shovat}, Snout \cite{Mikulskis2019}, Spicy \cite{DBLP:conf/acsac/SommerAH16}, SuperEye \cite{DBLP:conf/icccsec/LiYWLYH19}, TChecker \cite{DBLP:conf/ccs/Luo0022}, UE Security Reloaded \cite{10.1145/3558482.3590194}, VAPE-BRIDGE \cite{Vimala2022}, VERA \cite{blome2013vera}, VUDDY \cite{DBLP:conf/sp/KimWLO17}, VULCAN \cite{kamongi2013vulcan}, VulCNN \cite{10.1145/3510003.3510229}, VulDeePecker \cite{DBLP:conf/ndss/LiZXO0WDZ18}, VulPecker \cite{DBLP:conf/acsac/LiZXJQH16}, Vulnet \cite{8922605}, Vulnsloit \cite{10.1007/978-3-030-64881-7_6}, WAPTT \cite{dhuric2014waptt}, WebFuzz \cite{DBLP:conf/esorics/RooijCKPA21}, WebVIM \cite{DBLP:conf/iciis/RankothgeRS20}, ZGrab \cite{censys15} \tabularnewline 
\hline
Reconnaissance: Gather Victim Identity Information & DFBC \cite{ng2021dfbc} \tabularnewline 
\hline
Reconnaissance: Gather Victim Network Information & AVAIN \cite{Egert2019}, DELTA \cite{lee2017delta}, Gail-PT \cite{chen2023gail}, HILTI \cite{Sommer2014}, MaliceScript \cite{liu2018malicescript}, Mirage \cite{cayre2019mirage}, Puciuty \cite{muralidharan2023pyciuti}, RT-RCT \cite{fagroud2021rt}, ShoVAT \cite{genge2016shovat}, Snout \cite{Mikulskis2019}, Spicy \cite{DBLP:conf/acsac/SommerAH16} \tabularnewline 
\hline
Reconnaissance: Gather Victim Network Information: Network Topology & No Name (TTCN-3) \cite{DBLP:conf/isncc/LealT18} \tabularnewline 
\hline
Reconnaissance: Resource Development & HARMer \cite{DBLP:journals/access/EnochHMLAK20} \tabularnewline 
\hline
Resource Development: Develop Capabilities & HILTI \cite{Sommer2014}, PICT \cite{jain2015pjct}, Spicy \cite{DBLP:conf/acsac/SommerAH16} \tabularnewline 
\hline
Resource Development: Develop Capabilities: Exploits & ELAID \cite{DBLP:journals/cybersec/XuXLH20}, LAID \cite{DBLP:conf/cisc/XuLXLHMLH18}, Owfuzz \cite{10.1145/3558482.3590174}, Project Achilles \cite{DBLP:conf/kbse/SaccenteDDCX19}, UE Security Reloaded \cite{10.1145/3558482.3590194}, VulCNN \cite{10.1145/3510003.3510229} \tabularnewline 
\hline
Resource Development: Develop Capabilities: Malware & MAIT \cite{yucel2021mait} \tabularnewline 
\hline
Resource Development: Obtain Capabilities: Exploits & AIBugHunter \cite{Fu2023}, Autosploit \cite{Moscovich2020}, Chucky \cite{DBLP:conf/ccs/YamaguchiWGR13}, CuPerFuzzer \cite{DBLP:conf/sp/LiDLDG21}, PICT \cite{jain2015pjct}, Revealer \cite{DBLP:conf/sp/LiuZM21}, VUDDY \cite{DBLP:conf/sp/KimWLO17}, VulPecker \cite{DBLP:conf/acsac/LiZXJQH16} \tabularnewline 
\hline
				\caption{Mitre ATT\&CK classification (Details)\label{tab:mitre-details}}
				\end{longtable}

\clearpage
				\begin{longtable}{|>{\RaggedRight}p{0.50\textwidth}|>{\RaggedRight}p{0.50\textwidth}|}
				\hline 
				\textit{CyBOK} & \textit{Tools}\tabularnewline
\hline
\endhead
Attacks \& Defences:  Adversarial Behaviours & Cairis \cite{Faily2020}, ESASCF \cite{Ghanem2023}, ESSecA \cite{rak2022esseca}, HARMer \cite{DBLP:journals/access/EnochHMLAK20}, Lore \cite{DBLP:journals/tdsc/Holm23}, MAL \cite{johnson2018meta}, PenQuest \cite{DBLP:journals/virology/LuhTTSJ20}, PenQuest \cite{DBLP:journals/virology/LuhTTSJ20}, SVED \cite{DBLP:conf/milcom/HolmS16}, TAMELESS \cite{DBLP:journals/tdsc/ValenzaKSL23} \tabularnewline 
\hline
Attacks \& Defences: Malware \& Attack Technology: Malware Analysis: Analysis Techniques: Static Analysis/Dynamic Analysis & MAIT \cite{yucel2021mait} \tabularnewline 
\hline
Human, Organisational \& Regulatory Aspects: Human Factors & ESSecA \cite{rak2022esseca}, TAMELESS \cite{DBLP:journals/tdsc/ValenzaKSL23} \tabularnewline 
\hline
Human, Organisational \& Regulatory Aspects: Privacy \& Online Rights: Privacy Engineering: Privacy Evaluation & DFBC \cite{ng2021dfbc} \tabularnewline 
\hline
Infrastructure Security: Applied Cryptography: Cryptographic Implementation: Api Design For Cryptographic Libraries & CryptoGuard \cite{DBLP:conf/ccs/RahamanXASTFKY19} \tabularnewline 
\hline
Infrastructure Security: Applied Cryptography: Cryptographic Implementation: Cryptographic Libraries & Firmaster \cite{8457340} \tabularnewline 
\hline
Infrastructure Security: Cyber Physical Systems & ESSecA \cite{rak2022esseca}, TAMELESS \cite{DBLP:journals/tdsc/ValenzaKSL23} \tabularnewline 
\hline
Infrastructure Security: Network Security & AVAIN \cite{Egert2019}, Cairis \cite{Faily2020}, Delta \cite{lee2017delta}, ESASCF \cite{Ghanem2023}, Gail-PT \cite{chen2023gail}, HARMer \cite{DBLP:journals/access/EnochHMLAK20}, HILTI \cite{Sommer2014}, Lore \cite{DBLP:journals/tdsc/Holm23}, Masat \cite{mjihil2015masat}, NetCAT \cite{Kurth2020}, SVED \cite{DBLP:conf/milcom/HolmS16}, Spicy \cite{DBLP:conf/acsac/SommerAH16} \tabularnewline 
\hline
Infrastructure Security: Network Security: Network Protocols And Their Security & OSV \cite{kasemsuwan2017osv}, SuperEye \cite{DBLP:conf/icccsec/LiYWLYH19}, Vulnsloit \cite{10.1007/978-3-030-64881-7_6} \tabularnewline 
\hline
Infrastructure Security: Network Security: Network Protocols And Their Security: Security At The Internet Layer & Bbuzz \cite{DBLP:conf/milcom/BlumbergsV17} \tabularnewline 
\hline
Infrastructure Security: Network Security: Network Protocols And Their Security: Security At The Internet Layer: Ipv6 Security & No Name (TTCN-3) \cite{DBLP:conf/isncc/LealT18} \tabularnewline 
\hline
Infrastructure Security: Network Security: Networking Applications & Vulcan \cite{kamongi2013vulcan} \tabularnewline 
\hline
Infrastructure Security: Network Security: Networking Applications: Local Area Networks & ESRFuzzer \cite{DBLP:journals/cybersec/ZhangHJSLZZL21}, Firmaster \cite{8457340}, HILTI \cite{Sommer2014}, No Name (TTCN-3) \cite{DBLP:conf/isncc/LealT18}, Pyciuti \cite{muralidharan2023pyciuti}, Spicy \cite{DBLP:conf/acsac/SommerAH16} \tabularnewline 
\hline
Infrastructure Security: Network Security: Networking Applications: Wireless Networks & ESRFuzzer \cite{DBLP:journals/cybersec/ZhangHJSLZZL21}, Firmaster \cite{8457340}, LTESniffer \cite{10.1145/3558482.3590196}, Owfuzz \cite{10.1145/3558482.3590174}, RT-RCT \cite{fagroud2021rt}, Snout \cite{Mikulskis2019}, UE Security Reloaded \cite{10.1145/3558482.3590194} \tabularnewline 
\hline
Infrastructure Security: Network Security: Other Network Security Topics: Cloud And Data Center Security & Masat \cite{mjihil2015masat}, Vulcan \cite{kamongi2013vulcan} \tabularnewline 
\hline
Infrastructure Security: Network Security: Software-Defined Networking And Network Function Virtualization & Delta \cite{lee2017delta} \tabularnewline 
\hline
Infrastructure Security: Physical Layer \& Telecommunications Security: Identification: Attacks On Physical Layer Identification & Snout \cite{Mikulskis2019} \tabularnewline 
\hline
Operating Systems \& Virtualization Security: Operating System Hardening & ROSploit \cite{DBLP:conf/irc/RiveraLS19} \tabularnewline 
\hline
Physical Layer \& Telecommunications Security: Physical Layer Security Of Selected Communication Technologies: Cellular Networks: 4G (Lte) & LTESniffer \cite{10.1145/3558482.3590196} \tabularnewline 
\hline
Physical Layer \& Telecommunications Security: Physical Layer Security Of Selected Communication Technologies: Cellular Networks: 5G & UE Security Reloaded \cite{10.1145/3558482.3590194} \tabularnewline 
\hline
Resource Development: Develop Capabilities: Exploits & ESASCF \cite{Ghanem2023} \tabularnewline 
\hline
Software And Platform Security: Software Security: Categories Of Vulnerabilities: Memory Management Vulnerabilities & ARMONY \cite{chen2013armory}, ELAID \cite{DBLP:journals/cybersec/XuXLH20}, IoTFuzzer \cite{DBLP:conf/ndss/ChenDZZL0LSYZ18}, LAID \cite{DBLP:conf/cisc/XuLXLHMLH18}, WAPTT \cite{dhuric2014waptt} \tabularnewline 
\hline
Software And Platform Security: Software Security: Detection Of Vulnerabilities & ARMONY \cite{chen2013armory}, AVAIN \cite{Egert2019}, Autosploit \cite{Moscovich2020}, Bbuzz \cite{DBLP:conf/milcom/BlumbergsV17}, Black Ostrich \cite{Eriksson2023}, Black Widow \cite{DBLP:conf/sp/ErikssonPS21}, Cairis \cite{Faily2020}, Censys \cite{censys15}, Chainsaw \cite{DBLP:conf/ccs/AlhuzaliEGV16}, Commix \cite{DBLP:journals/ijisec/StasinopoulosNX19}, CryptoGuard \cite{DBLP:conf/ccs/RahamanXASTFKY19}, Deemon \cite{DBLP:conf/ccs/PellegrinoJ0BR17}, EBF \cite{Aljaafari2021}, ESASCF \cite{Ghanem2023}, FUGIO \cite{DBLP:conf/uss/ParkKJS22}, FUSE \cite{DBLP:conf/ndss/LeeWLS20}, Firmaster \cite{8457340}, HILTI \cite{Sommer2014}, JCOMIX \cite{stallenberg_jcomix_2019}, Link \cite{lee2022link}, Mace \cite{monshizadeh2014mace}, MaliceScript \cite{liu2018malicescript}, Mirage \cite{cayre2019mirage}, Mitch \cite{DBLP:conf/eurosp/CalzavaraCFRT19}, MoScan \cite{Wei2021}, NAUTILUS \cite{DBLP:conf/uss/DengZLL00YW23}, NAVEX \cite{DBLP:conf/uss/AlhuzaliGEV18}, No Name (CSRF) \cite{9357029}, No Name (TTCN-3) \cite{DBLP:conf/isncc/LealT18}, NodeXP \cite{ntantogian2021nodexp}, OSV \cite{kasemsuwan2017osv}, ObjectMap \cite{DBLP:conf/pci/Koutroumpouchos19}, Owfuzz \cite{10.1145/3558482.3590174}, PJCT \cite{jain2015pjct}, PURITY \cite{bozic2015purity}, PentestGPT \cite{DBLP:journals/corr/abs-2308-06782}, Project Achilles \cite{DBLP:conf/kbse/SaccenteDDCX19}, Pyciuti \cite{muralidharan2023pyciuti}, RAT \cite{Amouei2022}, ROSploit \cite{DBLP:conf/irc/RiveraLS19}, RT-RCT \cite{fagroud2021rt}, Revealer \cite{DBLP:conf/sp/LiuZM21}, SOA-Scanner \cite{antunes2013soa}, Scanner++ \cite{10.1145/3517036}, SerialDetector \cite{DBLP:conf/ndss/ShcherbakovB21}, ShoVAT \cite{genge2016shovat}, Snout \cite{Mikulskis2019}, Spicy \cite{DBLP:conf/acsac/SommerAH16}, SuperEye \cite{DBLP:conf/icccsec/LiYWLYH19}, TChecker \cite{DBLP:conf/ccs/Luo0022}, TORPEDO \cite{olivo2015detecting}, UE Security Reloaded \cite{10.1145/3558482.3590194}, VAPE-BRIDGE \cite{Vimala2022}, VERA \cite{blome2013vera}, VulDeePecker \cite{DBLP:conf/ndss/LiZXO0WDZ18}, Vulcan \cite{kamongi2013vulcan}, Vulnet \cite{8922605}, Vulnsloit \cite{10.1007/978-3-030-64881-7_6}, WAPTT \cite{dhuric2014waptt}, WebFuzz \cite{DBLP:conf/esorics/RooijCKPA21}, WebVIM \cite{DBLP:conf/iciis/RankothgeRS20} \tabularnewline 
\hline
Software And Platform Security: Software Security: Detection Of Vulnerabilities: Dynamic Detection & Bbuzz \cite{DBLP:conf/milcom/BlumbergsV17}, Black Ostrich \cite{Eriksson2023}, CuPerFuzzer \cite{DBLP:conf/sp/LiDLDG21}, Diane \cite{redini2021diane}, EBF \cite{Aljaafari2021}, Project Achilles \cite{DBLP:conf/kbse/SaccenteDDCX19} \tabularnewline 
\hline
Software And Platform Security: Software Security: Detection Of Vulnerabilities: Dynamic Detection: Black-Box Fuzzing & Bleem \cite{Luo2023}, Delta \cite{lee2017delta}, IoTFuzzer \cite{DBLP:conf/ndss/ChenDZZL0LSYZ18}, Owfuzz \cite{10.1145/3558482.3590174} \tabularnewline 
\hline
Software And Platform Security: Software Security: Detection Of Vulnerabilities: Dynamic Detection: Generating Relevant Executions: Dynamic Symbolic Execution & RiscyROP \cite{10.1145/3545948.3545997} \tabularnewline 
\hline
Software And Platform Security: Software Security: Detection Of Vulnerabilities: Static Detection & AIBugHunter \cite{Fu2023}, Chucky \cite{DBLP:conf/ccs/YamaguchiWGR13}, ELAID \cite{DBLP:journals/cybersec/XuXLH20}, LAID \cite{DBLP:conf/cisc/XuLXLHMLH18}, PhpSAFE \cite{DBLP:conf/dsn/NunesFV15}, Untangle \cite{DBLP:conf/dimva/BertaniBBCZP23}, VUDDY \cite{DBLP:conf/sp/KimWLO17}, VulCNN \cite{10.1145/3510003.3510229}, VulPecker \cite{DBLP:conf/acsac/LiZXJQH16} \tabularnewline 
\hline
Software And Platform Security: Software Security: Dynamic Detection & WebFuzz \cite{DBLP:conf/esorics/RooijCKPA21} \tabularnewline 
\hline
Software And Platform Security: Software Security: Side-Channel Vulnerabilities & NetCAT \cite{Kurth2020} \tabularnewline 
\hline
Software And Platform Security: Web \& Mobile Security & Black Ostrich \cite{Eriksson2023}, EBF \cite{Aljaafari2021}, Mace \cite{monshizadeh2014mace}, MoScan \cite{Wei2021}, NAUTILUS \cite{DBLP:conf/uss/DengZLL00YW23}, NAVEX \cite{DBLP:conf/uss/AlhuzaliGEV18}, RAT \cite{Amouei2022}, Revealer \cite{DBLP:conf/sp/LiuZM21}, Robin \cite{DBLP:journals/corr/abs-2007-06629}, Scanner++ \cite{10.1145/3517036}, ShoVAT \cite{genge2016shovat}, VAPE-BRIDGE \cite{Vimala2022} \tabularnewline 
\hline
Software And Platform Security: Web \& Mobile Security: Client Side Vulnerabilities And Mitigations & MaliceScript \cite{liu2018malicescript} \tabularnewline 
\hline
Software And Platform Security: Web \& Mobile Security: Server Side Vulnerabilities And Mitigations & Censys \cite{censys15}, PURITY \cite{bozic2015purity}, Pyciuti \cite{muralidharan2023pyciuti}, Robin \cite{DBLP:journals/corr/abs-2007-06629}, SOA-Scanner \cite{antunes2013soa}, TORPEDO \cite{olivo2015detecting}, VERA \cite{blome2013vera}, Vulnet \cite{8922605} \tabularnewline 
\hline
Software And Platform Security: Web \& Mobile Security: Server Side Vulnerabilities And Mitigations: Injection Vulnerabilities & Commix \cite{DBLP:journals/ijisec/StasinopoulosNX19}, FUGIO \cite{DBLP:conf/uss/ParkKJS22}, ObjectMap \cite{DBLP:conf/pci/Koutroumpouchos19}, SerialDetector \cite{DBLP:conf/ndss/ShcherbakovB21} \tabularnewline 
\hline
Software And Platform Security: Web \& Mobile Security: Server Side Vulnerabilities And Mitigations: Injection Vulnerabilities: Command Injection & JCOMIX \cite{stallenberg_jcomix_2019}, NodeXP \cite{ntantogian2021nodexp} \tabularnewline 
\hline
Software And Platform Security: Web \& Mobile Security: Server Side Vulnerabilities And Mitigations: Injection Vulnerabilities: Cross-Site Request Forgery (Csrf) & Deemon \cite{DBLP:conf/ccs/PellegrinoJ0BR17}, Mitch \cite{DBLP:conf/eurosp/CalzavaraCFRT19}, No Name (CSRF) \cite{9357029} \tabularnewline 
\hline
Software And Platform Security: Web \& Mobile Security: Server Side Vulnerabilities And Mitigations: Injection Vulnerabilities: Cross-Site Scripting (Xss) & Black Widow \cite{DBLP:conf/sp/ErikssonPS21}, Chainsaw \cite{DBLP:conf/ccs/AlhuzaliEGV16}, PhpSAFE \cite{DBLP:conf/dsn/NunesFV15}, TChecker \cite{DBLP:conf/ccs/Luo0022}, WAPTT \cite{dhuric2014waptt}, WebFuzz \cite{DBLP:conf/esorics/RooijCKPA21} \tabularnewline 
\hline
Software And Platform Security: Web \& Mobile Security: Server Side Vulnerabilities And Mitigations: Injection Vulnerabilities: Cross-Site Scripting (Xss): Reflected Xss & Link \cite{lee2022link} \tabularnewline 
\hline
Software And Platform Security: Web \& Mobile Security: Server Side Vulnerabilities And Mitigations: Injection Vulnerabilities: Sql-Injection & Chainsaw \cite{DBLP:conf/ccs/AlhuzaliEGV16}, PhpSAFE \cite{DBLP:conf/dsn/NunesFV15}, TChecker \cite{DBLP:conf/ccs/Luo0022}, WAPTT \cite{dhuric2014waptt}, WebVIM \cite{DBLP:conf/iciis/RankothgeRS20} \tabularnewline 
\hline
Software And Platform Security: Web \& Mobile Security: Server Side Vulnerabilities And Mitigations: Injection Vulnerabilities: User Uploaded Files & FUSE \cite{DBLP:conf/ndss/LeeWLS20} \tabularnewline 
\hline
Systems Security: Authentication, Authorisation \& Accountability: Authentication: Passwords & ADaMs \cite{pasquini2021reducing}, GNPassGAN \cite{Yu2022}, NeuralNetworkCracking \cite{melicher2016fast}, NoCrack \cite{Chatterjee2015}, OMEN \cite{durmuth2015omen}, PassGAN \cite{DBLP:conf/acns/HitajGAP19}, PassGPT \cite{rando2023passgpt}, PasswordCrackingTraining \cite{DBLP:conf/esorics/CampiFL22}, SemanticGuesser \cite{veras2014semantic} \tabularnewline 
\hline
Systems Security: Distributed Systems Security & Cairis \cite{Faily2020}, MAL \cite{johnson2018meta}, PenQuest \cite{DBLP:journals/virology/LuhTTSJ20} \tabularnewline 
\hline
\caption{CyBOK classification\label{tab:cybok}}
		\end{longtable}

\clearpage
		\begin{longtable}{|>{\RaggedRight}p{0.50\textwidth}|>{\RaggedRight}p{0.50\textwidth}|}
			\hline 
			\textit{ACM CCS} & \textit{Tools}\tabularnewline
\hline
\endhead
\hline
Hardware: Emerging Technologies: Analysis And Design Of Emerging Devices And Systems: Emerging Architectures & AVAIN \cite{Egert2019}, Diane \cite{redini2021diane}, EBF \cite{Aljaafari2021}, IoTFuzzer \cite{DBLP:conf/ndss/ChenDZZL0LSYZ18}, Mirage \cite{cayre2019mirage}, ROSploit \cite{DBLP:conf/irc/RiveraLS19}, RT-RCT \cite{fagroud2021rt}, Snout \cite{Mikulskis2019} \tabularnewline 
\hline
Human-Centered Computing: Human Computer Interaction (Hci): Interactive Systems And Tools & TAMELESS \cite{DBLP:journals/tdsc/ValenzaKSL23} \tabularnewline 
\hline
Networks: Network Components: Intermediate Nodes: Routers & ESRFuzzer \cite{DBLP:journals/cybersec/ZhangHJSLZZL21} \tabularnewline 
\hline
Networks: Network Protocols: Network Layer Protocols: Routing Protocols & No Name (TTCN-3) \cite{DBLP:conf/isncc/LealT18}, OSV \cite{kasemsuwan2017osv} \tabularnewline 
\hline
Security And Privacy: Cryptography & EBF \cite{Aljaafari2021} \tabularnewline 
\hline
Security And Privacy: Human And Societal Aspects Of Security And Privacy & DFBC \cite{ng2021dfbc} \tabularnewline 
\hline
Security And Privacy: Intrusion/Anomaly Detection And Malware Mitigation: Malware And Its Mitigation & MAIT \cite{yucel2021mait} \tabularnewline 
\hline
Security And Privacy: Network Security & AVAIN \cite{Egert2019}, Bbuzz \cite{DBLP:conf/milcom/BlumbergsV17}, Censys \cite{censys15}, NetCAT \cite{Kurth2020}, No Name (TTCN-3) \cite{DBLP:conf/isncc/LealT18}, OSV \cite{kasemsuwan2017osv}, Pyciuti \cite{muralidharan2023pyciuti}, RT-RCT \cite{fagroud2021rt}, SuperEye \cite{DBLP:conf/icccsec/LiYWLYH19}, Vulcan \cite{kamongi2013vulcan}, Vulnsloit \cite{10.1007/978-3-030-64881-7_6} \tabularnewline 
\hline
Security And Privacy: Network Security: Mobile And Wireless Security & ESRFuzzer \cite{DBLP:journals/cybersec/ZhangHJSLZZL21}, Firmaster \cite{8457340}, LTESniffer \cite{10.1145/3558482.3590196}, Owfuzz \cite{10.1145/3558482.3590174}, Scanner++ \cite{10.1145/3517036}, Snout \cite{Mikulskis2019}, UE Security Reloaded \cite{10.1145/3558482.3590194} \tabularnewline 
\hline
Security And Privacy: Network Security: Security Protocols & HILTI \cite{Sommer2014}, No Name (TTCN-3) \cite{DBLP:conf/isncc/LealT18}, Spicy \cite{DBLP:conf/acsac/SommerAH16} \tabularnewline 
\hline
Security And Privacy: Network Security: Web Protocol Security & Bbuzz \cite{DBLP:conf/milcom/BlumbergsV17} \tabularnewline 
\hline
Security And Privacy: Security Services: Authorization & ADaMs \cite{pasquini2021reducing}, GNPassGAN \cite{Yu2022}, NeuralNetworkCracking \cite{melicher2016fast}, NoCrack \cite{Chatterjee2015}, OMEN \cite{durmuth2015omen}, PassGAN \cite{DBLP:conf/acns/HitajGAP19}, PassGPT \cite{rando2023passgpt}, PasswordCrackingTraining \cite{DBLP:conf/esorics/CampiFL22}, SemanticGuesser \cite{veras2014semantic} \tabularnewline 
\hline
Security And Privacy: Software And Application Security & CryptoGuard \cite{DBLP:conf/ccs/RahamanXASTFKY19}, VulDeePecker \cite{DBLP:conf/ndss/LiZXO0WDZ18} \tabularnewline 
\hline
Security And Privacy: Software And Application Security: Domain-Specific Security And Privacy Architectures & ESSecA \cite{rak2022esseca}, MAL \cite{johnson2018meta}, PenQuest \cite{DBLP:journals/virology/LuhTTSJ20} \tabularnewline 
\hline
Security And Privacy: Software And Application Security: Software Reverse Engineering & RiscyROP \cite{10.1145/3545948.3545997}, VulCNN \cite{10.1145/3510003.3510229} \tabularnewline 
\hline
Security And Privacy: Software And Application Security: Software Security Engineering & AIBugHunter \cite{Fu2023}, Chucky \cite{DBLP:conf/ccs/YamaguchiWGR13}, CuPerFuzzer \cite{DBLP:conf/sp/LiDLDG21}, ELAID \cite{DBLP:journals/cybersec/XuXLH20}, LAID \cite{DBLP:conf/cisc/XuLXLHMLH18}, PJCT \cite{jain2015pjct}, Project Achilles \cite{DBLP:conf/kbse/SaccenteDDCX19}, Untangle \cite{DBLP:conf/dimva/BertaniBBCZP23}, VUDDY \cite{DBLP:conf/sp/KimWLO17}, VulPecker \cite{DBLP:conf/acsac/LiZXJQH16} \tabularnewline 
\hline
Security And Privacy: Software And Application Security: Web Applications Security & Black Ostrich \cite{Eriksson2023}, Black Widow \cite{DBLP:conf/sp/ErikssonPS21}, Censys \cite{censys15}, Chainsaw \cite{DBLP:conf/ccs/AlhuzaliEGV16}, Commix \cite{DBLP:journals/ijisec/StasinopoulosNX19}, Deemon \cite{DBLP:conf/ccs/PellegrinoJ0BR17}, FUGIO \cite{DBLP:conf/uss/ParkKJS22}, FUSE \cite{DBLP:conf/ndss/LeeWLS20}, JCOMIX \cite{stallenberg_jcomix_2019}, Link \cite{lee2022link}, Mace \cite{monshizadeh2014mace}, MaliceScript \cite{liu2018malicescript}, Mitch \cite{DBLP:conf/eurosp/CalzavaraCFRT19}, MoScan \cite{Wei2021}, NAUTILUS \cite{DBLP:conf/uss/DengZLL00YW23}, NAVEX \cite{DBLP:conf/uss/AlhuzaliGEV18}, No Name (CSRF) \cite{9357029}, NodeXP \cite{ntantogian2021nodexp}, ObjectMap \cite{DBLP:conf/pci/Koutroumpouchos19}, PURITY \cite{bozic2015purity}, PhpSAFE \cite{DBLP:conf/dsn/NunesFV15}, Pyciuti \cite{muralidharan2023pyciuti}, RAT \cite{Amouei2022}, Robin \cite{DBLP:journals/corr/abs-2007-06629}, SOA-Scanner \cite{antunes2013soa}, SerialDetector \cite{DBLP:conf/ndss/ShcherbakovB21}, ShoVAT \cite{genge2016shovat}, TChecker \cite{DBLP:conf/ccs/Luo0022}, TORPEDO \cite{olivo2015detecting}, VAPE-BRIDGE \cite{Vimala2022}, VERA \cite{blome2013vera}, Vulnet \cite{8922605}, WAPTT \cite{dhuric2014waptt}, WebFuzz \cite{DBLP:conf/esorics/RooijCKPA21}, WebVIM \cite{DBLP:conf/iciis/RankothgeRS20} \tabularnewline 
\hline
Security And Privacy: Systems Security: Denial Of Service Attacks & Revealer \cite{DBLP:conf/sp/LiuZM21} \tabularnewline 
\hline
Security And Privacy: Systems Security: Distributed Systems Security & MAL \cite{johnson2018meta}, PenQuest \cite{DBLP:journals/virology/LuhTTSJ20} \tabularnewline 
\hline
Security And Privacy: Systems Security: Vulnerability Management: Penetration Testing & Cairis \cite{Faily2020}, Diane \cite{redini2021diane}, ESASCF \cite{Ghanem2023}, ESSecA \cite{rak2022esseca}, Gail-PT \cite{chen2023gail}, HARMer \cite{DBLP:journals/access/EnochHMLAK20}, Lore \cite{DBLP:journals/tdsc/Holm23}, MAL \cite{johnson2018meta}, Mirage \cite{cayre2019mirage}, PenQuest \cite{DBLP:journals/virology/LuhTTSJ20}, PentestGPT \cite{DBLP:journals/corr/abs-2308-06782}, Pyciuti \cite{muralidharan2023pyciuti}, SVED \cite{DBLP:conf/milcom/HolmS16}, TAMELESS \cite{DBLP:journals/tdsc/ValenzaKSL23} \tabularnewline 
\hline
Security And Privacy: Systems Security: Vulnerability Management: Vulnerability Scanners & AIBugHunter \cite{Fu2023}, ARMONY \cite{chen2013armory}, AVAIN \cite{Egert2019}, Autosploit \cite{Moscovich2020}, Black Ostrich \cite{Eriksson2023}, Black Widow \cite{DBLP:conf/sp/ErikssonPS21}, Bleem \cite{Luo2023}, Censys \cite{censys15}, Chainsaw \cite{DBLP:conf/ccs/AlhuzaliEGV16}, Chucky \cite{DBLP:conf/ccs/YamaguchiWGR13}, Commix \cite{DBLP:journals/ijisec/StasinopoulosNX19}, CryptoGuard \cite{DBLP:conf/ccs/RahamanXASTFKY19}, CuPerFuzzer \cite{DBLP:conf/sp/LiDLDG21}, Deemon \cite{DBLP:conf/ccs/PellegrinoJ0BR17}, Delta \cite{lee2017delta}, EBF \cite{Aljaafari2021}, ELAID \cite{DBLP:journals/cybersec/XuXLH20}, ESSecA \cite{rak2022esseca}, FUGIO \cite{DBLP:conf/uss/ParkKJS22}, FUSE \cite{DBLP:conf/ndss/LeeWLS20}, Firmaster \cite{8457340}, HILTI \cite{Sommer2014}, IoTFuzzer \cite{DBLP:conf/ndss/ChenDZZL0LSYZ18}, JCOMIX \cite{stallenberg_jcomix_2019}, LAID \cite{DBLP:conf/cisc/XuLXLHMLH18}, Link \cite{lee2022link}, Mace \cite{monshizadeh2014mace}, MaliceScript \cite{liu2018malicescript}, Masat \cite{mjihil2015masat}, Mitch \cite{DBLP:conf/eurosp/CalzavaraCFRT19}, MoScan \cite{Wei2021}, NAUTILUS \cite{DBLP:conf/uss/DengZLL00YW23}, NAVEX \cite{DBLP:conf/uss/AlhuzaliGEV18}, No Name (CSRF) \cite{9357029}, No Name (TTCN-3) \cite{DBLP:conf/isncc/LealT18}, NodeXP \cite{ntantogian2021nodexp}, OSV \cite{kasemsuwan2017osv}, ObjectMap \cite{DBLP:conf/pci/Koutroumpouchos19}, Owfuzz \cite{10.1145/3558482.3590174}, PJCT \cite{jain2015pjct}, PURITY \cite{bozic2015purity}, PhpSAFE \cite{DBLP:conf/dsn/NunesFV15}, Project Achilles \cite{DBLP:conf/kbse/SaccenteDDCX19}, Pyciuti \cite{muralidharan2023pyciuti}, RAT \cite{Amouei2022}, ROSploit \cite{DBLP:conf/irc/RiveraLS19}, RT-RCT \cite{fagroud2021rt}, Revealer \cite{DBLP:conf/sp/LiuZM21}, RiscyROP \cite{10.1145/3545948.3545997}, Robin \cite{DBLP:journals/corr/abs-2007-06629}, SOA-Scanner \cite{antunes2013soa}, Scanner++ \cite{10.1145/3517036}, SerialDetector \cite{DBLP:conf/ndss/ShcherbakovB21}, ShoVAT \cite{genge2016shovat}, Snout \cite{Mikulskis2019}, Spicy \cite{DBLP:conf/acsac/SommerAH16}, SuperEye \cite{DBLP:conf/icccsec/LiYWLYH19}, TChecker \cite{DBLP:conf/ccs/Luo0022}, TORPEDO \cite{olivo2015detecting}, UE Security Reloaded \cite{10.1145/3558482.3590194}, Untangle \cite{DBLP:conf/dimva/BertaniBBCZP23}, VAPE-BRIDGE \cite{Vimala2022}, VERA \cite{blome2013vera}, VUDDY \cite{DBLP:conf/sp/KimWLO17}, VulCNN \cite{10.1145/3510003.3510229}, VulDeePecker \cite{DBLP:conf/ndss/LiZXO0WDZ18}, VulPecker \cite{DBLP:conf/acsac/LiZXJQH16}, Vulcan \cite{kamongi2013vulcan}, Vulnet \cite{8922605}, Vulnsloit \cite{10.1007/978-3-030-64881-7_6}, WAPTT \cite{dhuric2014waptt}, WebFuzz \cite{DBLP:conf/esorics/RooijCKPA21}, WebVIM \cite{DBLP:conf/iciis/RankothgeRS20} \tabularnewline 
\hline
\caption{ACM CCS classification\label{tab:acm-ccs}}
\end{longtable}

\processifversion{LONG}{
\clearpage
\section{Review of Tools\label{sec:Review-of-Research}}  %
We present a review of the EH tools included in our survey, in alphabetical order according to their names.

\subsection{Reducing bias in modelling real-world password strength via deep learning and dynamic dictionaries}

\texttt{ADaMs} (2021, Pasquini et al. \cite{pasquini2021reducing}) is an advanced tool aimed at improving the precision of modelling real-world password strengths. This tool leverages deep neural networks to simulate the methodology used by adversaries when constructing attack configurations. The efficiency of AdaMs was demonstrated through case studies involving MyHeritage and Youku using the rules-set generated and RockYou as the external dictionary. For MyHeritage, ADaMs' attack matches the precision of the optimal dictionary approach and achieves a comparable number of successful guesses. Similarly, for Youku, Adams outperforms the optimal dictionary in terms of guessing speed, achieving faster results within the initial $(10^{11})$ guesses, demonstrating its effectiveness in real-world scenarios.

\subsection{AIBugHunter: A Practical tool for predicting, classifying and repairing software vulnerabilities}

\texttt{AIBugHunter} (2023, Fu et al. \cite{Fu2023}) is a machine learning-based approach to detecting and mitigating software security bugs within C and C++. This package is integrated into Visual Studio Code to better bridge the gap between real-world developers and academic contributions to vulnerability detection frameworks. The tool was around 10\% to 141\% more effective than the compared baseline models at predicting CWE-ID within evaluated code samples.

\subsection{ARMORY: An automatic security testing tool for buffer overflow defect detection}

\texttt{ARMONY} (2013, Chen et al. \cite{chen2013armory}) is a kernel-based tool for detecting zero-day Buffer Overflow vulnerabilities to prevent the overflowing of sensitive data structures by testing the system and recording the results in a dump file. The components of the software include a coordinator, a checker, and two message storages.

\subsection{Autosploit: A Fully Automated Framework for Evaluating the Exploitability of Security Vulnerabilities}

\texttt{Autosploit} (2020, Moscovich et al. \cite{Moscovich2020}) is a framework for virtualising and evaluating the components within a system that are required for exploitation via a given vector. Although in the paper the authors outline several factors that can affect exploitation, but the simulation agent only supports four actions: stopping services , deleting packages, changing system firewall rules, and modifying file permissions. The tool was tested against a Metasploitable 2 machine, which is designed to be intentionally vulnerable against a set number of attacks, and the attacker agent only supported Metasploit-based attacks, once again limiting what exploits can and cannot be assessed to those supported with a Metasploit module.

\subsection{AVAIN -- A Framework for Automated Vulnerability Indication for the IoT in IP-based Networks}

\texttt{AVAIN} (2019, Egert et al. \cite{Egert2019}) is a framework for Automated Vulnerability Indication in IP-based Internet of Things (IoT) networks. It aims at enhancing security practices of IoT system administrators by automatically deploying existing tools to generate results, presented according to the Common Vulnerability Scoring System (CVSS). The framework's modular design enables the user to integrate arbitrary vulnerability scanners and analysis methods. AVAIN includes four components: the \textit{Controller} processes user instructions to orchestrate the tasks. The \textit{Module Updater} keeps modules up to date, the \textit{Scanner} performs vulnerability scanning, and the \textit{Analyser} processes the collected data.

\subsection{Bbuzz: A bit-aware fuzzing framework for network protocol systematic reverse engineering and analysis}

\texttt{Bbuzz} (2017, Blumbergs et al. \cite{DBLP:conf/milcom/BlumbergsV17}) is an open-source bit-aware network protocol fuzzing framework designed for systematic reverse engineering and analysis of network protocols. Operating at Layer-2, Bbuzz facilitates rapid protocol assessment, automatic test case creation, and user-friendly fuzzing. In a proprietary NATO Link-1 protocol case study, researchers used Bbuzz to efficiently reverse engineer the protocol, revealing critical components within a single day.

\subsection{Black Ostrich: Web Application Scanning with String Solvers}

\texttt{Black Ostrich} (2023, Eriksson et al. \cite{Eriksson2023}) is a tool for crawling web applications with a content-aware approach. The tool is able to differentiate expected input formats such as email addresses, usernames, etc., using regex interpretation to discover more pages on web applications than other crawling approaches. This tool was shown to be effective when compared against existing web application crawlers such as ZAP, Enemy, and Arachni. This tool was shown to achieve 99\% coverage, achieving high effectiveness in crawling capability. The tool also showed a 52\% improvement over existing methods when discovering vulnerable patterns within those same web pages.

\subsection{Black Widow: Blackbox Data-driven Web Scanning}

\texttt{Black Widow} (2021, Eriksson et al. \cite{DBLP:conf/sp/ErikssonPS21}) uses a black box data-driven approach for deep crawling and scanning of modern web apps. It focuses on three core pillars: navigation modelling, traversing, and tracking inter-state dependencies. Black Widow demonstrates significant code coverage improvements compared to other crawlers. Moreover, it excels in vulnerability scanning, detecting more cross-site scripting vulnerabilities without false positives, identifying missed vulnerabilities in older applications, and uncovering new vulnerabilities in production software like HotCRP, osCommerce, PrestaShop, and WordPress.

\subsection{Bleem: Packet Sequence Oriented Fuzzing for Protocol Implementations}

\texttt{Bleem} (2023, Luo et al. \cite{Luo2023}) is a novel black-box fuzzer designed to enhance the vulnerability detection of protocol implementations through packet-sequence-oriented fuzzing. It features a noninvasive feedback mechanism that examines system outputs (packet sequences) to deduce the internal state transitions within the protocol implementation. This feedback guides the fuzzing process, including generating packet sequences that align with the protocol's logic. Bleem significantly outperforms state-of-the-art protocol fuzzers in terms of branch coverage and vulnerability detection. It achieves up to a 175\% increase in branch coverage within 24 hours compared to tools like Peach. Furthermore, Bleem discovered 15 security-critical vulnerabilities across prominent protocol implementations, resulting in 10 CVEs being attributed, showcasing its effectiveness in identifying previously undetected vulnerabilities.

\subsection{Contextualisation of Data Flow Diagrams for Security Analysis}

\texttt{Cairis} (2020, Faily et al. \cite{Faily2020}) is a novel tool for identifying tainted data flows through the contextualisation of Data Flow Diagrams (DFDs) with other models related to usability and requirements. This adaptation of taint analysis, traditionally employed in code analysis to detect insecure data handling, is applied to design-level analysis. The Cairis tool identifies potential taint sources from human interactions or system processes, as represented in DFDs. The practical viability and the possibility of incorporating it into this tool's current security analysis workflows are demonstrated through its implementation in an open-source software platform.

\subsection{A Search Engine Backed by Internet-Wide Scanning}

\texttt{Censys} (2015, Durumeric et al. \cite{censys15}) is a vulnerability scanning tool that leverages data from continuous Internet-wide scans. Censys uses existing tools, specifically ZMap \cite{durumeric2013zmap} and ZGrab, for hosting discovery scans across the IPv4 address spaces and application-layer handshakes. Case studies demonstrate Censys's application in analyzing Industrial Control Systems, vulnerabilities like Heartbleed and SSLv3, and institutional attack surfaces. The application of the tool shows Censys's ability to identify vulnerable devices and networks quickly, generate statistical reports on usage patterns, and provide insights into the security posture of devices across the internet.

\subsection{Chainsaw: Chained Automated Workflow-based Exploit Generation}

\texttt{Chainsaw} (2016, Alhuzali et al. \cite{DBLP:conf/ccs/AlhuzaliEGV16}) offers a solution for automated exploit generation in web applications, surpassing existing methods in identifying and exploiting web injection vulnerabilities. It adeptly handles challenges posed by diverse web app structures, user input, and database back-ends by constructing precise models of application workflows, schemes, and native functions. Evaluated across 9 open-source applications, Chainsaw produced over 199 high-quality first- and second-order injection exploits, showcasing its superiority over comparable approaches in exploit generation for web vulnerabilities.

\subsection{Chucky: exposing missing checks in source code for vulnerability discovery}

\texttt{Chucky} (2013, Yamaguchi et al. \cite{DBLP:conf/ccs/YamaguchiWGR13}) is a tool aimed at expediting the auditing of software by exposing missing checks in source code, particularly focusing on input validation. By using static tainting techniques, Chucky identifies anomalies and omitted security-critical conditions, effectively pinpointing 12 previously undiscovered vulnerabilities in Pidgin and LibTIFF among five popular open-source projects. This method enhances the process of discovering security flaws by highlighting crucial areas of missing input validation checks.

\subsection{Commix: automating evaluation and exploitation of command injection vulnerabilities in Web applications}

\texttt{Commix} (2019, Stasinopoulos et al. \cite{DBLP:journals/ijisec/StasinopoulosNX19}), short for COMMand Injection eXploiter, is an open-source tool aimed at automating the detection and exploitation of command injection flaws in web applications. Commix's methodology encompasses attack vector generation, a vulnerability detection module, and an exploitation module. The vulnerability detection module utilises the generated attack vectors to probe for potential command injection vulnerabilities within target web applications by injecting and attempting command execution. Subsequently, the exploitation module leverages any confirmed vulnerabilities, exploiting them with the successful attack vector to facilitate the execution of arbitrary commands by the attacker. Commix underwent threefold experimentation—in a virtual lab, against other tools, and within real-world applications—demonstrating its efficacy in identifying and exploiting command injection vulnerabilities across these scenarios, thereby demonstrating its significant value in cybersecurity practices.

\subsection{CryptoGuard: High Precision Detection of Cryptographic Vulnerabilities in Massive-sized Java Projects}

\texttt{CryptoGuard} (2019, Rahaman et al. \cite{DBLP:conf/ccs/RahamanXASTFKY19}) is a tool with efficient slicing algorithms in Java programs that reduce false alerts by 76\% to 80\% in cryptographic API misuse threats. Running on large-scale projects like Apache and Android apps, CRYPTOGUARD provided security insights and aided projects like Spark, Ranger, and Ofbiz in code hardening. The tool achieved 98.61\% precision by manually confirming 1,277 true positives out of 1,295 Apache alerts. Additionally, it established a benchmark with basic and advanced cases, extensively comparing with CrySL, SpotBugs, and Coverity.

\subsection{Android Custom Permissions Demystified: From Privilege Escalation to Design Shortcomings}

\texttt{CuPerFuzzer} (2021, Li et al. \cite{DBLP:conf/sp/LiDLDG21}) is an automatic fuzzing tool designed to detect vulnerabilities within the Android OS related to custom permissions. Through extensive testing, it uncovered 2,384 effective cases and identified 30 critical paths, exposing severe design shortcomings within the Android permission framework. These issues include dangling custom permissions, inconsistent permission group mapping, permission elevating, and inconsistent permission definition, potentially allowing malicious apps to gain unauthorised system permissions.

\subsection{Deemon: Detecting CSRF with Dynamic Analysis and Property Graphs}

\texttt{Deemon} (2017, Pellegrino et al. \cite{DBLP:conf/ccs/PellegrinoJ0BR17}) introduces an automated security testing framework for CSRF vulnerability detection. Utilising a novel modelling paradigm capturing web app aspects in a unified property graph, it autonomously constructs models from dynamic traces and identifies potentially vulnerable operations via graph traversals. Deemon's validation approach conducts security tests, and 14 previously unknown CSRF vulnerabilities across 10 open-source web applications have been discovered, showing its efficacy in detecting exploitable issues threatening user accounts and entire websites.

\subsection{Delta: A security assessment framework for software-defined networks.}

\texttt{Delta} (2017, Lee et al. \cite{lee2017delta}) is a novel vulnerability analysis tool that focuses specifically on Software Defined Networking (SDN) and utilizes known published attacks. The tool uses a fuzzing module that automatically detects zero-day vulnerabilities and has been designed specifically for OpenFlow controller platforms.

\subsection{DFBC Recon Tool: Digital Footprint and Breach Check Reconnaissance Tool}

\texttt{DFBC} (2021, Yusof et al. \cite{ng2021dfbc}) is a reconnaissance tool that provides a Digital Footprint and Breach Check. This software can extract user information that is publicly available and check the breach activity status for accounts with high speed. The tool has both a CLI and GUI and focuses on gathering data on social networks such as Facebook and X/Twitter, as well as email breaches.

\subsection{Diane: Identifying fuzzing triggers in apps to generate under-constrained inputs for iot devices}

\texttt{Diane} (2021, Redini et al. \cite{redini2021diane}) is a tool for vulnerability analysis in IoT targeting Android applications used to control IoT devices, at the network level and UI level by detecting fuzzing triggers. The software has been tested and has successfully identified nine zero-day vulnerabilities. This tool can be applied to discover vulnerabilities in Smart Home appliances such as door Smart Locks. Given its reliance on dynamic analysis, Diane lacks the capability to detect fuzzing triggers that are not executed by the application. Additionally, it is unable to perform fuzzing on nested Java objects.

\subsection{Finding Security Vulnerabilities in IoT Cryptographic Protocol and Concurrent Implementations}

\texttt{EBF} (2021, Aljaafari et al. \cite{Aljaafari2021}) is a tool used for the discovery of vulnerabilities within IoT devices. It uses static and dynamic analysis to discover issues surrounding memory safety, race conditions, thread leak and arithmetic overflow. The tool simulates a server and a client in order to detect such vulnerabilities within the implemented protocol. The tool demonstrated its effectiveness by detecting a previously unknown race condition bug within WOLFMQTT in only 15 minutes with 22 MB memory consumption. The tool was also verified against pre-existing vulnerabilities within OpenSSL and successfully identified the issues as expected.

\subsection{ELAID: detecting integer-Overflow-to-Buffer-Overflow vulnerabilities by light-weight and accurate static analysis}

\texttt{ELAID} (2020, Xu et al. \cite{DBLP:journals/cybersec/XuXLH20}) (Enhanced Lightweight and Accurate method of static IO2BO vulnerability Detection) is a tool designed for detecting Integer-Overflow-to-Buffer-Overflow (IO2BO) vulnerabilities. Built on LLVM \cite{Lattner2004}, ELAID's indirect call analysis aims to eliminate false positives. It has been tested on NIST's SAMATE Juliet 1.2 suite \cite{black2018juliet} and other real-world applications, demonstrating the ability to detect 152 vulnerabilities without false positives. However, the authors acknowledge the need for a combined symbolic execution and fuzzing approach to enhance practicality.

\subsection{ESASCF: Expertise Extraction, Generalization and Reply Framework for Optimized Automation of Network Security Compliance}

\texttt{ESASCF} (2023, Ghanem et al. \cite{Ghanem2023}) is a tool designed to address the resource-intensity and repetitiveness of the network security auditing process by automating the extraction, processing, storage, and reuse of expertise in similar scenarios or during periodic re-testing. ESASCF leverages industrial and open-source vulnerability assessment and penetration testing tools to automate the security compliance (SC) process. It is designed to autonomously handle SC re-testing, offloading human experts from repetitive SC segments, allowing them to focus on more critical tasks in ad-hoc compliance tests. The framework was tested on networks of different sizes, demonstrating time efficiency and testing effectiveness improvements. Specifically, ESASCF significantly reduces the time required for an expert to complete the first security compliance of typical corporate networks by 50\% and 20\% in re-testing scenarios.

\subsection{ESRFuzzer: an enhanced fuzzing framework for physical SOHO router devices to discover multi-Type vulnerabilities}

\texttt{ESRFuzzer} (2021, Zhang et al. \cite{DBLP:journals/cybersec/ZhangHJSLZZL21}) (Enhanced SOHO Router Fuzzing Framework) is as a vulnerability discovery tool tailored for small office and home office (SOHO) routers. It operates through an automated FWSR fuzzing framework, incorporating KEY-VALUE and CONF-READ semantic models with power management for testing environment recovery. Equipped with diverse mutation rules and monitoring mechanisms, the tool efficiently identifies various vulnerability types. ESRFuzzer excels in discovering CONF and READ operation issues, particularly in general and D-CONF modes. Testing on 10 routers revealed 136 issues, with 120 of them confirmed as 0-day vulnerabilities.

\texttt{ESRFuzzer} (2021, Zhang et al. \cite{DBLP:journals/cybersec/ZhangHJSLZZL21}) (Enhanced SOHO Router Fuzzing Framework) is a vulnerability discovery tool for small office and home office (SOHO) routers, employing an automated FWSR fuzzing framework. It uses KEY-VALUE and CONF-READ semantic models with power management for testing environment recovery. With diverse mutation rules and monitoring mechanisms, it identifies various vulnerability types. In general and D-CONF modes, ESRFuzzer excels in discovering CONF and READ operation issues. Testing on 10 routers unveiled 136 issues, 120 confirmed as 0-day vulnerabilities.

\subsection{ESSecA: An automated expert system for threat modelling and penetration testing for IoT ecosystems}

\texttt{ESSecA} (2022, Rak et al. \cite{rak2022esseca}) is an autonomous system for Threat Modelling. It employs an algorithm that correlates attacks to threats in the penetration testing of IoT systems using a threat catalogue as a query table. This approach involves mapping threats to attacks and filtering out those that are not relevant. The tool is designed with a modular architecture and produces two main outputs: a threat model and a testing plan table. It utilises databases, such as Cyber Threat Intelligence (CTI), to enhance its functionality.

\subsection{Firmaster: Analysis Tool for Home Router Firmware}

\texttt{Firmaster} (2018, Visoottiviseth et al. \cite{8457340}) is a tool designed to analyse home router firmware source code vulnerabilities. By emulating router firmware, Firmaster identifies, evaluates, and simulates potential vulnerabilities. Its functions include Password Cracking, SSL Scanning, and Web Static Analysis, addressing the OWASP Top 10 2014 IoT vulnerabilities. The tool validates the uploaded firmware, attempts to crack root passwords, and verifies secure connections. Furthermore, it conducts static and dynamic web analysis to identify PHP source code vulnerabilities.

\subsection{FUGIO: Automatic Exploit Generation for PHP Object Injection Vulnerabilities}

\texttt{FUGIO} (2022, Park et al. \cite{DBLP:conf/uss/ParkKJS22}) is a tool addressing PHP Object Injection (POI) vulnerabilities, enabling automatic exploit generation. It uses static and dynamic analyses to create gadget chains, serving as exploit blueprints. By running fuzzing campaigns, FUGIO successfully generated exploit objects, producing 68 exploits from 30 vulnerable applications without false positives. Additionally, it uncovered two previously unreported POI vulnerabilities and created five functional exploits, showing its effectiveness in automatic exploit creation for POI vulnerabilities.

\subsection{FUSE: Finding File Upload Bugs via Penetration Testing}

\texttt{FUSE} (2020, Lee et al. \cite{DBLP:conf/ndss/LeeWLS20}) is a penetration testing tool specifically created to uncover Unrestricted File Upload (UFU) and Unrestricted Executable File Upload (UEFU) vulnerabilities in PHP-based server-side web applications. It effectively generates exploit payloads via upload requests, overcoming content-filtering checks while preserving file execution semantics. FUSE identified 30 previously unreported UEFU vulnerabilities, including 15 CVEs, across 33 real-world web applications.

\subsection{GAIL-PT: An intelligent penetration testing framework with generative adversarial imitation learning}

\texttt{Gail-PT} (2023, Chen et al. \cite{chen2023gail}) as a framework to automate penetration testing. The tool provides advice to the penetration testers for enhanced decision making reducing the reliance on manual testing whilst utilising Generative Adversarial Imitation Learning (GAIL) based on Reinforcement Learning (RL) and Deep Reinforcement Learning (DRL) that innovates the testing process. The tool has been observed to achieve state-of-the-art results when applied to the Metasploitable2 penetration testing target VM.

\subsection{GNPassGAN: Improved Generative Adversarial Networks For Trawling Offline Password Guessing}

\texttt{GNPassGAN} (2022, Yu et al. \cite{Yu2022}) is designed to enhance offline password guessing by leveraging generative adversarial networks (GANs) to generate password guesses by training on datasets obtained from real-world breaches. The tool is benchmarked against PassGAN and other password-guessing methods, using datasets like Rockyou and phpbb for training and evaluation. GNPassGAN demonstrates a significant improvement over PassGAN, achieving 88.03\% more accurate password guesses and producing 31.69\% fewer duplicates when generating a large number of guesses ($10^{8}$).

\subsection{HARMer: Cyber-Attacks Automation and Evaluation}

\texttt{HARMer} (2020, Enoch et al. \cite{DBLP:journals/access/EnochHMLAK20}) is an automation framework for cyber-attack generation, that addresses limitations in manual attack execution by red teams. Using the Hierarchical Attack Representation Model (HARM), it outlines requirements, key phases, and security metrics-based attack planning strategies. Through experiments conducted in both enterprise networks and Amazon Web Services, HARMer demonstrates effective modelling of attackers' operations. This framework offers automated assessment capabilities, enabling security administrators to evaluate diverse threats and attacks systematically, facilitating a more efficient defence against cyber threats in networked systems.

\subsection{HILTI: an Abstract Execution Environment for Deep, Stateful Network Traffic Analysis}

\texttt{HILTI} (2014, Sommer et al. \cite{Sommer2014}) is a tool that employs an abstract machine model specifically designed for deep stateful network traffic analysis. It addresses the significant gap between the high-level conceptualisation of network analysis tasks (such as pattern searching in HTTP requests) and the intricate low-level implementation details required to execute these tasks efficiently and securely. The system facilitates the development of network traffic analysis applications by providing built-in support for common data types, state management, concurrency models, and a secure memory model. This foundation enables developers to create robust applications capable of handling the vast and varied data flows in network traffic, all while operating within real-time performance constraints.

\subsection{IoTFuzzer: Discovering Memory Corruptions in IoT Through App-based Fuzzing}

\texttt{IoTFuzzer} (2018, Chen et al. \cite{DBLP:conf/ndss/ChenDZZL0LSYZ18}) is a vulnerability scanning tool designed to detect memory corruption vulnerabilities in IoT devices. It utilises an app-based fuzzing framework that operates without the need for access to the devices' firmware images. The process involves UI analysis to identify network event triggers and data-flow analysis to monitor the movement of fields related to the application protocol. By altering the original fields, it generates fuzzing messages. The effectiveness of IoTFuzzer is demonstrated through experiments, where it successfully uncovered 15 serious vulnerabilities in 9 out of 17 evaluated IoT devices. These vulnerabilities range from stack-based and heap-based buffer overflows to null pointer de-references and unidentified crashes, highlighting the framework’s capability to reveal critical security flaws.

\subsection{JCOMIX: a search-based tool to detect XML injection vulnerabilities in web applications}

\texttt{JCOMIX} (2019, Stallenberg et al. \cite{stallenberg_jcomix_2019}) is a tool developed in Java that is designed to generate attacks (test objectives) to identify XML injection threats in web applications. It assesses information sanitation and validation in micro-service frameworks, aiding in the identification of vulnerabilities.

\subsection{A Light-Weight and Accurate Method of Static Integer-Overflow-to-Buffer-Overflow Vulnerability Detection}

\texttt{LAID} (2018, Hu et al. \cite{DBLP:conf/cisc/XuLXLHMLH18}) is framework designed to accurately identify potential Integer-Overflow-to-Buffer-Overflow (IO2BO) vulnerabilities in software. The framework combines inter-procedural dataflow analysis and taint analysis to identify potential vulnerabilities and employs a lightweight method for further filtering out false positives. %
The framework's effectiveness was evaluated using NIST’s SAMATE Juliet test suite \cite{black2018juliet} and six known IO2BO vulnerabilities in real-world applications. The framework effectively detected all known IO2BO vulnerabilities in the test suite without any false positives.

\subsection{Link: Black-box detection of cross-site scripting vulnerabilities using reinforcement learning}

\texttt{Link} (2022, Lee et al. \cite{lee2022link}) is an autonomous black-box web scanner that operates without any input from humans, using Reinforcement Learning (RL). This software has been observed to decrease the quantity of attack attempts and finds more true positives with fewer showing as false. Particular success has been observed in XSS vulnerability detection, which can be leveraged in penetration testing.

\subsection{Lore a Red Team Emulation Tool}

\texttt{Lore} (2023, Holm et al. \cite{DBLP:journals/tdsc/Holm23}) is a red team emulation tool utilising boolean logic and trained models for automated red team actions, avoiding manual approaches in cyber defence exercises. Empirical tests demonstrate its model accuracy, achieving twice the compromised machines compared to expert-defined models and five times more than random action selection. Lore's unique approach enhances red team automation, offering a more engaging and educational experience in cyber defence simulations.

\subsection{LTESniffer: An Open-Source LTE Downlink/Uplink Eavesdropper}

\texttt{LTESniffer} (2023, Hoang et al. \cite{10.1145/3558482.3590196}) is designed to capture both uplink and downlink LTE traffic passively. Its architecture encompasses key components such as the conversion of analogue signals to digital samples, identification of modulation schemes and other radio configurations, and data channel decoding, which processes uplink and downlink signals according to their configurations. The results demonstrate that LTESniffer significantly surpasses existing tools and commercial sniffers, achieving a success rate more than twice that of AirScope, particularly in decoding LTE packets within high-throughput scenarios and LTE-A (Advanced) environments.

\subsection{Mace: Detecting privilege escalation vulnerabilities in web applications}

\texttt{Mace} (2014, Monshizadeh et al. \cite{monshizadeh2014mace}) is an automatic privilege escalation vulnerability analysis tool for web applications. It analyses large code bases to discover zero-day vulnerabilities by observing inconsistencies in the authorisation context and understanding flaws within the application using fundamental abstractions. Implementing the tool has been observed to save weeks of labour-intensive work for security professionals in penetration testing.

\subsection{MAIT: Malware Analysis and Intelligence Tool}

\texttt{MAIT} (2021, Yucel et al.\cite{yucel2021mait}) (Malware Analysis and Intelligence Tool) utilises state-of-the-art static and dynamic malware analysers alongside open-source malware databases to generate malware signatures and intelligence reports. The tool offers chronological data for malicious files, revealing related vulnerabilities and providing insights into attribution, techniques, tactics, and procedures employed by Advanced Persistent Threat groups in attacks.

\subsection{A meta-language for threat modelling and attack simulations}

\texttt{MAL} (2018, Johnson et al. \cite{johnson2018meta}) is a tool designed to facilitate the creation of domain-specific attack languages for cybersecurity threat modelling and attack simulations. MAL enables the semi-automated generation and efficient computation of large attack graphs, distinguishing it from traditional, manual attack graph constructions.

\subsection{Malicescript: A novel browser-based intranet threat}

\texttt{MaliceScript} (2018, Liu et al. \cite{liu2018malicescript}) is a tool that introduces a novel browser-based Web attack model allowing browsers to collect intranet topology and infiltrate websites from within. The tool is designed to exploit vulnerabilities, insert foreground JavaScript code into malicious web pages, and monitor intranet topology to ensure successful infiltration. Its significance lies in its potential ease of deployment and the challenges associated with detection, emphasising the need for proactive security measures.

\subsection{Masat: Model-based automated security assessment tool for cloud computing}

\texttt{MASAT} (2015, Mjihil et al. \cite{mjihil2015masat}) (Model-based Automated Security Assessment Tool) is a tool designed to address security issues of cloud computing. Focused on adaptive security assessments, MASAT utilises distributed agents to evaluate the security of virtual machines at different virtualisation levels. These agents employ vulnerability scanners, representation tools for attack models like graphs, and a communication mechanism to share analysis results. MASAT's contributions include comprehensive security assessments for both the cloud infrastructure and virtual machines, efficient task distribution among agents, and parallelised subsystem assessments to reduce analysis time.

\subsection{Mirage: towards a Metasploit-like framework for IoT}

\texttt{Mirage} (2019, Cayre et al. \cite{cayre2019mirage}) is an open-source attack framework applied to IoT by exploiting wireless communication protocols such as Bluetooth Classic/BLE, Zigbee, Enhanced ShockBurst, Mosart, and Wi-Fi. The software is modular and has the potential for future expansion with additional functionality, which is useful for the development of new wireless protocols. The tool was tested using a Smart Bulb, demonstrating success in assessing the attack surface and reverse-engineering the wireless protocol.

\subsection{Mitch: A Machine Learning Approach to the Black-Box Detection of CSRF Vulnerabilities}

The study by \texttt{Mitch} (2019, Calzavara et al. \cite{DBLP:conf/eurosp/CalzavaraCFRT19}) introduced a browser extension that leverages supervised machine learning to identify Cross-Site Request Forgery (CSRF) vulnerabilities. This extension features an automated system that spots sensitive HTTP requests which need CSRF protection for enhanced security. It was trained using nearly 6,000 HTTP requests from popular websites, enabling it to surpass the effectiveness of existing detection methods. This advancement was evidenced by finding three CSRF vulnerabilities in production software that had previously gone unnoticed by the most advanced tools available.

\subsection{MoScan: a model-based vulnerability scanner for web single sign-on services}

\texttt{MoScan} (2021, Wei et al. \cite{Wei2021}) is a vulnerability scanning tool for identifying security vulnerabilities in Single Sign-On (SSO) implementations through model-based scanning. By capturing network traces during the execution of SSO services, MoScan incrementally constructs and refines the state machine. This refined state machine enables MoScan to generate specific payloads for testing protocol participants, aiming to identify security vulnerabilities. MoScan's adaptability is demonstrated by testing it against other SSO services, such as Twitter, LinkedIn, and GitHub's authentication plugin in Jenkins. Despite minor adjustments needed for parameter names, MoScan's primary state machine required minor modifications to accommodate different implementations of the OAuth 2.0 standard.

\subsection{NAUTILUS: Automated RESTful API Vulnerability Detection}

\texttt{NAUTILUS} (2023, Deng et al.~\cite{DBLP:conf/uss/DengZLL00YW23}) is proposed to improve RESTful API vulnerability scanning by addressing limitations in existing black box scanners. It utilises a novel annotation strategy to identify proper operation relations and generate meaningful sequences for vulnerability detection. Compared to four state-of-the-art tools, NAUTILUS demonstrates superior performance, detecting 141\% more vulnerabilities on average and covering 104\% more API operations across six tested services. In real-world scenarios, NAUTILUS detected 23 unique 0-day vulnerabilities, including a remote code execution flaw in Atlassian Confluence and high-risk issues in Microsoft Azure.

\subsection{NAVEX: Precise and Scalable Exploit Generation for Dynamic Web Applications}

\texttt{NAVEX} (2018, Alhuzali et al. \cite{DBLP:conf/uss/AlhuzaliGEV18}) is a tool addressing complex vulnerabilities in multi-tier web applications by combining dynamic and static analysis. It integrates both techniques to automatically detect vulnerabilities and create functional exploits. Evaluated over 3.2 million lines of PHP code, NAVEX successfully identified and created 204 exploits, showcasing its scalability and effectiveness in vulnerability analysis and exploit generation for large-scale applications.

\subsection{NetCAT: Practical Cache Attacks from the Network}

\texttt{NetCAT} (2020, Kurth et al. \cite{Kurth2020}) is a tool designed to exploit Data-Direct I/O (DDIO) to observe and manipulate Last-Level Cache (LLC) states, thereby leaking sensitive information from a remote target server without requiring local access or execution privileges. NetCAT reverse engineers the behaviour of DDIO on Intel processors and develops a network-based PRIME+PROBE cache attack technique by crafting specific network packets to manipulate and observe changes in the LLC's state. The efficiency of NetCAT is demonstrated in various attack scenarios, including creating covert channels between network clients and executing keystroke timing attacks to infer sensitive information from encrypted SSH sessions.

\subsection{Fast, lean, and accurate: modelling password guessability using neural networks}

\texttt{NeuralNetworkCracking} (2016, Melicher et al. \cite{melicher2016fast}) is a tool designed to predict the guessability of passwords accurately and efficiently in real-time. By harnessing the power of neural networks for generating sequential data, the model utilizes an Artificial Neural Network (ANN) to anticipate the next character in a password sequence, thereby estimating the password's guessability. The effectiveness of the ANN model was benchmarked against traditional password guessability models, where it demonstrated superior performance across various scenarios. Notably, the neural networks were able to guess 70\% of 4class8 passwords within 10151015 guesses, significantly outperforming the next best guessing method, which guessed 57\%.

\subsection{Identification and Mitigation Tool For Cross-Site Request Forgery (CSRF)}

\texttt{No Name (CSRF)} (2020, Rankothge and Randeniya \cite{9357029}) focuses on the automated detection and mitigation of Cross-Site Request Forgery (CSRF) vulnerabilities in PHP-based web applications \cite{9357029}. It identifies and mitigates CSRF vulnerabilities by scanning form tags and automatically adding security solutions. However, the evaluation is limited to a few test websites, and the tool is applicable exclusively to PHP.

\subsection{Development of a suite of IPv6 vulnerability scanning tests using the TTCN-3 language}

The \texttt{No Name (TTCN-3)} tool, (2018, Leal and Teixeira \cite{DBLP:conf/isncc/LealT18}), is a collection of tests designed to find security weaknesses in IPv6 setups. It uses TTCN-3, a language for defining and executing tests that is endorsed by the European Telecommunications Standards Institute. The tool was evaluated using the SAMATE Juliet 1.2 suite from NIST \cite{black2018juliet} and proved its effectiveness by identifying and exploiting flaws, particularly focusing on preventing DOS attacks on the IPv6 ICMPv6 protocol over Ethernet networks.

\subsection{Cracking-Resistant Password Vaults Using Natural Language Encoders}

\texttt{NoCrack} (2015, Chatterjee et al. \cite{Chatterjee2015}) is a password-cracking tool that overcomes the limitations of a previous design, Kamouflage \cite{bojinov2010kamouflage}, which was shown to degrade security. NoCrack utilizes Natural Language Encoder (NLE) as a scheme for secure encoding. This approach utilises natural language processing (n-gram models) with probabilistic context-free grammars, to construct NLEs. These encoders can create realistic decoy passwords, thereby significantly enhancing the resistance of password vaults against cracking efforts.

\subsection{NodeXP: NOde. js server-side JavaScript injection vulnerability DEtection and eXPloitation}

\texttt{NodeXP} (2021, Ntantogian et al. \cite{ntantogian2021nodexp}) is a tool addressing security vulnerabilities in web applications, particularly Server-Side JavaScript Injection (SSJI) threats in Node.js. It detects and automatically exploits SSJI vulnerabilities, using obfuscation for enhanced stealth. NodeXP employs dynamic analysis with result and blind-based injection techniques for detection and automated exploitation. Its advanced functionalities, including attack vector obfuscation, distinguish it from other tools, allowing evasion of filtering mechanisms and security measures. Thorough assessments show NodeXP outperforming existing scanners. Released as open-source, it aims to drive research in SSJI vulnerabilities. Contributions include SSJI analysis, a novel detection method, and the discovery of a 0-day SSJI.

\subsection{ObjectMap: detecting insecure object deserialization}

\texttt{ObjectMap} (2019, Koutroumpouchos et al. \cite{DBLP:conf/pci/Koutroumpouchos19}) is a tool designed to address serialization-based vulnerabilities prevalent in web applications, particularly in Java and PHP. It aims to fill the existing gap by detecting implementation-agnostic deserialization and object injection vulnerabilities. Additionally, it introduces the first deserialization test environment, serving as a platform for vulnerability detection tool evaluation and educational purposes. Both tools are highly extendable and represent a unique combination of features in this domain, potentially fostering further research and aiding the development of more comprehensive solutions.

\subsection{OMEN: Faster password guessing using an ordered Markov enumerator}

\texttt{OMEN} (2015, Durmuth et al. \cite{durmuth2015omen}) employs an advanced Markov model algorithm that generates passwords in order of decreasing probability, a departure from earlier approaches \cite{narayanan2005fast}. This method involves discretising probabilities into bins and iterating over them to identify all passwords corresponding to each bin's probability. OMEN has shown a significant improvement in guessing speed over existing methods, accurately guessing over 40\% of passwords within the first 90 million attempts, marking a notable efficiency gain against other methods such as John the Ripper (JtR) and probabilistic context-free grammar (PCFG)-based strategies \cite{weir2009password}.

\subsection{OSV: OSPF vulnerability checking tool}

\texttt{OSV} (2017, Kasemsuwan et al. \cite{kasemsuwan2017osv}) is a tool that targets OSPF (Open Shortest Path First) network vulnerabilities prevalent in enterprise networks. Typically, OSPF vulnerabilities in router implementations can be mitigated through firmware updates. The tool conducts penetration tests and generates reports, aiding network operators in identifying and rectifying security issues. The tool has been validated on Quagga and Cisco routers.

\subsection{Owfuzz: Discovering Wi-Fi Flaws in Modern Devices through Over-The-Air Fuzzing}

\texttt{Owfuzz} (2023, Cao et al. \cite{10.1145/3558482.3590174}) is a fuzzing tool designed for discovering security flaws in Wi-Fi protocols through over-the-air fuzzing methods. Owfuzz sets itself apart from other Wi-Fi fuzzers by offering the capability to conduct fuzzing tests on any Wi-Fi device, enabling the fuzzing of all three Wi-Fi frame types (management, control, and data) across every version of the 802.11 standards, and facilitating interactive testing for a variety of protocol models. In experiments conducted on over 40 contemporary Wi-Fi devices from 7 chipset manufacturers, Owfuzz identified 23 security issues, resulting in the assignment of 8 CVE IDs.

\subsection{PassGAN: A Deep Learning Approach for Password Guessing}

\texttt{PassGAN} (2019, Hitaj et al. \cite{DBLP:conf/acns/HitajGAP19}) utilises a Generative Adversarial Network (GAN) trained on real-world datasets to make accurate password guesses. By autonomously learning the distribution of real passwords from actual leaks, PassGAN eliminates the need for manual rule creation. When combined with HashCat, PassGAN's output successfully matches 51\%-73\% more passwords than HashCat alone. This demonstrates the tool's capacity to automatically extract nuanced password properties not encompassed by current state-of-the-art rules.

\subsection{PassGPT: Password modelling and (Guided) Generation with Large Language Models}

\texttt{PassGPT} (2023, Rando et al. \cite{rando2023passgpt}) is a tool utilizing GPT-2 architecture and trained on leaked passwords, aiming to improve password guessing and strength estimation. PassGPT incorporates vector quantization to enhance the complexity of password generation (PassVQT). Comprehensive tests were conducted to evaluate PassGPT's efficacy against current password-guessing tools and to assess its generalization capabilities across various datasets. PassGPT recovers 41.9\% of the test set among 109109 guesses, whereas state-of-the-art GAN models matched 23.33\%.

\subsection{The Revenge of Password Crackers: Automated Training of Password Cracking Tools}

\texttt{PasswordCrackingTraining} (2022, Di Campi et al. \cite{DBLP:conf/esorics/CampiFL22}) is a password-cracking trainer combining various password-cracking techniques, trained and tested on a dataset of over 700 million real passwords. Their methodology includes evaluating existing hashcat rules and dictionaries and developing efficient algorithms to simulate mask attacks without actual password cracking. This approach nearly doubles the success rate of password-cracking tools compared to off-the-shelf configurations, achieving over 70\% success in certain instances.

\subsection{PenQuest: a gamified attacker/defender meta model for cyber security assessment and education}

\texttt{PenQuest} (2020, Luh et al. \cite{DBLP:journals/virology/LuhTTSJ20}) is a tool designed as a dynamic, multiplayer game that captures the behaviour of attackers and defenders over time, using a rule set based on established information security sources such as STIX CAPEC, CVE/CWE, and NIST SP 800-53. The tool is designed to improve cybersecurity risk assessments and serve as a platform for simulating specific attack scenarios within an abstracted IT infrastructure. PenQuest is structured around three main layers: the service layer, the information layer, and the event layer, each contributing to a comprehensive representation of a cybersecurity scenario.

\subsection{PentestGPT: An LLM-empowered Automatic Penetration Testing Tool}

\texttt{PentestGPT} (2023, Deng et al. \cite{DBLP:journals/corr/abs-2308-06782}) utilises Large Language Models (LLMs) for penetration testing, revealing their proficiency in specific tasks but struggles in holistic understanding. Based on LLM-powered automatic penetration testing tool, PENTESTGPT  has been released with three modules addressing distinct sub-tasks. Evaluation demonstrates PentestGPT's 228.6\% improvement over GPT-3.5 and effectiveness in real-world challenges. The tool employs Reasoning, Generation, and Parsing Modules, simulating human-like behaviour and adopting a divide-and-conquer problem-solving approach.

\subsection{phpSAFE: A Security Analysis Tool for OOP Web Application Plugins}

\texttt{PhpSAFE} (2015, Nunes et al. \cite{DBLP:conf/dsn/NunesFV15}) is introduced as a solution to identify security vulnerabilities in PHP plugins developed with Object-Oriented Programming (OOP) for web applications. In contrast to existing free tools that lack OOP support, PhpSAFE excels in static code analysis, outperforming two well-known tools when evaluated with 35 plugins in a popular Content Management System. The results highlight the prevalence of vulnerabilities in these plugins, indicating an increasing trend over time, emphasizing the necessity of robust security measures in plugin development.

\subsection{PJCT: Penetration testing based JAVA code testing tool}

\texttt{PJCT}  (2015, Jain et al.  \cite{jain2015pjct}) is a tool designed to secure attributes in Java code. This addresses a crucial aspect of secure software development while emphasizing early detection of vulnerabilities during the development process. PJCT highlights seven essential security attributes for identifying vulnerabilities in Java applications. These attributes include network and security packages, techniques for handling exceptions, secure data packet transmission and multithreading.

\subsection{Project Achilles: A Prototype Tool for Static Method-Level Vulnerability Detection of Java Source Code Using a Recurrent Neural Network}

\texttt{Project Achilles} (2019, Saccente et al. \cite{DBLP:conf/kbse/SaccenteDDCX19}) is a prototype tool for static method-level vulnerability detection in Java source code, leveraging LSTM Recurrent Neural Networks. The tool utilizes NIST's Juliet Java Suite \cite{black2018juliet}, which includes several examples of defective Java methods for various vulnerabilities. Employing an array of LSTM networks, the tool achieves over 90\% accuracy for 24 out of 29 Common Weakness Enumeration (CWE) vulnerabilities in an evaluation with around 45,000 test cases.

\subsection{PURITY: a Planning-based secURITY testing tool}

\texttt{PURITY} (2015, Bozic and Wotawa \cite{bozic2015purity}) is a security testing tool that employs a planning-centric methodology to preemptively identify and rectify vulnerabilities during the software development cycle. By executing automated test cases, PURITY focuses on detecting prevalent security flaws like SQL injections and cross-site scripting, simulating malicious activities through predetermined sequences of actions. PURITY generates concrete test cases from plans based on specific initial values and predefined actions, mimicking the behaviours of potential attackers, providing a versatile platform for both automated and manual evaluation of web applications throughout the software development life cycle.

\subsection{Pyciuti: A Python Based Customizable and Flexible Cybersecurity Utility Tool for Penetration Testing}

\texttt{Pyciuti} (2023, Muralidharan et al. \cite{muralidharan2023pyciuti}) is a Python-based general-purpose tool that integrates custom scanners, crawlers, malware functionalities, and more, offering flexibility and customisation for both beginners and experienced professionals. Users can access various subdomains such as OSINT, network-based tools, web-based tools, malware tools, and documentation, streamlining the execution of tasks. The tool records findings, scans, and exploits, providing a consolidated report while prioritising accuracy and performance.

\subsection{RAT: Reinforcement-Learning-Driven and Adaptive Testing for Vulnerability Discovery in Web Application Firewalls}

\texttt{RAT} (2022, Amouei et al. \cite{Amouei2022}) (Reinforcement-Learning-Driven and Adaptive Testing) is a tool for discovering vulnerabilities, specifically SQL injection (SQLi) and Cross-site Scripting (XSS), in Web Application Firewalls (WAFs). The tool's methodology begins with clustering similar attack samples, followed by applying a reinforcement learning technique that efficiently identifies bypassing attack patterns. The approach is further refined by integrating an adaptive search algorithm, which aids in discovering almost all possible bypassing payloads with higher efficiency. RAT outperforms existing methods by significant margins (33.53\% and 63.16\% on average) when tested against well-configured WAFs.

\subsection{Revealer: Detecting and Exploiting Regular Expression Denial-of-Service Vulnerabilities}

\texttt{Revealer} (2021, Liu et al. \cite{DBLP:conf/sp/LiuZM21}) is a tool designed to detect and exploit Regular expression Denial-of-Service (ReDoS) vulnerabilities present in extended-featured regular expressions (regexes). Revealer employs a combined method using static and dynamic analyses to identify vulnerable regex structures and generate attack strings that induce recursive backtracking. In evaluation against 29,088 regexes and comparison with three state-of-the-art tools, Revealer outperformed existing solutions, detecting all known vulnerabilities, finding 213 new vulnerabilities whilst surpassing the highest performing tool by 140.64\%. Additionally, it detected 45 vulnerable regexes in real-world applications, demonstrating its effectiveness and efficiency in detecting and exploiting ReDoS vulnerabilities.

\subsection{RiscyROP: Automated Return-Oriented Programming Attacks on RISC-V and ARM64}

\texttt{RiscyROP} (2022, Cloosters et al. \cite{10.1145/3545948.3545997}) is an automated Return-Oriented Programming (ROP) tool specifically designed for RISC-V and ARM64 architectures. It employs symbolic execution to analyze available gadgets and autonomously generate complex multi-stage chains for arbitrary function calls. The tool's effectiveness is evidenced by its analysis of real-world software from public repositories, demonstrating its capability to identify usable gadgets for executing attacker-controlled function calls.

\subsection{Robin: A Web Security Tool}

\texttt{Robin} (2020, Girotto and Zorzo \cite{DBLP:journals/corr/abs-2007-06629}) is a web security tool that includes a Proxy module for listing, intercepting, and editing HTTP and HTTPS requests. Robin's capabilities extend to active scanning, brute-force attack simulations, encoding/decoding contents using various hashing patterns, and providing a detailed Wiki module for understanding and safeguarding against common vulnerabilities. The tool's potential is demonstrated through a real case study involving a news company's website.

\subsection{ROSploit: Cybersecurity Tool for ROS}

\texttt{ROSploit} (2019, Rivera et al. \cite{DBLP:conf/irc/RiveraLS19}) is a tool for the assessment of the security of the Robot Operating System (ROS). The tool covers reconnaissance and exploitation. Its reconnaissance component integrates with Nmap, offering two scripts for master node scans and wide port scans, identifying ROS nodes and services. The exploitation component, built in Python, mirrors Metasploit's modular design. Moreover, ROSploit enables simulated attacks without a complete ROS installation.

\subsection{RT-RCT: an online tool for real-time retrieval of connected things}

\texttt{RT-RCT} (2021, Fagroud et al. \cite{fagroud2021rt}) utilises network port scanning techniques \cite{pale2015mastering} to extract real-time data from connected devices within the IoT domain. The main goal is to introduce a retrieval tool that swiftly provides users with current information about each requested entity. Real-time device data is collected through network port scanning, with a specific emphasis on the Python-nmap library. The data collection process involves a series of scans, each targeting specific information that may take considerable time to retrieve. To reduce time, parallel scans are implemented, ensuring simultaneous and accelerated execution of all scans to improve overall retrieval speed and efficiency.

\subsection{Scanner++: Enhanced Vulnerability Detection of Web Applications with Attack Intent Synchronization}

\texttt{Scanner++} (2023, Yin et al. \cite{10.1145/3517036}) demonstrates an enhancement of vulnerability assessment scanning technology in web applications. By utilizing purification mechanisms to refine attack intents from the request packets in the base scanners. The scanner has a synchronisation mechanism for runtime intent for the scanner's detection spots. This has been evaluated against four practitioner tools 
including BurpSuite, AWVS, Arachni, and ZAP testing on real-world finance web apps where the tool demonstrated to have a higher coverage when being compared to the practitioner tools by approximately 15\%-70\%.

\subsection{On the Semantic Patterns of Passwords and their Security Impact}

\texttt{SemanticGuesser} (2014, Veras et al. \cite{veras2014semantic}) is a framework that uses Natural Language Processing (NLP) techniques for the segmentation, semantic classification, and generalisation of passwords. The methodology begins with password segmentation to isolate distinct words or elements found within the passwords. Following this, it employs part-of-speech tagging and semantic classification through NLP methodologies, primarily utilising resources such as WordNet to decipher the meanings of the segments. Experimental results have shown that this method can guess about 67\% more passwords from the LinkedIn data breach and 32\% more from the MySpace leak, demonstrating its effectiveness in understanding and exploiting the semantic structures of passwords.

\subsection{SerialDetector: Principled and Practical Exploration of Object Injection Vulnerabilities for the Web}

\texttt{SerialDetector} (2021, Shcherbakov et al. \cite{DBLP:conf/ndss/ShcherbakovB21}) employs taint-based dataflow analysis to automatically detect OIV (Object Injection Vulnerability) patterns in .NET assemblies. By identifying untrusted information flow from public entry points to sensitive methods, it uncovers vulnerabilities and matches them with available gadgets, validating the feasibility of OIV attacks. The evaluation conducted on Azure DevOps Server showcased SerialDetector's effectiveness by discovering three CVEs and demonstrating its capability in identifying remote code execution vulnerabilities. Additionally, it performed a broad security analysis of recent CVEs, affirming its efficiency and effectiveness in OIV detection.

\subsection{ShoVAT: Shodan-based vulnerability assessment tool for Internet-facing services}

\texttt{ShoVAT} (2016, Genge et al. \cite{genge2016shovat}) a vulnerability analysis tool based on Shodan. This software acquires the output of traditional queries in Shodan to analyse service-specific data by leveraging the search engine. By adding to existing modules this tool provides the opportunity to analyse other targets in the penetration test.

\subsection{Snout: An Extensible IoT Pen-Testing Tool}

\texttt{Snout} (2019, Mikulskis et al. \cite{Mikulskis2019}) is a Software-Defined Radio (SDR)-based utility toolkit for passive sniffing and interaction with various IoT protocols, including Zigbee, Bluetooth, and Wi-Fi. Available as a Python 3 package or as a Docker container, Snout enables enumeration and analysis of multiple wireless protocols, including non-IP IoT protocols.

\subsection{SOA-Scanner: an integrated tool to detect vulnerabilities in service-based infrastructures}

\texttt{SOA-Scanner} (2013, Antunes et al. \cite{antunes2013soa}) is a vulnerability analysis tool designed for service-oriented architectures (SOA) after the application has been deployed. The tool tests services based on their level of access using an iterative process to discover services, resources, and interactions in real-time. Additionally, it provides anomaly detection, categorising services based on whether they are within reach, under full control, or under partial control.

\subsection{Spicy: a unified deep packet inspection framework for safely dissecting all your data}

\texttt{Spicy} (2016, Sommer et al. \cite{DBLP:conf/acsac/SommerAH16}) includes a format specification language, a compiler toolchain, and an API to address the challenges of deep packet inspection (DPI) across diverse network protocols and file formats. The Spicy framework automates the dissection process, enhancing DPI efficiency and safety. Enabling developers to create specific dissectors for varied protocols, Spicy is a reliable DPI tool. Its flexibility makes it valuable for processing network data from untrusted sources in diverse formats.

\subsection{SuperEye: A Distributed Port Scanning System}

\texttt{SuperEye} (2019, Li et al. \cite{DBLP:conf/icccsec/LiYWLYH19}) is an advanced distributed port scanning system adopting a distributed structure with task redundancy and real-time state display. SuperEye's core control subsystem efficiently manages distributed nodes, tasks, and result processing, leveraging a custom protocol stack to optimise resource utilisation. The distributed architecture significantly boosts scanning speed, mitigating risks associated with Intrusion Detection Systems (IDS). Innovative features include a script for port scanning and visualisation tools offering real-time updates.

\subsection{SVED: Scanning, Vulnerabilities, Exploits and Detection}

\texttt{SVED} (2016, Holm et al. \cite{DBLP:conf/milcom/HolmS16}) is a tool designed for secure and replicable experiments, enabling controlled execution and logging of malicious activities, including software exploits and intrusion detection alerts. Its distributed architecture supports extensive experiments involving numerous attackers, sensors, and targets. SVED automatically integrates threat intelligence from diverse services, ensuring up-to-date information for enhanced experimentation and analysis in cybersecurity.

\subsection{A Hybrid Threat Model for Smart Systems}

\texttt{TAMELESS} (2023, Valenza et al. \cite{DBLP:journals/tdsc/ValenzaKSL23}) is an automated threat modelling tool that includes a threat modelling approach integrating cyber, physical, and human elements, and a threat analysis method designed to evaluate the security posture of system components. TAMELESS can analyse threats, verify security properties, and produce graphical outputs of its analyses, thereby assisting security architects in identifying optimal prevention and mitigation solutions. The efficiency and applicability of TAMELESS have been demonstrated through case study evaluations involving unauthorised access to safe boxes, web servers, and wind farms, showcasing its effectiveness in real-world scenarios.

\subsection{TChecker: Precise Static Inter-Procedural Analysis for Detecting Taint-Style Vulnerabilities in PHP Applications}

\texttt{TChecker} (2022, Luo et al. \cite{DBLP:conf/ccs/Luo0022}) introduces a context-sensitive inter-procedural static taint analysis tool specifically tailored for PHP applications, addressing the challenge of taint-style vulnerabilities like SQL injection and cross-site scripting. By modelling PHP objects and dynamic language features, TChecker conducts iterative data-flow analysis to refine object types and accurately identify call targets. Comprehensive evaluations across diverse modern PHP applications showcased TChecker's effectiveness, discovering 18 previously unknown vulnerabilities while outperforming existing static analysis tools in vulnerability detection. It not only detected more vulnerabilities but also maintained a relatively good precision, surpassing competitors while releasing its source code to foster further research in this domain.

\subsection{Detecting and exploiting second order denial-of-service vulnerabilities in web applications}

\texttt{TORPEDO} (2015, Olivo et al. \cite{olivo2015detecting}) is a second-order vulnerability scanning tool that detects Denial of Service (DoS), Cross-Site Scripting (XSS), and SQL Injection. The program searches for two-phased DoS attacks that work by polluting a database with junk entries and resource exhaustion. When applied to six highly used web apps, it detected thirty-seven vulnerabilities and eighteen false positives.

\subsection{UE Security Reloaded: Developing a 5G Standalone User-Side Security Testing Framework}

\texttt{UE Security Reloaded} (2023, Hoang et al. \cite{10.1145/3558482.3590194}) is an open-source security testing framework specifically developed for 5G Standalone (SA) User Equipment (UE). This tool enhances existing open-source suites (Open5GS and srsRAN) by creating an extensive range of test cases for both the 5G Non-Access Stratum (NAS) and Radio Resource Control (RRC) layers. Such an approach offers in-depth insights through experiments on 5G SA mobile phones. The framework allows for the transmission of 5G control-plane messages (NAS and RRC) to a UE and facilitates the modification of these messages to examine the UE's reactions under a variety of conditions.

\subsection{Untangle: Aiding Global Function Pointer Hijacking for Post-CET Binary Exploitation}

\texttt{Untangle} (2023, Bertani et al. \cite{DBLP:conf/dimva/BertaniBBCZP23}) is a tool
that exploit global function pointer hijacking in order to defeat Intel’s Control-Flow Enforcement Technology (CET) The method combines symbolic execution and static code analysis to identify global function pointers within C libraries, which, when compromised, facilitate control-flow hijacking attacks. Experimental results demonstrated the effectiveness of Untangle in identifying global function pointers across eight widely used open-source C libraries.

\subsection{VAPE-BRIDGE: Bridging OpenVAS Results for Automating Metasploit Framework}

\texttt{VAPE-BRIDGE} (2022, Vimala et al. \cite{Vimala2022}) is a tool designed to streamline the transition between vulnerability assessment (VA) and penetration testing (PenTest) processes by automating the conversion of scan results from the Open Vulnerability Assessment Scanner (OpenVAS) into executable scripts for the Metasploit Framework. The VAPE-BRIDGE system comprises three main components: Scan result extraction, responsible for parsing the VA scan results from OpenVAS; Target list repository, accountable for maintaining a database of identified vulnerabilities to be used in the PenTest process; and the Automated shell scripts exploitation, which generates shell scripts based on the extracted vulnerabilities, which are then executed within Metasploit to simulate attacks and test the system's resilience. 

\subsection{Vera: A flexible model-based vulnerability testing tool}

\texttt{VERA} (2013, Blome et al. \cite{blome2013vera}) is an automated tool that supports Penetration Testers to define attacker models (separating payloads and behaviour) using state machines for vulnerability analysis in web applications. The models acquired are then converted into libraries for specific vulnerability targeting. The tool is highly flexible with the availability to expand and integrate custom libraries to enhance functionality.

\subsection{VUDDY: A Scalable Approach for Vulnerable Code Clone Discovery}

\texttt{VUDDY} (2017, Kim et al. \cite{DBLP:conf/sp/KimWLO17}) aims to detect defective code in large open-source programs. In particular, it is capable to process a billion lines of code in under 15 hours and quickly identify code clones using function-level granularity and a length-filtering technique. 
The evaluation included comparison with four other code clone detection methods and VUDDY showed better scalability and accuracy. For example, it was possible to find zero-day vulnerabilities popular software like Apache Web Server and Ubuntu Linux OS.

\subsection{Vulcan: Vulnerability assessment framework for cloud computing}

\texttt{Vulcan} (2013, Kamongi et al. \cite{kamongi2013vulcan}) is a tool for vulnerability analysis and remediation in cloud and mobile computing. This tool provides software and zero-day vulnerability modelling and assessments. The tool is very flexible, presenting the opportunity to add original modules by developers and the integration of Vulcan into other vulnerability analysis tools that, for example, focus on web application vulnerabilities to expand their assessment to cloud and mobile technology.

\subsection{VulCNN: An Image-Inspired Scalable Vulnerability Detection System}

\texttt{VulCNN} (2022,(Wu et al. \cite{10.1145/3510003.3510229}) is designed to address the limitations of existing text-based and graph-based vulnerability detection methods. The tool converts the source code of functions into images that preserve program details and then uses these images to detect vulnerabilities through a Convolutional Neural Network (CNN) model. VulCNN was evaluated on a dataset of 13,687 vulnerable and 26,970 non-vulnerable functions. With an accuracy of 82\% and a True Positive Rate (TPR) of 94\%, VulCNN outperformed eight other state-of-the-art vulnerability detectors, including both commercial tools and deep learning-based approaches.

\subsection{VulDeePecker: A Deep Learning-Based System for Vulnerability Detection}

\texttt{VulDeePecker} (2018, Zhen et al. \cite{DBLP:conf/ndss/LiZXO0WDZ18}) employs deep learning for software vulnerability detection, aiming to reduce reliance on human-defined features and mitigate false negatives. It uses code gadgets to represent and transform programs into vectors suitable for deep learning. The system's evaluation, using the first vulnerability dataset for deep learning, demonstrates significantly fewer false negatives compared to other methods, with reasonable false positives. VulDeePecker successfully detects four previously unreported vulnerabilities in Xen, Seamonkey, and Libav, unnoticed by other detection systems, highlighting its efficacy in uncovering vulnerabilities missed by existing approaches.

\subsection{An Intelligent and Automated WCMS Vulnerability-Discovery Tool: The Current State of the Web}

\texttt{Vulnet} (2019, Cigoj and Blazic \cite{8922605}) is characterized by its capability to conduct automated, rapid, and dynamic vulnerability scans across a wide array of internet websites. Specifically, VulNet focuses on those utilising the WordPress Web Content Management Systems (WCMS) and its associated plugins. A crucial aspect of the tool involves the application of a scoring mechanism tailored to evaluate known vulnerabilities. It's important to note that VulNet's vulnerability detection is limited to WordPress web applications and their associated plugins.

\subsection{Vulsploit: A Module for Semi-automatic Exploitation of Vulnerabilities}

\texttt{Vulnsloit} (2020, Castiglione et al. \cite{10.1007/978-3-030-64881-7_6}) is a semi-automatic penetration testing tool that collects vulnerability data using existing tools like the Nmap Scripting Engine (NSE) and the Vulscan scanner \cite{pale2015mastering}. This data is then processed to identify relevant exploits from various repositories, including local and remote sources. In preliminary testing on Metasploitable2, Vulnsloit identified 23 open ports and approximately 220,000 vulnerabilities.

\subsection{VulPecker: an automated vulnerability detection system based on code similarity analysis}

\texttt{VulPecker} (2016, Li et al. \cite{DBLP:conf/acsac/LiZXJQH16}) automatically detects specific vulnerabilities within software source code. Leveraging a set of defined features characterizing patches and utilizing code-similarity algorithms tailored for different vulnerability types, VulPecker successfully identifies 40 vulnerabilities not listed in the National Vulnerability Database (NVD). Among these, 18 previously unknown vulnerabilities (anonymized for ethical considerations) are confirmed, while 22 vulnerabilities have been patched silently by vendors in later product releases.

\subsection{WAPTT-Web application penetration testing tool}

\texttt{WAPTT} (2014, Duric et al. \cite{dhuric2014waptt}) is designed for web application penetration testing using page similarity detection. The structure of the tool is modular and when compared to tools such as Nikto, Vega, and ZAP, the tool detected similar or greater quantity of vulnerabilities in the area of XSS, but at the price of increasing detection time.

\subsection{webFuzz: Grey-Box Fuzzing for Web Applications}

\texttt{WebFuzz} (2021, van Rooij et al. \cite{DBLP:conf/esorics/RooijCKPA21}) is a gray-box fuzzing tool designed for web applications, with a focus on discovering vulnerabilities like cross-site scripting (XSS). It effectively utilises instrumentation, outperforming black-box fuzzers by identifying XSS vulnerabilities swiftly and covering more code. WebFuzz has demonstrated its capability by discovering one zero-day vulnerability in WordPress and five in CE-Phoenix.

\subsection{Identification and Mitigation Tool for Sql Injection Attacks (SQLIA)}

\texttt{WebVIM} (2020, Rankothge et al. \cite{DBLP:conf/iciis/RankothgeRS20}) is a tool designed for identifying SQL injection vulnerabilities in PHP-based web applications during the development phase. When vulnerabilities are detected, WebVIM automatically adds security solutions to the source code. However, the evaluation is very limited, and the tool focuses exclusively on PHP applications.

}
\newpage
\bibliographystyle{elsarticle-num}
\bibliography{literature}
\end{document}